\documentclass[aps,secnumarabic,nobalancelastpage,amsmath,amssymb,nofootinbib]{revtex4}

\usepackage{hyperref,graphicx}
\usepackage{dcolumn}
\usepackage{bm}
\usepackage{bbm}
\usepackage{bbold} 
\usepackage{tikz}
\usepackage{physics}
\usepackage{enumitem}
\usepackage{amssymb,amsfonts}
\usepackage{epsfig}
\usepackage{epstopdf}
\usepackage[all]{xy}
\usepackage{amsthm}
\usepackage{dcolumn}
\usepackage{hyperref}
\usepackage{url}
\usepackage{dsfont}
\usepackage{slashed}
\usepackage{mathrsfs}
\usepackage{soul}
\usepackage{float}
\DeclareMathAlphabet{\mathpzc}{OT1}{pzc}{m}{it}

\newcommand{\be}{\begin{equation}}
\newcommand{\ee}{\end{equation}}
\newcommand{\bea}{\begin{eqnarray}}

\newcommand{\eea}{\end{eqnarray}}

\newcommand{\vsi}{\varsigma}

\newcommand{\ds}{\displaystyle}

\def\inbar{\,\vrule height1.5ex width.4pt depth0pt}
\def\IR{\relax{\rm I\kern-.18em R}}
\def\IC{\relax\hbox{$\inbar\kern-.3em{\rm C}$}}

\def\H{\mathbb{H}}
\def\R{\mathbb{R}}
\def\N{\mathbb{N}}
\def\C{\mathbb{C}}
\def\Z{\mathbb{Z}}

\def\vrh{\varrho}

\def\ii{\mathrm{i}}
\def\ud{\mathrm{d}}

\def\pz{\pmb{z}}
\def\pr{\pmb{r}}

\def\bpz{\overline{\pmb{z}}}
\def\px{\pmb{x}}
\def\py{\pmb{y}}

\def\bu{\mathbbm{1}}

\def\C{{\mathbb C}}
\def\E{{\mathcal E}}

\def\N{{\mathbb N}}
\def\R{{\mathbb R}}
\def\U{{\mathcal U}}
\def\Z{{\mathbb Z}}

\def\a{{\alpha}}
\def\b{{\beta}}

\def\w{{\omega}}
\def\var{{\varepsilon}}
\def\av{{\mathbf a}}
\def\xv{{\mathbf x}}
\def\vv{{\mathbf v}}
\def\vq{{\mathbf q}}
\def\vp{{\mathbf p}}
\def\vk{{\mathbf k}}
\def\vz{{\mathbf z}}
\def\vj{{\mathbf j}}
\def\Pv{{\mathbf P}}
\def\Kv{{\mathbf K}}
\def\Jv{{\mathbf J}}

\definecolor{hervecolor}{rgb}{0.8,0,0.7}

\begin{document}

\title{Anti-de Sitterian ``massive" elementary systems\\and their Minkowskian and Newton-Hooke contraction limits}

\author{Mohammad Enayati$^1$}

\author{Jean-Pierre Gazeau$^{2}$\footnote{gazeau@apc.in2p3.fr}}

\author{Mariano A. del Olmo$^3$\footnote{marianoantonio.olmo@uva.es}}

\author{Hamed Pejhan$^{4}$\footnote{pejhan@math.bas.bg}}

\affiliation{$^1$Department of Physics, Razi University, Kermanshah 6741414971, Iran\\
$^2$Universit\'e Paris Cit\'{e}, CNRS, Astroparticule et Cosmologie, F-75013 Paris, France\\
$^3$Departamento de F\'{\i}sica Te\'orica and IMUVA, Universidad de Valladolid, E-47011, Valladolid, Spain\\
$^4$Institute of Mathematics and Informatics, Bulgarian Academy of Sciences, Acad. G. Bonchev Str. Bl. 8, 1113, Sofia, Bulgaria}

\date{\today}

\begin{abstract}
We elaborate the definition and properties of ``massive" elementary systems in the $(1+3)$-dimensional Anti-de Sitter (AdS$_4$) spacetime, on both classical and quantum levels. We fully exploit the symmetry group {isomorphic to}  Sp$(4,\R)$, that is, the two-fold covering of SO$_0(2,3)$ (Sp$(4,\R) \sim$ SO$_0(2,3)\times \Z_2$), recognized as the relativity/kinematical group of motions in AdS$_4$ spacetime. In particular, we discuss that the group coset Sp$(4,\R)/\mathrm{S}\big(\mathrm{U}(1)\times \mathrm{SU}(2)\big)$, as one of the Cartan classical domains, can be interpreted as a phase space for the set of free motions of a test massive particle on AdS$_4$ spacetime; technically, in order to facilitate the computations, the whole process is carried out in terms of complex quaternions. The (projective) unitary irreducible representations (UIRs) of the Sp$(4,\R)$ group, describing the quantum version of such motions, are found in the discrete series of the Sp$(4,\R)$ UIRs. We also describe the null-curvature (Poincar\'{e}) and non-relativistic (Newton-Hooke) contraction limits of such systems, on both classical and quantum levels. On this basis, we unveil the dual nature of ``massive" elementary systems living in AdS$_4$ spacetime, as each being a combination of a Minkowskian-like elementary system {with positive proper mass}, with an isotropic harmonic oscillator arising from the AdS$_4$ curvature and viewed as a Newton-Hooke elementary system. This matter-vibration duality will take its whole importance in the quantum regime (in the context of the validity of the equipartition theorem) in view of its possible r\^{o}le in the explanation of the current existence of dark matter.
\end{abstract}

\maketitle

\setcounter{equation}{0} \section{Introduction}
For a physical system, like a ``free" elementary system living in (A)dS$_4$ spacetime \cite{de Sitter}, for which the global and local symmetries of its classical phase space are respectively given by a Lie group $G$ and its Lie algebra $\mathfrak{g}$, the phase space can be naturally identified with an orbit under the co-adjoint action of $G$ in the dual linear space to $\mathfrak{g}$ (denoted here by $\mathfrak{g}^\circledast_{}$); such orbits (simply, say co-adjoint orbits) are symplectic manifolds so that each of them carries a natural $G$-invariant (Liouville) measure, and is a homogeneous space homeomorphic to an even-dimensional group coset $G/{\cal{S}}$, where ${\cal{S}}$, being a (closed) subgroup of $G$, stabilizes some orbit point \cite{Kirillov1, Kirillov2}. On the other hand, co-adjoint orbits, enjoying very rich analytic structures, \emph{underlie}\footnote{Here, one may consider a comprehensive program of quantization of functions (or distributions) by using all resources of covariant integral quantization as it is defined, for instance, in Refs. \cite{GazeauWiley,aagbook13,bergaz14,gazeauAP16,gazmur16}.} (projective) Hilbert spaces carrying UIRs of the respective symmetry group, here referred to as $G$. In the sense that was first brought up by Wigner \cite{Wigner, Newton/Wigner} in the context of Poincar\'{e}-Einstein relativity and then developed by In\"{o}n\"{u} \cite{Wigner1952}, L\'{e}vy-Leblond \cite{Levy-Leblond}, and Voisin \cite{Voisin} to Galilean systems, and by G\"{u}rsey \cite{Gursey1963} and Fronsdal \cite{Fronsdal 1, Fronsdal 2} respectively to dS$_4$ and AdS$_4$ systems, the (projective) Hilbert spaces identify (in some restricted sense) the corresponding quantum (``one-particle") states spaces of the given systems with the symmetry $G$; then, the invariant parameters labeling the (projective) UIRs would represent the basic quantum numbers labeling the states of the respective physical systems. Note that, in such a formulation, a ``smooth" transition from classical physics to quantum physics is guaranteed by construction.

In the above sense, we intend in this paper to elaborate the definition and properties of elementary systems in AdS$_4$ spacetime, on both classical and quantum levels. This work is indeed the natural continuation of a previous article \cite{olmogaz19} by del Olmo and Gazeau, which was devoted to such a study in the $(1+1)$-dimensional Anti-de Sitter (AdS$_2$) spacetime. In a broader picture, however, these two papers are both parts of a series of books/papers serving a wider research plan (see Ref. \cite{Gazeau2022} and references therein) that attempts to develop a consistent formulation (in both classical and quantum senses) of elementary systems in the global structure of dS$_4$ and AdS$_4$ spacetimes, or in other words, to depict (A)dS$_4$ relativity versus Poincar\'{e}-Einstein relativity. The main motive behind this attempt originates from a set of conceptual considerations/worries that we now briefly explain.

First of all, one must notice that both field theoretical formulation and phenomenological treatment of an elementary system, on the level of interpretation in particular, rest on the concepts of energy, momentum, mass, and spin, whose existence is due to invariance principles, strictly speaking, the principle of invariance under the Poincar\'{e} group (the relativity group of flat Minkowski spacetime); the rest mass $m$ and the spin $s$ of a (Minkowskian) elementary system living in flat Minkowski spacetime are the two invariants that specify the respective UIR of the Poincar\'{e} group \cite{Wigner, Newton/Wigner}.

In a curved spacetime, however, any interpretation with reference to the relativity group of \emph{flat} Minkowski spacetime is physically irrelevant. On the other hand, in curved spacetimes generally (with the exception of dS$_4$ and AdS$_4$ spacetimes)\footnote{Note that, trivially, this exception holds for any $1+n$-dimensional dS and AdS spacetimes ($n=1,2,...$), but, since in the current paper we are interested in $(1+3)$ dimension, we merely point out dS$_4$ and AdS$_4$ spacetimes throughout the paper.}, \emph{no} non-trivial groups of motion, and consequently, \emph{no} literal or unique extension of the aforementioned physical concepts exists. Therefore, although, one may simply generalize the important differential equations (such as Klein-Gordon and Dirac equations) to forms that enjoy general covariance in curved spacetimes, in the end, such mathematical constructions cannot be associated with physical elementary systems in the sense given in the previous paragraph.

Nevertheless, the dS$_4$ and AdS$_4$ cases constitute a particular family of curved spacetimes, in which the path to generalizations of the aforementioned concepts is well marked, in the sense given by Fronsdal \cite{Fronsdal 1}: \emph{``A physical theory that treats spacetime as Minkowskian flat must be obtainable as a well-defined limit of a more general physical theory, for which the assumption of flatness is not essential"}. The dS$_4$ and AdS$_4$ are indeed the two curved spacetimes of constant curvature (respectively, of negative and positive curvatures), which like Minkowski (zero curvature) spacetime, admit continuous groups of motion of maximal symmetry such that, as Minkowski spacetime is the zero-curvature limit of dS$_4$ and AdS$_4$ spacetimes, the Poincar\'{e} group can be realized by a contraction limit of either the dS$_4$ relativity group SO$_0(1,4)$ or the AdS$_4$ one SO$_0(2,3)$. On the representation level, and quite similar to the Poincar\'{e} case, the (A)dS$_4$ UIRs are labeled by two invariant parameters of the spin and energy scales (the latter, in the AdS$_4$ case, is actually the rest energy) \cite{evans67, Fronsdal 2,baelgagi92, Gazeau2022, Dobrev1, Dobrev2, 1Thomas, 2Newton, 3Takahashi, 4Dixmier}. From the point of view of a local (``tangent") Minkowskian observer, the (A)dS$_4$ UIRs fall basically into three sets: the set of (A)dS$_4$ ``massive" UIRs, in the sense that they contract to the Poincar\'{e} massive ($m^2 >0$) UIRs and exhaust the whole set of the latter \cite{evans67,5Mickelsson,6Garidi}; the set of (A)dS$_4$ ``massless" UIRs constituting by those (A)dS$_4$ UIRs with a unique extension to the conformal group (SO$_0(2,4)$) UIRs, while that extension is equivalent to the conformal extension of the Poincar\'{e} massless UIRs (of course, this correspondence exhaust the whole set of the Poincar\'{e} massless UIRs) \cite{7Barut,8Mack,9Angelopoulos}; and finally, the set of those (A)dS$_4$ UIRs with either \emph{no} physical Poincar\'{e} contraction limit or \emph{no} (nontrivial) Poincar\'{e} contraction limit at all.\footnote{In the context of the latter category, specifically in the AdS$_4$ scenario pertinent to our study, there exist particular UIRs of the universal covering group of SO$_0(2, 3)$, which contract to massive yet tachyonic ($m^2<0$) representations of the Poincar\'{e} group \cite{evans67, Ehrman}. These representations are classified as type $III_{|m|}$ according to the scheme outlined by Angelopoulos and Laoues in Ref. \cite{Laoues}. Owing to their tachyonic nature and consequently the acausal propagation of their corresponding fields, we categorize them within this latter category, representing AdS$_4$ UIRs devoid of a physical Poincar\'{e} contraction limit. {Moreover, the present work solely focuses on the physical Wigner massive representations of the Poincar\'e group (with positive mass and positive energy). These representations are demonstrated to be contraction limits of the (extended) holomorphic discrete series of the AdS group and its universal covering.} }

Considering the above and following our goal of developing the relativity of (A)dS$_4$, we here invoke the Sp$(4,\R)$ group, which is {isomorphic to} the two-fold covering of SO$_0(2,3)$ and can also be interpreted as the kinematical/relativity group for AdS$_4$ spacetime. We first, in section \ref{Sec. AdS geom+ gr}, present all needed material about AdS$_4$ spacetime and its relativity group Sp$(4,\R)$. We also elaborate the geometry of the domain $\mathcal{D}^{(3)}$, that is, a Sp$(4,\R)$ left coset issued from the Cartan decomposition of the group, with the group action on it in the usual manner; $\mathcal{D}^{(3)} \sim \mathrm{Sp}(4,\R)/\mathrm{S}(\mathrm{U}(2)\times\mathrm{U}(1))$. This discussion is complemented by studying the K\"{a}hlerian structure of the Cartan domain $\mathcal{D}^{(3)}$. Then, in section \ref{Sec lieAdS}, we show that this domain can be viewed as the phase space for the set of free motions of a test massive particle living in AdS$_4$ spacetime. The identification of the ten basic (classical) observables with the Sp$(4,\R)$ generators and of course the Poincar\'{e} and the Newton-Hooke contraction limits of this AdS$_4$ phase-space structure are given. In section \ref{Sec AdS4 reps}, we introduce the discrete series (in a wide sense) of the Sp$(4,\R)$ representations which act on the (Segal-)Bargmann-Fock Hilbert spaces of holomorphic functions in the Cartan domain $\mathcal{D}^{(3)}$. The corresponding infinitesimal generators, as first-order differential operators acting on the holomorphic functions in $\mathcal{D}^{(3)}$, are also presented. On the quantum level, the Poincar\'{e} and the Newton-Hooke contraction limits (in the Wigner-In\"{o}n\"{u} sense \cite{Wigner1952} on the respective Lie algebra level, and on the Mickelsson-Niederle \cite{5Mickelsson} and Dooley-Rice \cite{dooley83, dooley-rice83, dooley-rice85, renaud96, aagbook13} sense on the UIR Lie group level) of the representations are discussed as well. Our results reveal the dual nature of any ``massive" elementary system in AdS$_4$ as a combination of a Minkowskian-like massive elementary system with an isotropic harmonic oscillator due to the curvature and viewed as a {Newton-Hooke} elementary system. Finally, we summarize our results in section \ref{Sec. Conclusion}. {The isomorphism between the two-fold covering  of SO$_0(2,3)$ and  Sp$(4,\R)$ is explained in Appendix \ref{appendix:sp4R}.} A reminder of the Poincar\'{e} and Newton-Hooke groups and their respective UIRs, relevant to the present work, is given in Appendices \ref{appendix:WigPoin} and \ref{appendix:Newton-Hooke}. Further technical details of the presented mathematical material are also developed in the other appendices.

\setcounter{equation}{0} 
\section{AdS$_4$ spacetime and its relativity group} \label{Sec. AdS geom+ gr}
AdS$_4$ spacetime is the unique maximally symmetric solution to the vacuum Einstein's equations with negative cosmological constant $\Lambda$. This constant is linked to the (constant) Ricci curvature $4 \Lambda$ of this spacetime. In this context, usually three \emph{equivalent} fundamental/universal concepts are taken into account in the literature: a fundamental length $\ell_{\Lambda} \equiv \sqrt{3/\vert\Lambda \vert}$; a universal frequency $\omega_{\Lambda}=c/\ell_{\Lambda}$ ($c$ is the speed of light); and finally, a universal positive curvature $\varkappa = \sqrt{\vert\Lambda \vert/3}$. The most convenient way to visualize the AdS$_4$ manifold is to embed it into $\R^{2,3}$. It then appears as the pseudo-sphere:
\begin{eqnarray}\label{adsps}
\mathrm{AdS}_4 \equiv \Big\{y=(y^\alpha) \in \R^{2,3}\;;\; (y)^2 \equiv \eta_{\alpha \beta}y^{\alpha}y^{\beta} &=& (y^5)^2 + (y^0)^2 - (y^1)^2 - (y^2)^2 - (y^3)^2 \nonumber\\
&=& (y^5)^2 + (y^0)^2 - \Vert \py\Vert^2 = {\varkappa}^{-2} \Big\}\,,
\end{eqnarray}
where $\eta_{\alpha\beta}=\mathrm{diag}(+1,+1,-1,-1,-1)$ and the indices $\alpha,\beta,\dotsc$ run on the values $\{5,0,1,2,3\}$.\footnote{Note that, in the sequel, the Minkowski indices $\mu,\nu,\dotsc$ will be reserved to the subset $\{0,1,2,3\}$, the spatial indices $i,j,\dotsc$ to the subset $\{1,2,3\}$, and the indices $a,b,\dotsc$ to the subset $\{1,2,3,5\}$.} [The number `$4$' is traditionally left out for possible extensions towards conformal theories.] Note that: (i) The embedding Eq. \eqref{adsps} makes manifest the O$(2,3)$ symmetry group of AdS$_4$ spacetime. (ii) The AdS$_4$ metric is the induced metric from that of flat Minkowski spacetime on the embedding space. In this way, several systems of coordinates can be defined on the AdS$_4$ manifold. For instance, AdS$_4$ can be (almost) described by the global coordinates $\{x^{\mu}\}$, which are tensorial with respect to the Minkowski metric  $\eta_{\mu\nu}=\mathrm{diag}(+1,-1,-1,-1)$:
\begin{eqnarray}
\label{cylcoord5} y^5 & = & \left({\varkappa}^{-2} + \Vert\px\Vert^2\right)^{1/2}\cos {\varkappa} x^0\,, \quad\quad -\pi \leqslant {\varkappa} x^0 < \pi\,, \\
\label{cylcoord0} y^0 & = & \left({\varkappa}^{-2} + \Vert\px\Vert^2\right)^{1/2}\sin {\varkappa} x^0\,, \\
\label{cylcoordi} \py &=& (y^i) = (x^i) = \px \,, \quad\quad x^i\in \R\,.
\end{eqnarray}
Another system of coordinates (sometimes referred to as horospherical coordinates \cite{Jadid1, Jadid2}), which will be useful for the sequel, is obtained by a combination of hyperbolic and spherical coordinates:
\begin{eqnarray}
\label{cylcoord5''} y^5 & = & {\varkappa}^{-1}\cosh \vrh\,\cos \theta \,, \quad\quad \vrh \in \R^+\,,\; -\pi \leqslant \theta < \pi\,,\\
\label{cylcoord0''} y^0 & = & {\varkappa}^{-1}\cosh \vrh\,\sin \theta\,,\\
\label{cylcoordi''} \py & = & {\varkappa}^{-1}\sinh \vrh\, \hat{\pmb{y}}\, , \quad\quad \hat{\pmb{y}}\in \mathbb{S}^2\,,
\end{eqnarray}
where $\mathbb{S}^2$ is the $2$-sphere.

The relativity/kinematical group of AdS$_4$ spacetime is SO$_0(2,3)$, that is, the connected subgroup of the identity of the symmetry group O$(2,3)$.\footnote{Technically, the connected component of O$(2,3)$, SO$_0(2,3)$,  containing the identity consists of all linear transformations in $\R^{2,3}$ that leave invariant the form of the metric $\eta_{\alpha\beta}$, have determinant unity, and also preserve the orientation of the ``time" variable $\mathrm{t}$, being defined through the relation $y^5\pm\ii y^0\equiv \sqrt{(y^5)^2+(y^0)^2}\, \exp(\pm\ii \mathrm{t})$ ($-\pi\leqslant \mathrm{t}< \pi$). Remember that O$(2,3)$ and SO$(2,3)$ have four and two connected components, respectively.} The two-fold covering of the SO$_0(2,3)$ group is isomorphic to  the real symplectic group Sp$(4,\R)$ {(see Appendix \ref{appendix:sp4R})}. Below, we will study the homomorphism between these two groups. But, before that, let us give a brief outline on complex quaternions, which become handy when we delve more deeply into the mathematical details in the coming discussions.

\subsection{Complex quaternions}
Note that most of the material and condensed complex quaternionic notations, that are discussed here, are borrowed from Ref. \cite{baelgagi92}.

A complex quaternion $z \; \big( z\in\H_{\mathbb{C}} \big)$ is the scalar-vector pair $(z^4,\pmb{z})$ written as:
\begin{equation}\label{comquat}
z=(z^4,\pmb{z}) \equiv z^4 + z^1\pmb{e}_1 + z^2\pmb{e}_2 + z^3\pmb{e}_3\,,
\end{equation}
where $\left(z^1,z^2,z^3,z^4\right) \in \C^4$, and $\{\pmb{e}_1, \pmb{e}_2, \pmb{e}_3\}$ satisfy the quaternionic algebra:
\begin{equation}\label{quatal}
\pmb{e}_i^2 = -1\,, \quad \pmb{e}_i\,\pmb{e}_j = \epsilon_{ij}^{\,\,\,k} \pmb{e}_k\,,
\end{equation}
where $\epsilon_{ij}^{\,\,\,k}$ is the three-dimensional totally antisymmetric Levi-Civita symbol. [Note that above we have employed the abusive identification $1\equiv (1,\pmb{0})$.] In the sequel, we denote as well $z^4$ by $(z)_\mathrm{s}$ (`$\mathrm{s}$' for scalar), and also make use of the decomposition of $z$ into real and imaginary parts, i.e., $z=x+\ii y$, while $x$ and $y$ are real quaternions ($x,y \in \H$).\footnote{For the real quaternions and the related discussions, readers can go to Ref. \cite{Gazeau2022}.} The \emph{complex conjugate}, the \emph{quaternionic conjugate}, and the \emph{adjoint} of $z$ are respectively defined by:
\begin{equation}\label{ccqcad}
\overline z = \left(\overline{z^4},\overline{\pmb{z}} \right)\,, \quad \widetilde{z} = (z^4, -\pmb{z})\,, \quad z^{\ast} = \overline{\widetilde z} = \widetilde{\overline z} = \left(\overline{z^4}, -\overline{\pmb{z}}\right)\,,
\end{equation}
or equivalently by:
\begin{equation}\label{ccqcadr}
\overline z = x-\ii y\,, \quad \widetilde{z} = \widetilde x + \ii \widetilde y\,, \quad z^{\ast} = \widetilde x - \ii \widetilde y\,.
\end{equation} 
The product of two complex quaternions is given by:
\begin{equation}\label{zzp}
zz^{\prime} = (z^4,\pmb{z})({z^{\prime}}^4,\pmb{z}^{\prime}) = \left(z^4{z^{\prime}}^4 -\pmb{z}\cdot\pmb{z}^{\prime}, z^4\pmb{z}^{\prime} + {z^{\prime}}^4\pmb{z} + \pmb{z}\times\pmb{z}^{\prime}\right)\,,
\end{equation}
where $\pmb{z}\cdot\pmb{z}^{\prime}$ and $\pmb{z}\times\pmb{z}^{\prime}$ are the analytic continuations of respectively the Euclidean inner product and the cross product in $\R^3$. Note that $\overline{{zz^\prime}} = \overline z \, \overline{{z^\prime}}$ and $\widetilde{{zz^\prime}} = \widetilde{{z^\prime}} \widetilde z$, and hence, $({zz^\prime})^\ast = {z^{\prime\ast}} z^\ast$. The \textit{determinant} of a complex quaternion $z$ is defined by the complex scalar:
\begin{equation}\label{detz}
\det z = \det \widetilde z = z\widetilde z = \widetilde z z = (z^1)^2 + (z^2)^2 + (z^3)^2 + (z^4)^2 = \Vert x \Vert^2 - \Vert y \Vert^2 + 2 \ii x\cdot y\,,
\end{equation}
where $\Vert x\Vert^2 = x\cdot x$ and `$\cdot$' is the Euclidean inner product in $\R^4\sim \H$. From Eq. \eqref{detz}, one derives the expression of the inverse $z^{-1}$ of $z$:
\begin{equation} \label{invcq}
z^{-1} = \frac{\widetilde z}{\det z}\,,
\end{equation}
which exists whenever $\det z \neq 0$. Note that $\big( z z^\prime\big)^{-1} = {z^\prime}^{-1} z^{-1}$.

For later use, it is also useful to point out that for a generic $2\times 2$-matrix $M=\begin{pmatrix} a & b \\ c & d \end{pmatrix}$, with complex quaternionic {entries} $a,b,c,d \in \H_{\C}$, the determinant of the matrix in terms of the determinant of its quaternionic components reads \cite{Aslaksen1996, Cohen2000}:
\begin{eqnarray}
\label{det 1} \det M &=& \big(\det a\big) \det (d-c\,a^{-1}b)\,, \\
\label{det 2} &=& \big(\det b\big) \det (c-d\,b^{-1}a)\,, \\
\label{det 3} &=& \big(\det c\big) \det (b-a\,c^{-1}d)\,, \\
\label{det 4} &=& \big(\det d\big) \det (a-b\,d^{-1}c)\,,
\end{eqnarray}
where these expressions are properly extended in case $a,b,c$, {or} $d$ are {non-invertible}.

The isomorphism between the algebra $\H_{\C} \sim \H \otimes_\R \C$ of complex quaternions and the algebra $\mathcal{M}_2(\C)$ of $2\times2$-complex matrices is defined through the correspondences:
\begin{equation} \label{esigma}
(1,\pmb{0}) \equiv 1 \mapsto \bu_2\, , \qquad \pmb{e}_1\mapsto \ii \,\sigma_1\, , \qquad \pmb{e}_2\mapsto -\ii \,\sigma_2\, , \qquad \pmb{e}_3\mapsto \ii\, \sigma_3\,,
\end{equation} 
with $\sigma_i$ the usual Pauli matrices. Hence:
\begin{equation} \label{cqumat}
\H_{\C}\ni z \;\mapsto\; Z(z) =
\begin{pmatrix} z^4 + \ii z^3 & \ii z^1 -z^2 \\\ii z^1+z^2 & z^4 - \ii z^3 \end{pmatrix} \equiv Z \in\mathcal{M}_2(\C)\,, \quad \det Z = \det z\,.
\end{equation}
In view of this matrix representation of $z\in \H_\C$, one also introduces another complex conjugate of $z$, denoted by  $z^{\mathrm{cc}}$, and defined by the relation:
\begin{equation} \label{ccz}
Z\left(z^{\mathrm{cc}}\right)= Z^{\mathrm{cc}}(z)= \begin{pmatrix} \overline{z^4} - \ii \overline{z^3} & -\ii \overline{z^1} -\overline{z^2} \\-\ii \overline{z^1}+\overline{z^2} & \overline{z^4} + \ii \overline{z^3} \end{pmatrix}\,,  
\end{equation}
where $\overline{z^\texttt{a}}$ is the conjugate of the complex number $z^\texttt{a}$, with $\texttt{a}=1,2,3,4$. 
We then have the important relation between the two notions \eqref{ccqcadr} and \eqref{ccz} of complex conjugates of $z\in \H_\C$:
\begin{equation} \label{ccovl}
z^{\mathrm{cc}} = - \pmb{e}_2\,\overline{z}\,\pmb{e}_2\ \Leftrightarrow\  \overline{z}= - \pmb{e}_2\,z^{\mathrm{cc}}\,\pmb{e}_2\,. 
\end{equation}
Let us end our brief introduction to the (complex) quaternions by pointing out a property that will be used in the sequel frequently, namely, the fact that the multiplicative subgroup of real quaternions of \emph{norm}\footnote{For a real quaternion $x=(x^4,\textbf{x})\in\H$, the norm {$\Vert x \Vert$}  is given by: ${\Vert x \Vert^2}= \det x = x\widetilde x = (x^1)^2 + (x^2)^2 + (x^3)^2 + (x^4)^2 \in \R^+$. It is zero if and only if all the components are zero.} $1$ is isomorphic to the SU$(2)$ group ($\sim \mathbb{S}^3$); see the details in Ref. \cite{Gazeau2022}.

\subsection{Homomorphism between SO$_0(2,3)$ and Sp$(4,\R)$}\label{Subsubsec. homo SO and Sp}
Let us write the elements of the group $SL(4,\C)$ as $2 \times 2$-complex quaternionic matrices. For simplicity, we do not distinguish between $z\in \mathbb{H}_\C$ and its matrix representation $Z(z)$:
$\begin{pmatrix}
   a   & b \\
   c   &  d
\end{pmatrix}$, with $a,b,c,d \in \H_{\C}$. We now consider the subset of $SL(4,\C)$ defined as:
\begin{equation}\label{sp4Rdef1}
\mathcal{S}=\left \{\mathrm{SL}(4,\C) \ni g =
\begin{pmatrix} a & b \\ -\overline b & \overline a \end{pmatrix}, \quad a, b \in \H_{\C}\, , \, g\,\gamma^0_\mathrm{W}\,{}^{\mathrm{t}}\widetilde{g}= \gamma^0_\mathrm{W}\right\}\,,
\end{equation}
where:
\begin{equation}
\label{defgttgam}
{}^{\mathrm{t}}\widetilde{g}= \begin{pmatrix} \widetilde{a} & -b^{\ast} \\ \widetilde{b} & a^{\ast} \end{pmatrix}\, ,
\end{equation}
and $\gamma^0_\mathrm{W}= \begin{pmatrix}
   0   & 1   \\
     1 &  0
\end{pmatrix}$ is the Weyl representation of the Dirac $\gamma^0$ matrix. One easily proves that the set $\mathcal{S}$ is a subgroup of $SL(4,\C)$. In particular, from \eqref{defgttgam} one infers that the inverse $g^{-1}$ is given by:
\begin{equation}\label{sp4inv}
g^{-1} = \begin{pmatrix} 0 & 1 \\ 1 & 0 \end{pmatrix}\,{}^{\mathrm{t}}\widetilde{g}\,
\begin{pmatrix} 0 & 1 \\ 1 & 0 \end{pmatrix} =
\begin{pmatrix} a^{\ast} & \widetilde b \\ -b^{\ast} & \widetilde a \end{pmatrix}\,. 
\end{equation}
Accordingly, the complex quaternionic entries of $g$ have to obey:
\begin{equation}\label{ggm1}
aa^{\ast}-bb^{\ast} = 1\,, \quad \mbox{and} \quad a\widetilde b = -b\widetilde a\,,
\end{equation}
or equivalently (since $g^{-1}g=\mathsf{e}$):
\begin{equation}\label{ggm2}
a^{\ast}a-\widetilde b\, \overline b = 1\,, \quad \mbox{and} \quad a^{\ast} b = -\widetilde b\overline a\,.
\end{equation}
An immediate consequence of the above relations is that $a\widetilde b$ and $a^{\ast} b$ are \emph{pure-vector} complex quaternions since, for instance, in the case of $a\widetilde b$, we have:
\begin{equation}
\widetilde{\big(a\widetilde b\big)} = \widetilde{\big(-b\widetilde a\big)} = - a\widetilde b\,.
\end{equation}
A similar relation also holds for the $a^{\ast} b$ case.
 
The group $\mathcal{S}$ is isomorphic to the real symplectic group Sp$(4,\mathbb{R})$. In this complex form \cite{zhang22} (see Appendix \ref{appendix:sp4R} for more details), the latter is defined as: 
\begin{equation} \label{Sp4RC}
\mathrm{Sp}(4,\R)= \left\{\mathrm{SL}(4,\C) \ni g =\begin{pmatrix}
   a   &   b \\
    b^{\mathrm{cc}}  &  a^{\mathrm{cc}} 
\end{pmatrix}\, , \, a, b \in \mathrm{M}(2,\C)\, , \, g\,\mathcal{J}\,{}^{\mathrm{t}}g= \mathcal{J}\right\}\, , 
\end{equation}
where $\mathcal{J}= \begin{pmatrix}
   0   &  1  \\
 -1   &  0
\end{pmatrix}$. 
The relations \eqref{ccovl} allow us to establish the isomorphism of $\mathcal{S}$ with Sp$(4,\mathbb{R})$:
\begin{equation} \label{isospS}
\mathcal{S} \ni g= \begin{pmatrix} a & b \\ -\overline b & \overline a \end{pmatrix} \mapsto g^{\prime}= \begin{pmatrix}
1   & 0   \\
   0   & \ii  \pmb{e}_2 
\end{pmatrix} \begin{pmatrix} a & b \\ -\overline b & \overline a \end{pmatrix}\begin{pmatrix}
  1    &   0 \\
  0    &  \ii  \pmb{e}_2
\end{pmatrix}= \begin{pmatrix}
 a     &  \ii b   \pmb{e}_2\\
  -\ii  \pmb{e}_2\overline{b}    &  - \pmb{e}_2\overline{a}  \pmb{e}_2
\end{pmatrix}= \begin{pmatrix}
  a    & \ii b   \pmb{e}_2  \\
  [\ii b   \pmb{e}_2]^{\mathrm{cc}}    &  a^{\mathrm{cc}}
\end{pmatrix}\in \mathrm{Sp}(4,\R)\,.
\end{equation}
For simplicity, we will also refer to the group $\mathcal{S}$ as Sp$(4,\mathbb{R})$.


In order to display the homomorphism between SO$_0(2,3)$ and its two-fold covering Sp$(4,\R)$, one needs to associate the following $4\times 4$-complex matrix with any $5$-tuple $y^{\alpha}$ in $\R^{2,3}$:
\begin{equation}\label{Gammay}
y = \left(y^{\alpha}\right) \;\mapsto\; \Gamma(y)=
\begin{pmatrix} y_+ & \pmb{Y} \\ - \pmb{Y} & y_- \end{pmatrix}, \quad y_{\pm} = y^5\pm \ii y^0\,, \quad \pmb{Y}=
\begin{pmatrix} \ii y^3 & \ii y^1-y^2 \\ \ii y^1+y^2 & - \ii y^3 \end{pmatrix}.
\end{equation}
The above map reads in terms of five basic $4\times 4$ matrices $\Gamma_\alpha$ as  $\Gamma(y) = y^{\alpha}\Gamma_\alpha$, where:
\begin{equation}\label{Gammalp}
 \Gamma_5 = \bu_4\,,\quad
 \Gamma_0 = \begin{pmatrix} \ii  & 0 \\ 0 & -\ii  \end{pmatrix}, 
 \quad \Gamma_1 = \begin{pmatrix} 0 & \ii \sigma_1 \\ -\ii \sigma_1 & 0 \end{pmatrix},
\quad\Gamma_2 = \begin{pmatrix} 0 & -\ii \sigma_2 \\ \ii \sigma_2 & 0 \end{pmatrix},
\quad \Gamma_3 = \begin{pmatrix} 0 & \ii \sigma_3 \\ -\ii \sigma_3 & 0 \end{pmatrix}.
\end{equation}
One can check that $\det\Gamma(y) = \big(y_{\alpha}y^{\alpha}\big)^2$. On the other hand, we may also express $\Gamma(y)$ in terms of the complex quaternions $(y^5 \pm \ii y^0,y^1,y^2,y^3)=(y_{\pm}, \pmb{y})$ and view it as the element of $\mathcal{M}_2\left(\H_{\C}\right)$:
\begin{equation} \label{GyM2Hc}
\Gamma(y) \equiv \begin{pmatrix} y_+ & \pmb{y} \\ - \pmb{y} & y_- \end{pmatrix},
\end{equation}
such that, for $y\in \mathrm{AdS}_4$, the following relation holds:
\begin{eqnarray}\label{second identity}
\Gamma(y)\begin{pmatrix} 0 & 1 \\ 1 & 0 \end{pmatrix}\Gamma(y)
\begin{pmatrix} 0 & 1 \\ 1 & 0 \end{pmatrix} = \big(y_{\alpha}y^{\alpha}\big) \bu_4 = {\varkappa}^{-2} \bu_4\,.
\end{eqnarray}
Note that, for $y\in \mathrm{AdS}_4$, we have ${\varkappa}\Gamma(y) \in \mathrm{Sp}(4,\R)$, since, on one hand, $y=(y^\alpha)\in \R^{2,3}$, and therefore, $\overline{\pmb{y}}=\pmb{y}$, and on the other hand, $y_- = \overline{y_+}$.

Having the map \eqref{Gammay} in mind, the action of the Sp$(4,\R)$ group on $\R^{2,3}$ is given by:\footnote{A full justification of this action will be provided in the next subsection.}
\begin{equation}\label{spads}
\mathrm{Sp}(4,\R) \ni g \,:\, \Gamma(y) \; \mapsto \; \Gamma(y^{\prime}) = g\,\Gamma(y)\,{}^{\mathrm{t}}\widetilde{g}\,,
\end{equation}
explicitly:
\begin{eqnarray}\label{spadss}
\Gamma(y^{\prime}) & = &
\begin{pmatrix} a & b \\ -\overline b & \overline a \end{pmatrix}
\begin{pmatrix} y_+ & \pmb{y} \\ - \pmb{y} & y_- \end{pmatrix}
\begin{pmatrix} \widetilde a & -b^\ast \\ \widetilde b & a^\ast \end{pmatrix} \nonumber\\
& = & \begin{pmatrix} y_+ \det a + y_- \det b + 2(a\pmb{y}\widetilde b)_{\mathrm{s}} & & -y_+ ab^{\ast} +y_- ba^{\ast} + a\pmb{y}a^{\ast} + b\pmb{y}b^{\ast} \\ & & \\ -\big(-y_- \overline{a}\widetilde{b} + y_+ \overline{b}\widetilde{a} + \overline{a}\pmb{y}\widetilde{a} + \overline{b}\pmb{y}\widetilde{b}\big) & & y_- \det \overline a + y_+ \det \overline b + 2(\overline a\pmb{y} b^\ast)_{\mathrm{s}} \end{pmatrix} \equiv \begin{pmatrix} y_+^\prime & \pmb{y}^\prime \\ - \pmb{y}^\prime & y_-^\prime \end{pmatrix}.
\end{eqnarray}
The transformed matrix $\Gamma(y^{\prime})$ corresponds to a point $y^{\prime}$ in $\R^{2,3}$ (the proof will be given later, by Eq.~\eqref{actj}). On the other hand, the matrix elements of $\Gamma(y^{\prime})$ are linearly expressed in terms of the elements of $\Gamma(y)$, which means that the transformation \eqref{spads} is linear. Moreover, this linear transformation preserves the quadratic form $\big(y^2\big)^2=\big(y_{\alpha}y^{\alpha}\big)^2$  since:
\begin{eqnarray}
\det\Gamma(y^{\prime}) = \det\big( g\,\Gamma(y)\,{}^{\mathrm{t}}\widetilde{g} \big) = \det\Gamma(y) = \big(y_{\alpha}y^{\alpha}\big)^2\,.
\end{eqnarray}
More accurately, it preserves the form $(y)^2=y_{\alpha}y^{\alpha}$, because from Eqs.~\eqref{second identity} and \eqref{sp4inv} we have:
\begin{eqnarray}\label{eee}
\Gamma(y^\prime)\begin{pmatrix} 0 & 1 \\ 1 & 0 \end{pmatrix}\Gamma(y^\prime)
\begin{pmatrix} 0 & 1 \\ 1 & 0 \end{pmatrix} &=& g\,\Gamma(y)\,{}^{\mathrm{t}}\widetilde{g}\begin{pmatrix} 0 & 1 \\ 1 & 0 \end{pmatrix}g\,\Gamma(y)\,{}^{\mathrm{t}}\widetilde{g} \begin{pmatrix} 0 & 1 \\ 1 & 0 \end{pmatrix}    \nonumber\\
&=& g\,\Gamma(y)\begin{pmatrix} 0 & 1 \\ 1 & 0 \end{pmatrix}\, \underbrace{\begin{pmatrix} 0 & 1 \\ 1 & 0 \end{pmatrix} {}^{\mathrm{t}}\widetilde{g}\begin{pmatrix} 0 & 1 \\ 1 & 0 \end{pmatrix}}_{g^{-1}} g\,\Gamma(y) \begin{pmatrix} 0 & 1 \\ 1 & 0 \end{pmatrix}\, \underbrace{\begin{pmatrix} 0 & 1 \\ 1 & 0 \end{pmatrix}{}^{\mathrm{t}}\widetilde{g} \begin{pmatrix} 0 & 1 \\ 1 & 0 \end{pmatrix}}_{g^{-1}} \nonumber\\
&=& g\, \left[\Gamma(y)\begin{pmatrix} 0 & 1 \\ 1 & 0 \end{pmatrix}\Gamma(y)
\begin{pmatrix} 0 & 1 \\ 1 & 0 \end{pmatrix}\right]\, g^{-1} \nonumber\\
&=& g\,\left[\big(y_{\alpha}y^{\alpha}\big) \bu_4\right]\, g^{-1} = \big(y_{\alpha}y^{\alpha}\big) \bu_4\,.
\end{eqnarray}
The transformation \eqref{spads} also preserves the ``time" orientation. To see this property, let us consider the point in
$\R^{2,3}$ as $y_o=(y_o^5=0,y_o^0=1,0,0,0)$, that is, $\Gamma({y_o}) \equiv \begin{pmatrix} \ii & 0 \\ 0 & -\ii \end{pmatrix}$, for which the corresponding ``time" parameter $\mathrm{t}_o$, being defined through the relation 
$y_o^5\pm\ii y_o^0 = 0\pm\ii = \exp(\pm\ii \mathrm{t}_o)$, is clearly positive ($\mathrm{t}_o=\pi/2$). After the transformation, we have:
\begin{eqnarray}\label{preservest}
\Gamma({y_o}^{\prime}) =
\begin{pmatrix} a & b \\ -\overline b & \overline a \end{pmatrix}
\begin{pmatrix} \ii & \pmb{0} \\ \pmb{0} & -\ii \end{pmatrix}
\begin{pmatrix} \widetilde a & -b^\ast \\ \widetilde b & a^\ast \end{pmatrix} &=&
\begin{pmatrix} \ii\det a - \ii\det b & -\ii ab^{\ast} - \ii ba^{\ast} \\ \ii ab^{\ast} + \ii ba^{\ast} & -\ii\det \overline a + \ii\det \overline b \end{pmatrix} \equiv \begin{pmatrix} {y_o}_{+}^\prime & \pmb{{y}}_o^\prime \\ - \pmb{{y}}_o^\prime & {y_o}_{-}^\prime \end{pmatrix},
\end{eqnarray}
hence:
\begin{eqnarray}\label{preservest1}
{y_o}^{\prime 5} = \frac{{y_o}_{+}^\prime + {y_o}_-^\prime}{2} = -\Im( \det a - \det b )\,, \quad\quad {y_o}^{\prime 0} = \frac{{y_o}_{+}^\prime - {y_o}_-^\prime}{2\ii} = \Re( \det a - \det b )\,.
\end{eqnarray}
Examining the ratio ${y_o}^{\prime 0}/{y_o}^{\prime 5}$ individually for each subgroups, involved in the time-space-Lorentz decomposition (see the next subsection \ref{Subsec space-time-Lorentz}), of Sp$(4,\R)$ one shows that the ratio is always non-negative. Take a closer look at the procedure also shows that ${y_o}^{\prime 0}\geqslant0$. Therefore, the corresponding ``time" parameter 
$\mathrm{t}_o^\prime$ $\left(y_o^{\prime 5}\pm\ii y_o^{\prime 0} = \sqrt{(y_o^{\prime 5})^2+(y_o^{\prime 0})^2} \, \exp({\pm\ii \mathrm{t}_o^\prime})\right)$ remains non-negative, as well. The above arguments clearly reveal that the transformation \eqref{spads} also belongs to the $\mathrm{SO}_0(2,3)$ group. Strictly speaking, for any transformation in $\mathrm{SO}_0(2,3)$, there are two transformations in $\mathrm{Sp}(4,\R)$:
\begin{equation}\label{SpSO}
\pm g\in \mathrm{Sp}(4,\R) \;\mapsto\; R_g \in \mathrm{SO}_0(2,3) \;:\; \Gamma(y^{\prime}) = \big(\pm g\big)\,\Gamma(y)\,\big(\pm{}^{\mathrm{t}}\widetilde{g}\big) = \Gamma(R_gy)\,.
\end{equation}
The $\mathrm{Sp}(4,\R)$ group then is two-to-one homomorphic to $\mathrm{SO}_0(2,3)$, or in other words, $\mathrm{Sp}(4,\R)$ is two-fold covering of $\mathrm{SO}_0(2,3)$. The kernel of this homomorphism $\{\pm\bu_2\}$ is isomorphic to $\Z_2$, that is,  SO$_0(2,3) \sim $ Sp$(4,\R)/ \Z_2$.

\subsection{Time-space-Lorentz decomposition}\label{Subsec space-time-Lorentz}
With respect to the group involution:
\begin{eqnarray}
\mathfrak{i}(g) \;:\; g \mapsto {}^{\mathrm{t}}\widetilde{g}\,,
\end{eqnarray}
any element $g$ of the Sp$(4,\R)$ group can be decomposed into:
\begin{eqnarray}\label{jl}
g = \begin{pmatrix} a & b \\ -\overline b & \overline a \end{pmatrix} = j \; l \,,
\end{eqnarray}
in such a way that $l$ belongs to the subgroup:
\begin{equation}\label{lorentz}
L = \Big\{ l \in \mathrm{Sp}(4,\R) \;;\; \mathfrak{i}(l) = l^{-1} \Big\}\,.
\end{equation}
Adjusting ${l} = \begin{pmatrix} a^{}_l\;\; & b^{}_l \\ -\overline b^{}_l & \overline a^{}_l \end{pmatrix}$, $a^{}_l$ and $b^{}_l$ being 
complex quaternions, the above definition together with \eqref{sp4inv} imply that:
\begin{eqnarray} \label{a_l b_l}
a^{}_l = \overline a^{}_l\,, \quad b^{}_l = -\overline b^{}_l\,.
\end{eqnarray}
Obviously, since $l$ is an element of the $\mathrm{Sp}(4,\R)$ group, the components of $l$ also need to obey the conditions \eqref{ggm1}:
\begin{equation}\label{ggm1 for l}
a^{}_l a^{\ast}_l-b^{}_l b^{\ast}_l = 1\,, \quad \mbox{and} \quad a^{}_l \widetilde b^{}_l = -b^{}_l \widetilde a^{}_l\,.
\end{equation}
A possible solution to this set of equations is:
\begin{eqnarray}
a^{}_l = (\cosh\textstyle\frac{\nu}{2}, \pmb{0})\,, \quad b^{}_l = (0, \ii \hat{\pmb{v}} \sinh\textstyle\frac{\nu}{2})\,,
\end{eqnarray}
where $\nu \in \R^+$ and $\hat{\pmb{v}}$ is a pure-vector real quaternion belonging to SU$(2)$, that is, $\hat{\pmb{v}}\equiv v^1\pmb{e}_1 + v^2\pmb{e}_2 + v^3\pmb{e}_3$, while $v^1,v^2,v^3\in\R$ and $(v^1)^2+(v^2)^2+(v^3)^2=1$, and therefore in a more accurate sense, $\hat{\pmb{v}}\in\mathbb{S}^2 \subset \mathbb{S}^3 \sim \mathrm{SU}(2)$. Such a solution can also be easily generalized by resetting $a^{}_l \mapsto \xi a^{}_l$ and $b^{}_l \mapsto \xi b^{}_l$, where $\xi\in\H$ and $\xi\widetilde{\xi}= \widetilde{\xi}\xi=1$, that is, $\xi\in\mathrm{SU}(2)$. On this basis, a generic matrix representation of $l$ in the (complex) quaternionic notations reads:
\begin{eqnarray}\label{solu lorentz}
{l} = \begin{pmatrix} \xi & 0 \\ 0 & \xi \end{pmatrix}
\begin{pmatrix} \cosh \frac{\nu}{2} & \ii\, \hat{\pmb{v}} \, \sinh \frac{\nu}{2} \\ \ii\,\hat{\pmb{v}} \sinh \frac{\nu}{2} \, & \cosh \frac{\nu}{2} \end{pmatrix} \equiv r(\xi) \, \lambda(\nu,\hat{\pmb{v}}) \,.
\end{eqnarray}
In the sequel, we will discuss that the subgroup $L$, with the generic element $l$, is indeed the Lorentz subgroup of Sp$(4,\R)$. Note that the above decomposition is referred to as the Cartan decomposition of the Lorentz subgroup.

Now, we concentrate on the other factor involved in the group decomposition \eqref{jl}, that is, $j$. Considering the definition \eqref{lorentz} and \eqref{jl}, the following identity holds:
\begin{eqnarray}\label{ji(j)}
j \; \mathfrak{i}(j) = g \; \mathfrak{i}(g) =
\begin{pmatrix} \det a + \det b & & -a b^\ast + b a^\ast \\ -\overline b \widetilde a + \overline a \widetilde b & & \det \overline{a} + \det \overline{b} \end{pmatrix}
\equiv
\begin{pmatrix} e^{\ii\theta} \cosh{\varrho} & & \hat{\pmb{u}} \, \sinh{\vrh} \\ - \hat{\pmb{u}} \, \sinh{\varrho} & & e^{-\ii\theta} \cosh{\varrho} \end{pmatrix}\,,
\end{eqnarray}
where $-\pi \leqslant \theta < \pi$, $\vrh \in \R^+$, and $\hat{\pmb{u}} \in \mathbb{S}^2$. Then, a possible solution reads as:
\begin{eqnarray} \label{j solu}
j = \begin{pmatrix} e^{\ii \theta/2} & 0 \\ 0 & e^{-\ii \theta/2} \end{pmatrix}
\begin{pmatrix} \cosh \frac{\vrh}{2} & \hat{\pmb{u}} \, \sinh \frac{\vrh}{2} \\ - \hat{\pmb{u}} \, \sinh \frac{\vrh}{2} & \cosh \frac{\vrh}{2} \end{pmatrix} \equiv h(\theta) \, s(\vrh, \hat{\pmb{u}}) \,.
\end{eqnarray}

In conclusion, the following non-unique decomposition of Sp$(4,\R)$ holds {(at least in a neighbourhood of the identity)}:
\begin{eqnarray} \label{Spstdec}
\mathrm{Sp}(4,\R) \ni g = j(\theta,\vrh,\hat{\pmb{u}})\,l(\xi,\nu,\hat{\pmb{v}}) &=& h(\theta)\,s(\vrh,\hat{\pmb{u}})\,r(\xi)\,\lambda(\nu,\hat{\pmb{v}}) \nonumber\\
&=& \exp(\theta {X}_0) \exp(\sum_{i=1}^3\varrho^i{X}_i) \exp(\sum_{i=1}^3\vartheta^i{Y}_i) \exp(\sum_{i=1}^3\nu^i{Z}_i)\,,
\end{eqnarray}
where ${X}_0, {X}_i, {Y}_i$ and ${Z}_i$ stand for the associated infinitesimal generators:
\begin{eqnarray}
\label{X0} {X}_0 &=& \frac{\mathrm{d}}{\mathrm{d}\theta} \begin{pmatrix} e^{\ii \theta/2} & 0 \\ 0 & e^{-\ii \theta/2} \end{pmatrix}\Big|_{\theta=0} = \frac{1}{2} \begin{pmatrix} \ii & 0 \\ 0 & -\ii \end{pmatrix}, \\
\label{Xi} {X}_i &=& \frac{\mathrm{d}}{\mathrm{d}\varrho}
\begin{pmatrix} \cosh \frac{\vrh}{2} & \pmb{e}_i \, \sinh \frac{\vrh}{2} \\ - \pmb{e}_i \, \sinh \frac{\vrh}{2} & \cosh \frac{\vrh}{2} \end{pmatrix}\Big|_{\varrho^{}=0} =
\frac{1}{2} \begin{pmatrix} 0 & \pmb{e}_i \\ -\pmb{e}_i & 0 \end{pmatrix}, \\
\label{Yi} {Y}_i &=& \frac{\mathrm{d}}{\mathrm{d}\vartheta}
\begin{pmatrix} \xi_i & 0 \\ 0 & \xi_i \end{pmatrix}\Big|_{\vartheta=0} =
\frac{1}{2} \begin{pmatrix} {\pmb{e}}^{}_i & 0 \\ 0 & {\pmb{e}}^{}_i \end{pmatrix}, \quad \mbox{where}\,\, \xi_i \equiv \big( \cos\frac{\vartheta}{2}, \pmb{e}_i \sin\frac{\vartheta}{2} \big)\in \mathrm{SU}(2)\,,\\
\label{Zi} {Z}_i &=& \frac{\mathrm{d}}{\mathrm{d}\nu_i}
\begin{pmatrix} \cosh \frac{\nu}{2} & \ii\, \pmb{e}_i \, \sinh \frac{\nu}{2} \\ \ii\,\pmb{e}_i \sinh \frac{\nu}{2} \, & \cosh \frac{\nu}{2} \end{pmatrix}\Big|_{\nu=0} =
\frac{1}{2} \begin{pmatrix} 0 & \ii\,\pmb{e}_i \\ \ii\,\pmb{e}_i & 0 \end{pmatrix}\,.
\end{eqnarray}
The infinitesimal generators verify the commutation relations:
\begin{equation}\begin{array}{llll}\label{commutation relations AdS4}
&\big[{Y}_i,{Y}_j\big] = {{\epsilon}_{ij}}^{k} \;{Y}_k\,,\qquad
&\big[{Y}_i,{X}_j\big] = {{\epsilon}_{ij}}^{k} \;{X}_k\,,\qquad
&\big[{X}_i,{X}_j\big] = - {{\epsilon}_{ij}}^{k} \;{Y}_k\,,\\[0.4cm]
&\big[{Y}_i,{Z}_j\big] = {{\epsilon}_{ij}}^{k} \;{Z}_k\,,\qquad
&\big[{X}_i,{Z}_j\big] = -\delta^{}_{ij} \;{X}_0\,,\qquad
&\big[{Z}_i,{Z}_j\big] = -{{\epsilon}_{ij}}^{k} \;{Y}_k\,, \\[0.4cm]
&\big[{X}_0,{X}_i\big] = {Z}_i\,,\qquad
&\big[{X}_0,{Z}_i\big] = -{X}_i\,,\qquad &\big[{X}_0,{Y}_i\big] = 0\,. 
\end{array}\end{equation}
One can bring these commutation rules into a form that explicitly displays the AdS$_4$ Lie algebra $\mathfrak{sp}(4,\R)$, by defining:
\begin{eqnarray}
N_{5i} \equiv {X}_i\,, \quad N_{50} \equiv {X}_0\,, \quad N_{ij} \equiv -{{\epsilon}_{ij}}^{k} \;{Y}_k\,, \quad N_{0i} \equiv {Z}_i\,,
\end{eqnarray}
based upon which, we have:
\begin{eqnarray} \label{AdS4 algebra}
[N_{\alpha \beta},N_{\varsigma \rho}] = \eta^{}_{\alpha \varsigma} {N^{}_{\beta \rho}} + \eta^{}_{\beta \rho} {N^{}_{\alpha \varsigma}} - \eta^{}_{\alpha \rho} {N^{}_{\beta \varsigma}} - \eta^{}_{\beta \varsigma} {N^{}_{\alpha \rho}} \,.
\end{eqnarray}
Note that $N_{\alpha\beta} = - N_{\beta\alpha}$.

Let us clarify the reason for naming `time-space-Lorentz decomposition' for the group decomposition \eqref{jl}. We begin by pointing out that the involved factor $j\equiv j(\theta,\vrh,\hat{\pmb{u}})$ is actually a kind of ``spacetime" square root, which provides a global coordinates system for AdS$_4$ spacetime. We make this point apparent by invoking the aforementioned map $\R^{2,3} \ni y \mapsto \Gamma(y)$ (see  \eqref{Gammay}), and then defining a coordinates system in $\R^{2,3}$ such that:
\begin{eqnarray}\label{j coordinates}
{\varkappa}^{-1} j \; \mathfrak{i}(j) =
{\varkappa}^{-1}  \begin{pmatrix} e^{\ii\theta} \cosh{\varrho} & & \hat{\pmb{u}} \, \sinh{\vrh} \\ - \hat{\pmb{u}} \, \sinh{\varrho} & & e^{-\ii\theta} \cosh{\varrho} \end{pmatrix} \equiv
\begin{pmatrix} y_+ & \pmb{y} \\ - \pmb{y} & y_- \end{pmatrix} = \Gamma(y) \,,
\end{eqnarray}
where ${\varkappa}\in \R^+$. One can now understand the action \eqref{spads} as the left action of the group on the set of matrices $j$:
\begin{equation}\label{actj}
\mathrm{Sp}(4,\R) \ni g \,: \, j\mapsto j^{\prime}\,, \quad gj = j^{\prime}\,l^{\prime} \;\Leftrightarrow\; \Gamma(y^{\prime})
={\varkappa}^{-1} j^{\prime}\,{}^{\mathrm{t}}\widetilde{j}^{\prime}
= {\varkappa}^{-1}g\, \big( j\,{}^{\mathrm{t}}\widetilde{j}\big) \,{}^{\mathrm{t}}\widetilde{g} = g\, \Gamma(y) \,{}^{\mathrm{t}}\widetilde{g}\,.
\end{equation}
This group action preserves the determinant:
\begin{eqnarray}
\det \big(\Gamma(y^{\prime})\big) = \det \big(\Gamma(y)\big) = \big(y_\alpha y^\alpha \big)^2 = \big({\varkappa}^{-2}\big)^{2}\,,
\end{eqnarray}
and also, having Eq. \eqref{eee} in mind, the identity $y_\alpha y^\alpha = {\varkappa}^{-2}$. The group decomposition \eqref{jl} therefore shows each point of the associated AdS$_4$ manifold is in one-to-one correspondence with each class of the left coset $\mathrm{Sp}(4,\R)/L$; topologically, $\mathrm{Sp}(4,\R) = \mbox{AdS}_4 \times L$.

We are now in a position to discuss the physical meaning of each involved subgroup in the above group factorization. Technically, according to the group action \eqref{actj}, the generic element $l$ (see \eqref{solu lorentz}) of the subgroup $L$ leaves invariant the point $y^{}_\odot = (y_5={\varkappa}^{-1},0,0,0,0)$ (or correspondingly, $\Gamma(y^{}_\odot)= {\varkappa}^{-1} \bu_2$), taken as the origin of the AdS$_4$ manifold.\footnote{Note that the selection of the origin is only a matter of choice because all the points on the AdS$_4$ manifold are equivalent (recall that AdS$_4$ is actually a homogeneous space of  Sp$(4,R)$, as we have seen above). Of course, if one was to deal with, for instance, the unit-sphere $\mathbb{S}^4$ as representing a four-dimensional manifold in $\R^{5}$, one has no hesitation to acknowledge such a property. Dealing with the representation of the AdS$_4$ hyperboloid embedded in $\R^{2,3}$ might be however misleading in this respect due to its deformed shape.} The tangent space to the AdS$_4$ manifold on $y^{}_\odot$ (i.e., the hyperplane $\{y\in\R^{2,3} \; ;\; y^5 = {\varkappa}^{-1} \}$) is taken into account as the $(1+3)$-dimensional Minkowski spacetime (equipped with the pseudo-metric $\mathrm{d} y_0^2 - \mathrm{d} y_1^2 - \mathrm{d} y_2^2 - \mathrm{d} y_3^2$) onto which AdS$_4$ spacetime is contracted when the curvature goes to zero. Accordingly, the subgroup $L \sim \mathrm{SL}(2,\C)$, being the stabilizer subgroup of a given point of the AdS$_4$ manifold, is interpreted as {the universal covering} of the Lorentz group of the tangent Minkowski spacetime; this guarantees that the neighborhood of any point of AdS$_4$ spacetime acts like flat Minkowski spacetime of special relativity. This subsequently clarifies the interpretation of the associated infinitesimal transformations ${Y}_i$ and ${Z}_i$ 
(see Eqs.~\eqref{Yi} and \eqref{Zi}) as the  ``space rotations" and ``boost transformations", respectively. The parameter $\xi$ then is presumed to carry the meaning of space rotation, $\hat{\pmb{v}}$ the boost velocity direction, and $\nu$ the rapidity.

On the other hand, the set of matrices $j$ maps the taken base point $y^{}_\odot$ to any point of the AdS$_4$ manifold:
\begin{eqnarray}\label{454545}
{j} \begin{pmatrix} {\varkappa}^{-1} & 0 \\ 0 & {\varkappa}^{-1} \end{pmatrix} {}^{\mathrm{t}}\widetilde{j} =
{\varkappa}^{-1}  \begin{pmatrix} e^{\ii\theta} \cosh{\varrho} & & \hat{\pmb{u}} \, \sinh{\vrh} \\ - \hat{\pmb{u}} \, \sinh{\varrho} & & e^{-\ii\theta} \cosh{\varrho} \end{pmatrix} \equiv
\begin{pmatrix} y_+ & \pmb{y} \\ - \pmb{y} & y_- \end{pmatrix} = \Gamma(y) \,.
\end{eqnarray}
The set $(\theta,\varrho,\hat{\pmb{u}})$ then introduces a global coordinates system for AdS$_4$ spacetime as:
\begin{eqnarray}\label{22222}
y^5= \frac{y_+ + y_-}{2} = {\varkappa}^{-1} \cos\theta\cosh{\varrho} \,, \quad y^0= \frac{y_+ - y_-}{2\ii} = {\varkappa}^{-1} \sin\theta\cosh{\varrho} \,, \quad \pmb{y}= {\varkappa}^{-1} \,\hat{\pmb{u}} \, \sinh{\vrh} \,.
\end{eqnarray}
Hence, the associated infinitesimal generators ${X}_i$ and ${X}_0$ (see Eqs.~\eqref{Xi} and \eqref{X0}) are respectively interpreted as the ``space translations" and ``time translations".

\subsection{Cartan decomposition}\label{Subsec Cartan AdS4}
As any simple Lie group, Sp$(4,\R)$ admits a Cartan factorization \cite{helgason78}:
\begin{equation} \label{cartansp4r}
\mathrm{Sp}(4,\R) = P\,K \,, \quad \mathrm{Sp}(4,\R) \ni g = \begin{pmatrix} a & b \\ -\overline b & \overline a \end{pmatrix} = p\,k\,,\qquad p\in P\,,k\in K\,.
\end{equation}
This decomposition is associated with the Cartan involution $\mathfrak{i}(g) : g \;\mapsto\; \left( g^{\dag} \right)^{-1}$, where $g^{\dag} \equiv {{}^{\,\mathrm{t}}g^{\ast}}$. The Cartan pair $(P,K)$ is made of all elements $p,k\in \mathrm{Sp}(4,\R)$ in such a way that $\mathfrak{i}(p)=p^{-1}$, that is, $p$  is Hermitian $\left(p=p^\dag\right)$, and $\mathfrak{i}(k)=k$, that is, $k$ 
 is unitary $\left(k^\dag=k^{-1}\right)$. This implies that $p^2=g\,g^\dag$.

We now give an explicit form of the involved factors $p$ and $k$. Using the fact that any complex quaternion $z$ can be written in the polar form as:
\begin{equation}\label{apol}
z = \left(zz^{\ast}\right)^{1/2}u \,, \quad uu^{\ast} = u^{\ast} u = 1\,,
\end{equation}
we perform the aforementioned decomposition on the quaternionic representation of Sp$(4,\R)$ as:
\begin{equation}\label{carinvg}
g = \begin{pmatrix} a & b \\ -\overline b & \overline a \end{pmatrix} =
\begin{pmatrix} \left(aa^{\ast}\right)^{1/2}u & b \\ -\overline b & \left(\overline a \widetilde a\right)^{1/2}\overline u \end{pmatrix}=
\underbrace{\begin{pmatrix} \left(aa^{\ast}\right)^{1/2} & b {\overline u}^{-1} \\ -\overline b u^{-1} & \left(\overline a \widetilde a\right)^{1/2} \end{pmatrix}}_{\equiv\, p }
 \underbrace{\begin{pmatrix} u & 0 \\ 0 & \overline u \end{pmatrix}}_{\equiv\, k }\,.
\end{equation}
Utilizing the identities given so far, one checks that the matrix $p$ is indeed Hermitian and $k$ unitary.

Note that the subgroup $K$ is isomorphic to $\mathrm{S}\big(\mathrm{U}(1)\times \mathrm{SU}(2)\big)$ and it is indeed the maximal compact subgroup of Sp$(4,\R)$. Having Eqs.~\eqref{solu lorentz} and \eqref{j solu} in mind, a realization of $K$ is achieved by:
\begin{equation}\label{sgK}
K = \Big\{k\big(u(\theta,\xi)\big) = h(\theta)\, r(\xi) \,, \, -\pi \leqslant \theta < \pi\,,\, \xi \in \mathrm{SU}(2) \Big\} \sim \mathrm{S}\big(\mathrm{U}(1)\times \mathrm{SU}(2)\big)\,,
\end{equation}
where:
\begin{equation}\label{sgK1}
h(\theta) = \begin{pmatrix} e^{\ii \theta/2} & 0 \\ 0 & e^{-\ii \theta/2} \end{pmatrix} ,\;\;
-\pi \leqslant \theta < \pi\,, \quad\mbox{and}\quad
r(\xi)=\begin{pmatrix} \xi & 0 \\ 0 & \xi \end{pmatrix}, \;\; \xi \in \mathrm{SU}(2)\,.
\end{equation}

The subset of Hermitian matrices $P$, on the other hand, is in one-to-one correspondence with the group coset Sp$(4,\R)/K$. The latter, as we further explain now, is in turn homeomorphic to the classical domain $\mathcal{D}^{(3)}$ \cite{borel52,hua63} whose definition is given in Eq.~\eqref{domC3}. Technically, as a consequence of 
Eq.~\eqref{ggm1}, $aa^{\ast}= 1 + bb^{\ast}$ is positive definite and so is invertible, and of course the same holds for $a$. Then, it results in that:
\begin{eqnarray}\label{aa-1bb}
\left(aa^{\ast}\right)^{-1} &=& 1- \left(\overline b a^{-1}\right)^{\ast}\left( \overline ba^{-1}\right) \nonumber\\
&=& 1- \left(b \overline{a}^{-1}\right)\left( b \overline{a}^{-1}\right)^{\ast}.
\end{eqnarray}
Note that the proof of the first line requires the use of the first identity in \eqref{ggm2}, and that of the second line requires the use of the fact that the term $b \overline{a}^{-1}$ is a pure-vector (complex) quaternion. The latter property is actually issued from a by-product of the second identity in \eqref{ggm2}:
\begin{equation}\label{ggm3}
b \overline{a}^{-1} = - (a^{\ast})^{-1}\widetilde b = -\widetilde{b \overline{a}^{-1}}.
\end{equation}
From now on, we denote the pure-vector (complex) quaternion $b \overline{a}^{-1}$ by $\pmb{z}$ (by the abusive identification in $\H_{\C}$, $(0,\pmb{z})\equiv\pmb{z}$). It follows from Eqs.~\eqref{apol}, \eqref{aa-1bb} and \eqref{ggm3} the relations:
\begin{equation}\label{aabbz}
aa^{\ast}= (1 + \pz\bpz)^{-1}\,, \quad u = (1 + \pz\bpz)^{1/2}a \,.
\end{equation}
Hence, the factor $p$ in the Cartan factorization \eqref{carinvg} can be rewritten as:
\begin{equation} \label{cartanpz}
p(\pz) = \begin{pmatrix} (1 + \pz\bpz)^{-1/2} & \pz (1 + \bpz\pz)^{-1/2} \\ -\bpz (1 + \pz\bpz)^{-1/2} & (1 + \bpz\pz)^{-1/2} \end{pmatrix},
\end{equation}
with $\pz\bpz = \big(-\pz\cdot\bpz,\;\pz\times\bpz\big) = \big(-\Vert \pz\Vert^2,\; \pz\times\bpz\big)$ and $\bpz\pz = \big(-\Vert \pz\Vert^2,\; -\pz\times\bpz\big)$. Note that: (i) For an arbitrary complex vector quaternion $\pmb{z}$, the term $\pz\times\bpz$ results in a \emph{purely imaginary} vector quaternion. (ii) Having in mind the fact that $p $ is Hermitian, we get the useful commutation relation $\pz (1 + \bpz\pz)^{-1/2} = (1 + \pz\bpz)^{-1/2} \pz$. (iii) From the latter identity and \eqref{sp4inv}, we obtain:
\begin{equation} \label{invpz}
\big(p(\pz)\big)^{-1} = p(-\pz)\,.
\end{equation}

It is clear from \eqref{aabbz} that the variable $\pz$ is confined to lie in the bounded domain of $\C^3$ defined by:
\begin{equation}\label{domC3}
\mathcal{D}^{(3)}= \Big\{ \pz\,;\, \bu_2 - \pmb{Z}\pmb{Z}^{\dag} =(a a^{\ast})^{-1} >0 \Big\}\,,
\end{equation}
where the positiveness condition is naturally understood in terms of the matrix representation (see \eqref{cqumat}) of complex quaternions:
\begin{equation}\label{D3mat}
\pz \mapsto \pmb{Z}\,, \quad  1+ \pz\bpz = 1-\pz\pz^\ast >0 \;\;\Leftrightarrow\;\; \bu_2 - \pmb{Z}\pmb{Z}^{\dag}>0 \,.
\end{equation}
This means that the spectral radius or the largest eigenvalue $\rho^+_{\pmb{Z}}$ of the \emph{non-negative} Hermitian matrix $\pmb{Z}\pmb{Z}^{\dag}$ is smaller than $1$. This condition reads in terms of $\pz = \pmb{x} + \ii \pmb{y}$ as:
\begin{equation} \label{D3vect}
\rho^+_{\pmb{Z}} = \Vert \pz \Vert^2 + \Vert \pz \times \bpz\Vert = \Vert \pmb{x} \Vert^2 +\Vert \pmb{y} \Vert^2 + 2 \Vert \pmb{x} \times \pmb{y} \Vert = \Vert \pmb{x} \Vert^2 + \Vert \pmb{y} \Vert^2 + 2 \Vert \pmb{x} \Vert \Vert\pmb{y} \Vert \vert\sin(\pmb{x},\pmb{y})\vert < 1 \,.
\end{equation}
The other eigenvalue is $\rho^-_{\pmb{Z}}= \Vert \pz \Vert^2 - \Vert \pz \times \bpz\Vert$. Hence, the expression of $\det (1 + \pz\bpz)$ is given by:
\begin{eqnarray}\label{det1zz}
\det (1 + \pz\bpz) &=& \det (1 + \bpz\pz) = \det \left(\bu_2 -\pmb{Z}\pmb{Z}^{\dag} \right) \nonumber\\
&=& \left(1- \Vert \pz \Vert^2 - \Vert \pz \times \bpz\Vert\right) \left(1- \Vert \pz \Vert^2 + \Vert \pz \times \bpz\Vert\right) = \sigma^{+}_{\pmb{Z}} \sigma^{-}_{\pmb{Z}} \,,
\end{eqnarray}
where, for convenience, we introduced the notations $\sigma^{\pm}_{\pmb{Z}} = 1-\rho^{\pm}_{\pmb{Z}}$. Also, note the alternative formula:
\begin{equation} \label{det1zbz}
\det (1 + \pz\bpz) = 1 + \vert \det\pz\vert^2 - 2 \Vert \pz\Vert^2 \,,
\end{equation}
where we have used the identity:
\begin{equation} \label{deta+b}
\det(a+b)= \det a + \det b + 2(a\widetilde{b})_s\,, \quad \mbox{for} \,\,\, a,b \in \mathbb{H}_{\C}\,.
\end{equation}
The domain $\mathcal{D}^{(3)}$ is an irreducible bounded symmetric domain or classical domain \cite{helgason78,hua63,borel52}. Since $\Vert \pz \Vert^2 \leqslant \rho^+_{\pmb{Z}} < 1$, it is strictly included in the unit ball in $\R^6$. Its Shilov boundary is diffeomorphic to $\mathbb{S}^1\times_{\mathbb{Z}_2} \mathbb{S}^2$, where $\times_{\mathbb{Z}_2}$ is the Cartesian product modulo a $\pm1$-factor.

We end the current discussion by pointing out that the left group coset Sp$(4,\R)/\mathrm{S}\big(\mathrm{U}(1)\times \mathrm{SU}(2)\big)\sim \mathcal{D}^{(3)} $ is homogeneous space for the left action of Sp$(4,\R)$:
\begin{equation} \label{gpzk}
\mathrm{Sp}(4,\R) \ni g \,:\,\, p(\pz) \,\mapsto\, p(\pz^\prime) \equiv p(g\diamond\pz) \,, \quad gp(\pz) = p(\pz^{\prime}) k^\prime \,,
\end{equation}
with:
\begin{eqnarray} \label{gpz}
\pz^{\prime} \equiv g\diamond\pz &=& (a\pz +b)(-\overline b \pz + \overline a)^{-1} \nonumber\\
&=& (\pz b^{\ast} + a^\mathrm{\ast})^{-1}(\pz \widetilde a - \widetilde b)\,.
\end{eqnarray}
The second line can be simply obtained from the first one by noticing the fact that $\pz^{\prime}$, which belongs to $\mathcal{D}^{(3)}$, is a pure-vector (complex) quaternion, and hence, $\pz^\prime=-\widetilde{\,\pz^\prime}$.

\subsubsection{K\"{a}hlerian structure of the domain $\mathcal{D}^{(3)}$} \label{Subsub sec kaehler}
The domain $\mathcal{D}^{(3)}$ is K\"{a}hlerian, and it has $G$-invariant metric and $G$-invariant $2$-form with respect to the analytic diffeomorphism \eqref{gpz}. Both arise from the K\"{a}hlerian potential $\ln \left[K(\pz,\bpz)\right]$, where $K(\pz,\bpz)$ is the Bergman kernel:
\begin{equation}\label{Kpzbpz}
K(\pz,\bpz) = \frac{1}{V}\big[\det (1 + \pz\bpz)\big]^{-3} =  \frac{1}{V}\left[\sigma^{+}_{\pmb{Z}} \sigma^{-}_{\pmb{Z}}\right]^{-3} \,,
\end{equation}
in which, $V$ is the Euclidean volume of $\mathcal{D}^{(3)}$ \cite{hua63}:
\begin{equation}\label{VD3}
V = \int_{\mathcal{D}^{(3)}}d{\pz} = \frac{\pi^3}{24}\,, \qquad d{\pz} \equiv \ud^3 \pmb{x}\, \ud^3 \pmb{y}\,.
\end{equation}
Note that the volume $V$ is a quarter of the volume of the unit ball in $\mathbb{R}^6$.

The Riemannian metric and the closed $2$-form on $\mathcal{D}^{(3)}$ are derived from the Bergman kernel:
\begin{eqnarray}
\label{ds2} \ud s^2 &=& -h_{i j}\, \ud z^i \ud \overline{z^j} = \ud \overline\ud  \,K(\pz,\bpz) = -\left(\frac{\partial}{\partial z^i}\frac{\partial}{\partial \overline{{z}^j}}\,\ln \left[K(\pz,\bpz)\right] \right)\,\ud z^i \ud \overline{{z}^j}\,, \\
\label{omD3} \pmb{\omega} &=& -\frac{\ii}{2} h_{i j}\, \ud z^i \wedge\ud \overline{{z}^j}\, , \quad \mbox{with}\quad \ud\pmb{\omega}=0\,,
\end{eqnarray}
where the minus sign in the metric is necessary since the spatial part of the AdS$_4$ metric is negative. We derive from \eqref{Kpzbpz} and \eqref{ds2} the expression of the tensor $h_{ij}$:
\begin{eqnarray}\label{hijz}
h_{i j} &=& -3\frac{\partial}{\partial z^i}\frac{\partial}{\partial \overline{{z}^j}}\,\ln\big[\det (1 + \pz\bpz)\big] \nonumber\\
&=& \frac{12}{\big[\det(1 + \pz\bpz)\big]^{2}}\left\lbrace
\begin{array}{lr}
\overline{z^i} z^j- z^i\overline{z^j}- z^iz^j\,\overline{\pz\cdot \pz} -\overline{z^iz^j} \,\pz\cdot \pz + 2 z^i\overline{z^j}\,\Vert \pz\Vert^2\,, & \quad i\neq j \\ \\
\frac{1}{2}\det (1 + \pz\bpz) - \left(z^i\right)^2\overline{\pz\cdot \pz} -\left(\overline{z^i}\right)^2 \pz\cdot \pz + 2 \vert z^i\vert^2\Vert \pz\Vert^2\,, & \quad i=j
\end{array}\right. \,,
\end{eqnarray}
and in terms of real and imaginary parts of $\pz (\,=x+\ii y)$:
\begin{equation}\label{hijxy}
h_{i j} = \frac{24}{\left[\sigma^{+}_{\pmb{Z}} \sigma^{-}_{\pmb{Z}} \right]^2}\left\lbrace
\begin{array}{lr}
2 x^ix^j \,\Vert \pmb{y}\Vert^2 + 2y^i y^j \Vert \pmb{x}\Vert^2 - 2\left(x^iy^j +x^jy^i \right)\,\pmb{x}\cdot\pmb{y} & \\ \qquad+\ii\left(x^iy^j-x^jy^i \right) \left(1-\Vert \pmb{x}\Vert^2- \Vert \pmb{y}\Vert^2\right)\,, & \quad i\neq j \\ \\
\frac{1}{4}  \sigma^{+}_{\pmb{Z}} \sigma^{-}_{\pmb{Z}} +2(x^i)^2\Vert \pmb{y}\Vert^2 +2(y^i )^2\Vert \pmb{x}\Vert^2 - 4x^iy^i \pmb{x}\cdot \pmb{y} \,, & \quad i=j
\end{array}\right. \,.
\end{equation}
The existence of the above symplectic structure confirms the phase-space nature of the Cartan domain $\mathcal{D}^{(3)}$ with regards to the kinematic group SO$_0(2,3)$ and its double covering Sp$(4,\R)$ for AdS$_4$ spacetime. Finally, let us give the explicit form of the invariant measure on $\mathcal{D}^{(3)}$ with respect to the group action \eqref{gpz}:
\begin{equation} \label{invmeas}
\mathrm{Sp}(4,\R)\ni g \;:\; \ud\mu(g\diamond\pz) = \ud\mu(\pz) = \frac{1}{V} \big[\det (1 + \pz\bpz)\big]^{-3} \,\dot{\pz}\,.
\end{equation}
Note that a key element for this is the transformation law:
\begin{equation}
\mathrm{Sp}(4,\R)\ni g = \begin{pmatrix} a & b \\ -\overline b & \overline a \end{pmatrix} \;:\; (1 + g\diamond\pz\,\overline{g\diamond\pz}) = \left(\pz b^{\ast}+  a^{\ast}\right)^{-1} (1 + \pz\bpz)\left(-b \bpz +a\right)^{-1}\,,
\end{equation}
and consequently:
\begin{equation}\label{footnote 12}
\det (1 + g\diamond\pz\,\overline{g\diamond\pz}) = \left\vert\det( -\overline{b}\pz +\overline{a})\right\vert^{-2} \, \det (1 + \pz\bpz)\,.
\end{equation}

\setcounter{equation}{0} 
\section{AdS$_4$ Lie algebra, ``massive'' co-adjoint orbits, and moment map} \label{Sec lieAdS}
A Lie group $G$ has natural actions on its Lie algebra $\mathfrak{g}$ and on the dual linear space to $\mathfrak{g}$, denoted here by $\mathfrak{g}^\circledast_{}$. These actions are respectively referred to as adjoint and co-adjoint actions. Below, we first give an explicit realization of these two actions, then discuss how they are related to the subject of the current study.

Technically, considering any $g \in G$, the conjugation $g^\prime\mapsto g\;g^\prime g^{-1}$ defines a differentiable map from $G$ to itself (inner automorphism, $i_g$, such that $i_g(g' )=g g^\prime g^{-1}$, which leaves invariant the identity element of the group, i.e., if $g'=\mathsf{e}$ then $i_g(\mathsf{e})=\mathsf{e}$. The adjoint action, symbolically denoted here by $\mbox{Ad}_g$, is nothing but the derivative of this map at $g^\prime=\mathsf{e}$, which defines an invertible linear transformation of $\mathfrak{g}$ onto itself; for $r \in (-\varepsilon,\varepsilon)\subset \R$, with $\varepsilon> 0$, in such a way that $e^{rX}\in G$ and the infinitesimal generator $X\in \mathfrak{g}$, the adjoint action is given by:
\begin{eqnarray}
\mbox{Ad}_g(X) \equiv \frac{\mathrm{d}}{\mathrm{d} r}\Big[ g \; e^{rX} g^{-1} \Big] \Bigg|_{r=0}\,.
\end{eqnarray}
Geometrically, $\mbox{Ad}_g(X)$ can be understood as a tangent vector in the tangent space at the identity, $T_\mathsf{e}G=\mathfrak{g}$. Note that, in case $G$ being a matrix group, the adjoint action would be simply matrix conjugation:
\begin{eqnarray}\label{samad}
\mbox{Ad}_g(X) = g \,X g^{-1}\,.
\end{eqnarray}

The corresponding co-adjoint action of $g \in G$, symbolized here by $\mbox{Ad}^{\circledast}_g$, is found by dualization; $\mbox{Ad}^{\circledast}_g$ acts on the dual linear space to $\mathfrak{g}$, that is, ${\mathfrak{g}}_{}^{\circledast}$, as follows:
\begin{eqnarray}\label{Co Ad_g}
\langle \mbox{Ad}_g^{\circledast}({X}_{}^{\circledast}) \;;\; X\rangle = \langle {X}_{}^{\circledast} \;;\; \mbox{Ad}_{g^{-1}}(X) \rangle\,,\quad {X}_{}^{\circledast} \in {\mathfrak{g}}_{}^{\circledast}\,,\quad X\in \mathfrak{g}\,,
\end{eqnarray}
in which $\langle \cdot \;;\; \cdot \rangle$ refers to the pairing between $\mathfrak{g}$ and its dual ${\mathfrak{g}}_{}^{\circledast}$. Note that, under the co-adjoint action, the vector space ${\mathfrak{g}}_{}^{\circledast}$ is divided into a union of mutually disjoint orbits (simply, say co-adjoint orbits). The co-adjoint orbit of ${X}_{}^{\circledast} \in {\mathfrak{g}}_{}^{\circledast}$, as a homogeneous space for the co-adjoint action of $G$, reads as:
\begin{eqnarray}\label{gcoadorbit}
\mathcal O({X}_{}^{\circledast}) \equiv \Big\{\mbox{Ad}^{\circledast}_g({X}_{}^{\circledast}) \;;\; g \in G \Big\}\,.
\end{eqnarray}
The adjoint and co-adjoint actions of a group $G$ are inequivalent unless the algebra $\mathfrak{g}$ admits a non-degenerate bilinear form (which is the case, for instance, for semi-simple Lie groups \cite{Kirillov1, Kirillov2}), then these two actions are equivalent.

Co-adjoint orbits are physically of great importance since for a physical system, like a (``free") AdS$_4$ elementary system, for which the global and local symmetries of its classical phase space are respectively given by a Lie group $G$ and its Lie algebra $\mathfrak{g}$. The phase space can be naturally identified with a co-adjoint orbit of $G$ in the dual linear space to ${\mathfrak{g}}$, say, ${\mathfrak{g}}_{}^{\circledast}$. As a matter of fact, such orbits are symplectic manifolds so that each of them carries a natural $G$-invariant (Liouville) measure, and is a homogeneous space homeomorphic to an even-dimensional group coset $G/S$, where $S$, being a (closed) subgroup of $G$, stabilizes some orbit point \cite{Kirillov1, Kirillov2}.

Now, we focus on our case, that is, the AdS$_4$ group Sp$(4,\R)$. A realization of the AdS$_4$ Lie algebra $\mathfrak{sp}(4,\R)$ can be achieved by taking into account the $10$ infinitesimal generators $X_0,X_i,Y_i$, and $Z_i$ ($i=1,2,3$) of the one-parameter subgroups of Sp$(4,\R)$ involved in the time-space-Lorentz decomposition of the group \eqref{Spstdec}. These generators actually constitute a basis of $\mathfrak{sp}(4,\R)$:
\begin{align} \label{sp4RX}
&\mathfrak{sp}(4,\R) = \bigg\{ 2k^0X_0 + 2\varsigma^iY_i + 2\alpha^iX_i + 2\beta^iZ_i = \begin{pmatrix} \ii k^0 + \varsigma^i\pmb{e}_i & (\alpha^i+\ii\beta^i)\pmb{e}_i \\ (-\alpha^i+\ii\beta^i)\pmb{e}_i & - \ii k^0 + \varsigma^i\pmb{e}_i \end{pmatrix} = \begin{pmatrix} \ii k^0 + \pmb{\varsigma} & \pmb{k} \\ -\overline{\pmb{k}} & - \ii k^0 + \pmb{\varsigma} \end{pmatrix}, \nonumber\\
& \hspace{1cm} k^0\in \R\,, \quad \pmb{\varsigma}\equiv \big(0,\pmb{\varsigma}\big) = \big(0,(\varsigma^i)\big)\in \H\,, \quad \pmb{k}= \pmb{\alpha}+\ii\pmb{\beta} \equiv \big(0, \pmb{\alpha}+\ii\pmb{\beta} \big) = \big(0, (\alpha^i+\ii\beta^i) \big) \in \H_{\C}\,,\, (\pmb{\alpha},\pmb{\beta}\in\H) \bigg\}\,.
\end{align}
The $\mathfrak{sp}(4,\R)$ Lie algebra is simple. It admits the symmetric bilinear form:
\begin{eqnarray}\label{su11bifo}
X, X^\prime \in \mathfrak{sp}(4,\R) \;:\; \langle X ; X^\prime \rangle \equiv \mbox{tr}(X X^\prime) = 2\Big( \pmb{\varsigma}{\pmb{\varsigma}}^\prime - k^0k^{\prime 0} - \Re\big({\pmb{k}}\overline{{\pmb{k}}^\prime}\big) \Big)\,,
\end{eqnarray}
which is non-degenerate (of course, this result was already expected because the algebra is simple).\footnote{Note that the form \eqref{su11bifo} is proportional to the Killing form for  $\mathfrak{sp}(4,\R)$  \cite{Gazeau2022}, that is, $K(X, X^\prime ) \equiv \mbox{tr}({\mbox{ad}}_X {\mbox{ad}}_{X^\prime})$, where ${\mbox{ad}}_X$ stands for the adjoint action of  $\mathfrak{sp}(4,\R)$  on itself:
$
\mathfrak{sp}(4,\R) \ni X^\prime \;\mapsto\; {\mbox{ad}}_X (X^\prime) \equiv \left[X,X^\prime\right].
\,$
This action is nothing but the derivative of the respective adjoint action of Sp$(4,\R)$ on  $\mathfrak{sp}(4,\R)$ (as given by \eqref{Ad_g AdS4}). See more details in Ref. \cite{Gazeau2022}.} Therefore, as already pointed out, the classification of its co-adjoint and adjoint orbits is exactly equivalent and can be realized through the following {adjoint} action:
\begin{eqnarray}\label{Ad_g AdS4}
g \in \mathrm{Sp}(4,\R)\,, \; X \in \mathfrak{sp}(4,\R) \; : \; \mbox{Ad}_{{g}}(X) = {g} X {g}^{-1}\,,
\end{eqnarray}
or equivalently through the  co-adjoint action \eqref{Co Ad_g}:
\begin{eqnarray}\label{Co Ad_g1}
\langle \mbox{Ad}_g^{\circledast}({X}_{}^{\circledast}) \;;\; X\rangle = \langle {X}_{}^{\circledast} \;;\; {g}^{-1} X {g}\rangle\,,\qquad 
g \in \mathrm{Sp}(4,\R)\,,\;{X}_{}^{\circledast} \in\mathfrak{sp}(4,\R)^{\circledast}\,,\; X\in \mathfrak{sp}(4,\R)\,.
\end{eqnarray}

Below, we will study those AdS$_4$ (co-)adjoint orbits relevant to the set of free motions on AdS$_4$ spacetime, with fixed ``energy" at rest, i.e., AdS$_4$ ``massive" (co-)adjoint orbits. For a comparison with the dS$_4$ case, readers are referred to Ref. \cite{Gazeau2022}.

\subsection{``Massive" scalar (co-)adjoint orbits} \label{Subsec co-A orbit}
We discuss here a specific class of the Sp$(4,\R)$ (co-)adjoint orbits, each being relevant to the transport of the element 
$2 \ell X_0 = \ell \begin{pmatrix} \ii & 0 \\ 0 & -\ii \end{pmatrix}$ of $\mathfrak{sp}(4,\R)$, with a given $\ell \in \R^+$, under the (co-)adjoint action \eqref{Ad_g AdS4}. The subgroup  stabilizing this element, $S_{2 \ell X_0}$,  consists of the \emph{time-translations and space-rotations subgroups}\footnote{Note that the space-rotations generators $Y_i$s commute with the time-translations one $X_0$ (see \eqref{commutation relations AdS4}).} already involved in the time-space-Lorentz decomposition of Sp$(4,\R)$ (see subsection \ref{Subsec space-time-Lorentz}) and coincides with the subgroup $K$ \eqref{sgK} of the Cartan decomposition of Sp$(4,\R)$ (see subsection~\ref{Subsec Cartan AdS4}).

Accordingly, this (co-)adjoint orbit class admits a homogeneous space realization identified with the group coset 
$\mathcal O(2 \ell {X}_0) \sim \mathrm{Sp}(4,\R)/\mathrm{S}\big(\mathrm{U}(1)\times \mathrm{SU}(2)\big) \sim \mathcal{D}^{(3)}$. Having in mind the time-space-Lorentz factorization \eqref{Spstdec} of the group, this realization can be technically understood by transporting the element $2 \ell {X}_0$ under the (co-)adjoint action \eqref{Ad_g AdS4}, when $g$ involved in the action corresponds to space translations and Lorentz boosts, i.e, $g$ belongs to the subgroup: 
\begin{equation}\label{spaceboots0}
L_{sb}=\Big\{s(\vrh,\hat{\pmb{u}})\, \lambda(\nu,\hat{\pmb{v}}) \;;\; \vrh,\nu \in \R^+, \; \hat{\pmb{u}},\hat{\pmb{v}} \in \mathbb{S}^2 \Big\}\,,
\end{equation} 
such that:
\begin{equation}\label{spaceboots}
s(\vrh,\hat{\pmb{u}})=\begin{pmatrix} \cosh\frac{\vrh}{2} & \hat{\pmb{u}} \, \sinh \frac{\vrh}{2} \\ - \hat{\pmb{u}} \, \sinh \frac{\vrh}{2} & \cosh \frac{\vrh}{2} \end{pmatrix} \,,\qquad
\lambda(\nu,\hat{\pmb{v}})=\begin{pmatrix} \cosh \frac{\nu}{2} & \ii\, \hat{\pmb{v}} \, \sinh \frac{\nu}{2} \\ \ii\,\hat{\pmb{v}} \sinh \frac{\nu}{2} \, & \cosh \frac{\nu}{2} \end{pmatrix}\,.
\end{equation}
Then, we have:
\begin{align}\label{zzzzz}
&\mbox{Ad}_g(2 \ell {X}_0) = s(\vrh,\hat{\pmb{u}}) \; \lambda(\nu,\hat{\pmb{v}}) \; \Big( 2 \ell {X}_0 \Big)\; \lambda^{-1}(\nu,\hat{\pmb{v}}) \; s^{-1}(\vrh,\hat{\pmb{u}}) \nonumber\\
&\;\;\;= \ell\begin{pmatrix} \ii\cosh{\varrho}\cosh\nu + \frac{1}{2}(\hat{\pmb{v}}\hat{\pmb{u}} - \hat{\pmb{u}}\hat{\pmb{v}})\sinh\varrho\sinh\nu & \hat{\pmb{v}}\cosh^2\frac{\varrho}{2}\sinh\nu + \hat{\pmb{u}}\hat{\pmb{v}}\hat{\pmb{u}}\sinh^2\frac{\varrho}{2}\sinh\nu - \ii \hat{\pmb{u}}\sinh{\varrho}\cosh\nu \\ & \\ -\hat{\pmb{v}}\cosh^2\frac{\varrho}{2}\sinh\nu - \hat{\pmb{u}}\hat{\pmb{v}}\hat{\pmb{u}}\sinh^2\frac{\varrho}{2}\sinh\nu - \ii \hat{\pmb{u}}\sinh{\varrho}\cosh\nu  & - \ii\cosh{\varrho}\cosh\nu + \frac{1}{2}(\hat{\pmb{v}}\hat{\pmb{u}} - \hat{\pmb{u}}\hat{\pmb{v}})\sinh\varrho\sinh\nu
\end{pmatrix} \nonumber\\
&\;\;\;\equiv \begin{pmatrix} \ii k^0 + \pmb{\varsigma} & \pmb{k}\, \big(=\pmb{\alpha}+\ii\pmb{\beta}\big) \\ -\overline{\pmb{k}}\, \big(=-\pmb{\alpha}+\ii\pmb{\beta}\big) & - \ii k^0 + \pmb{\varsigma} \end{pmatrix} = \mathsf{X}( k^0, \pmb{\varsigma}, \pmb{\alpha}, \pmb{\beta})\,.
\end{align}
Considering the relations that hold between the {entries} of $\mathsf{X}( k^0, \pmb{\varsigma}, \pmb{\alpha}, \pmb{\beta})$, as the generic element of the (co-)adjoint orbit class $\mathcal O(2 \ell {X}_0)$, one can simply show that:\footnote{One notices here that the generic element $\mathsf{X}( k^0, \pmb{\varsigma}, \pmb{\alpha}, \pmb{\beta})$ is kind of reminiscent of the form assumed by the elements of Sp$(4,\R)$, and subsequently, the identities given in \eqref{neut} of the identities already given in \eqref{ggm1}. Having this point in mind, one can show that the determinant of the generic element $\mathsf{X}( k^0, \pmb{\varsigma}, \pmb{\alpha}, \pmb{\beta})$ is fixed and equal to $(\ell^2)^2$. This is indeed the result already expected from the fact that the (co-)adjoint action \eqref{Ad_g AdS4} is determinant preserving since
$\det(\mathsf{X}) = \det(2 \ell {X}_0) = (\ell^2)^2$.\\
Another interesting point to be noticed here is that the first identity in \eqref{neut} is consistent with the constraint issued from the Killing form of the algebra for the generic element $\mathsf{X}$ of the (co-)adjoint orbit class, i.e. $k(\mathsf{X},\mathsf{X}) \propto\mbox{tr}(\mathsf{X}\mathsf{X})= 2\left( \pmb{k}\pmb{k}^\ast - (\ii k^0 + \pmb{\varsigma})(\ii k^0 + \pmb{\varsigma})^\ast \right)_{\mathrm{s}} =-2\ell^2$, where we have used the notation $(z)_{\mathrm{s}}$ for the scalar part of the complex quaternion $z$ and the fact that $(\ii k^0 + \pmb{\varsigma})^\ast = - (\ii k^0 + \pmb{\varsigma})$ and that $\pmb{k}^\ast = - \overline{\pmb{k}}$.}
\begin{eqnarray}\label{neut}
\ell^2 = (\ii k^0 + \pmb{\varsigma})(\ii k^0 + \pmb{\varsigma})^{\ast} - \pmb{k} \pmb{k}^{\ast}\,, \quad \big( \ii k^0 + \pmb{\varsigma} \big) \,\widetilde{\pmb{k}} = - \pmb{k}\, \widetilde{\big( \ii k^0 + \pmb{\varsigma} \big)}\,.\,
\end{eqnarray}
These two identities together yield:
\begin{eqnarray}\label{poori}
2\ii k^0 \pmb{\varsigma} = \pmb{k}\times\overline{{\pmb{k}}}\,, \quad \ell^2 = (k^0)^2 + \Vert \pmb{\varsigma} \Vert^2- \Vert \pmb{k} \Vert^2\,,
\end{eqnarray}
and in terms of the real and imaginary parts of $\pmb{k}= \pmb{\alpha}+\ii \pmb{\beta}$ ($\pmb{\alpha}, \pmb{\beta} \in \H$):
\begin{eqnarray} \label{vctcondr}
\pmb{\varsigma} = -\frac{1}{k^0}\pmb{\alpha}\times\pmb{\beta}\,, \quad \ell^2 = (k^0)^2 + \Vert \pmb{\varsigma} \Vert^2 - \Vert \pmb{\alpha} \Vert^2 -\Vert \pmb{\beta} \Vert^2\,.
\end{eqnarray}
The two relations given in \eqref{vctcondr} characterize the (co-)adjoint orbit class $\mathcal O(2\ell X_0)$ of Sp$(4,\R)$ in the (dual of) its Lie algebra $\mathfrak{sp}(4,\R)$:
\begin{equation} \label{omell}
O(2\ell X_0) = \left\{( k^0, \pmb{\varsigma}, \pmb{\alpha}, \pmb{\beta})\,;\, \pmb{\varsigma} = -\frac{1}{ k^0}\pmb{\alpha}\times\pmb{\beta} \,,\, \ell^2 = (k^0)^2 + \Vert \pmb{\varsigma} \Vert^2 - \Vert \pmb{\alpha} \Vert^2 - \Vert \pmb{\beta} \Vert^2 \right\}\,.
\end{equation}

As already pointed out in passing, the orbit class $\mathcal O(2\ell X_0)$ is homeomorphic to the domain $\mathcal{D}^{(3)}$ ($\ell \in\R^+$). To see the point, we first invoke from the Cartan decomposition of the group (see subsection \ref{Subsec Cartan AdS4}) the set of matrices $p$ (more accurately, their squared form $p^2$), representing the coset space $\mathrm{Sp}(4,\R)/\mathrm{S}\big(\mathrm{U}(1)\times \mathrm{SU}(2)\big) \sim \mathcal{D}^{(3)}$; from the definition \eqref{cartanpz} of these matrices, the term $\ii\ell p^2$ reads as:
\begin{equation}
\ii\ell p^2 = \frac{\ii\ell}{\det(1+ \pz\bpz)} \begin{pmatrix} \big(1-\vert\det \pz\vert^2\big) - 2 \pz \times \bpz & 2 \big(\pz-\bpz\det \pz\big) \\ & \\ -2 \big(\bpz-\pz\det \bpz\big) & \big(1-\vert\det \pz\vert^2\big) + 2 \pz \times \bpz \end{pmatrix} \equiv \begin{pmatrix} \ii k^0_p + \pmb{\varsigma}_p & \pmb{k}_p \\ -\overline{\pmb{k}}_p & \ii k^0_p - \pmb{\varsigma}_p \end{pmatrix},
\end{equation}
where $k^0_p\in \R,\, \pmb{\varsigma}_p\equiv \big(0,\pmb{\varsigma}_p\big) \in \H$, and $\pmb{k}_p \equiv \big(0, \pmb{k}_p \big) \in \H_{\C}$; we also recall the expressions  $\det(1+ \pz\bpz) = \det(1+ \bpz\pz)$,  $\pz (1 + \bpz\pz)^{-1/2} = (1 + \pz\bpz)^{-1/2} \pz$, and that $\pz \times \bpz$ is a purely imaginary vector quaternion. Clearly, the matrix representation of $\ii\ell p^2$ is a particular case of the form assumed by the elements of Sp$(4,\R)$. Strictly speaking, it leaves invariant the left-hand side of the first equality in \eqref{ggm1} and fulfills the second equality:
\begin{eqnarray}
\ell^2 = (\ii k^0_p + \pmb{\varsigma}_p)(\ii k^0_p + \pmb{\varsigma}_p)^{\ast} - \pmb{k}_p \pmb{k}^{\ast}_p\,, \quad \big( \ii k^0_p + \pmb{\varsigma}_p \big) \,\widetilde{\pmb{k}}_p = - \pmb{k}_p\, \widetilde{\big( \ii k^0_p + \pmb{\varsigma}_p \big)}\,.\,
\end{eqnarray}
which together result in:
\begin{eqnarray}
2\ii k^0_p \pmb{\varsigma}_p = \pmb{k}_p\times\overline{{\pmb{k}}_p}\,, \quad \ell^2 = (k^0_p)^2 + \Vert \pmb{\varsigma}_p \Vert^2- \Vert \pmb{k}_p \Vert^2\,.
\end{eqnarray}
Then, comparing the above with \eqref{poori} characterizing the (co-)adjoint orbit class $\mathcal O(2\ell X_0)$, we put forward the following one-to-one differentiable map $\mathcal{D}^{(3)} \ni \pmb{z} \mapsto ( k^0, \pmb{\varsigma}, \pmb{\alpha}, \pmb{\beta}) \in O(2\ell X_0)$:
\begin{align}
\label{tauz} k^0 &= k^0_p = \ell\, \frac{1-\vert\det \pz\vert^2}{\det(1+ \pz\bpz)}\,,\\
\label{zetz} \pmb{\varsigma} &= \pmb{\varsigma}_p = -2\ii \ell\, \frac{\pz \times \bpz}{\det(1+ \pz\bpz)}\,,\\
\label{betaz} \pmb{k} &= - \, \pmb{k}_p = - \, 2\ii\ell\, \frac{\pz-\bpz\det \pz}{\det(1+ \pz\bpz)}\,.
\end{align}
Note that the two global minus signs that appear in Eq. \eqref{betaz} are chosen here to keep our results comparable with those already given by Onofri in his seminal paper \cite{onofri76}. The inverse of this ``moment'' map is given by:
\begin{equation}\label{invmom}
\mathcal{D}^{(3)}\ni \pz = \frac{\ii}{2 k^0}\left(\frac{\det \pmb{k}}{2 k^0( k^0+\ell) - \pmb{k}\pmb{k}^\ast}\,\overline{\pmb{k}} +\pmb{k} \right)\,, \quad \det \pmb{k} = \Vert \pmb{\alpha}\Vert^2 - \Vert \pmb{\beta}\Vert^2 + 2\ii \pmb{\alpha}\cdot \pmb{\beta}\,,
\end{equation}
together with the constraint \eqref{omell}. Again, we would like to point out that the above results exactly meet their counterparts already given by Onofri in Ref. \cite{onofri76}.

\subsubsection{The phase-space interpretation} \label{Subsec phase-space inter}
Here, we attempt to understand the physical meaning of the two identities, given in \eqref{vctcondr}, characterizing the (co-)adjoint orbit class $\mathcal O(2\ell X_0)$. They are indeed conservation laws for the elementary systems associated with such orbits. To clarify this point, we first introduce physical constants, namely, a mass $m$, the speed of light $c$, without forgetting the AdS$_4$ radius of curvature ${\varkappa}^{-1}$. For $\ell\in\R^+$, these physical constants allow us to give appropriate physical dimensions to the variables $(k^0,\pmb{\alpha}, \pmb{\beta})$ as follows:
\begin{eqnarray}
\label{dimk0} k^0 &=& \ell \, \frac{E}{mc^2} \,, \\
\label{dimal} \pmb{\alpha} &=& \ell \,{\varkappa} \,\pmb{q} \,,\\
\label{dimbet} \pmb{\beta} &=& \ell\,\frac{\pmb{p}}{mc} \,.
\end{eqnarray}
There results from \eqref{vctcondr} the two physical equations:
\begin{eqnarray}
\label{zetaL} \pmb{\varsigma} &=& -\ell \frac{{\varkappa} c}{E}\,\pmb{l}\,, \quad \pmb{l}\equiv \pmb{q}\times\pmb{p}\ \mbox{angular momentum}\,, \\
\label{E4E2} 0 &=& E^4 -\left( m^2c^4 + c^2\Vert \pmb{p}\Vert^2 +m^2c^4 {\varkappa}^2\Vert \pmb{q}\Vert^2 \right) \,E^2 + m^2c^6{\varkappa}^2 \Vert \pmb{l}\Vert^2\,.
\end{eqnarray}
Note that the latter equation, in the second order, gives us two solutions:
\begin{eqnarray}
E &\approx& mc^2 + \frac{1}{2m}\,\Vert \pmb{p}\Vert^2 + \frac{1}{2}\, m\omega^2 \Vert \pmb{q}\Vert^2 
- \frac{1}{2mc^2}\,\omega^2 \Vert \pmb{l}\Vert^2\,,\\
E &\approx& \omega \Vert \pmb{l}\Vert\,,
\end{eqnarray}
where $c\,{\varkappa} \equiv \omega$. Clearly, the first solution consists of the mass at rest energy and the energy of a harmonic oscillator with frequency arising from the AdS$_4$ curvature. We will see that this dual nature survives during the non-relativistic (Newtonian-Hooke) contraction limit of the system. The second solution, on the other hand, provides a kind of rotational energy!

Let us examine the flat limit AdS$_4$ $\longrightarrow$ Minkowski, at ${\varkappa}\rightarrow 0$. Clearly, Eq. \eqref{E4E2} yields the mass shell hyperboloid:\footnote{The co-adjoint orbit class describing massive elementary systems for the Poincar\'e kinematical symmetry is described in Ref.~\cite{carinena90}.}
\begin{equation} \label{masshel}
E^2 - \Vert \pmb{p}\Vert^2c^2 = m^2c^ 4\,.
\end{equation}
The relations \eqref{dimk0}, \eqref{dimal}, \eqref{dimbet} then become:
\begin{eqnarray}
\label{dimk00} k^0_\circ &=& \ell \, \sqrt{1+ \frac{\Vert \pmb{p}\Vert^2}{m^2c^2}} \,, \\
\label{dimal0} \pmb{\alpha}_\circ &=& \pmb{0} \,,\\
\label{dimbet0} \pmb{\beta}_\circ &=& \ell\,\frac{\pmb{p}}{mc} \,,
\end{eqnarray}
and the complex quaternion $\pz \; \big(\equiv \textbf{x}+\ii\textbf{y}, \; \textbf{x},\textbf{y}\in\H\big)$ in \eqref{invmom}, becomes a purely real quaternion as:
\begin{equation}\label{invmom0}
\textbf{x}_\circ(p) = - \frac{\pmb{p}c}{E + mc^2}= - \frac{\pmb{p}}{mc}\frac{1}{1 + \sqrt{1+ \frac{\Vert \pmb{p}\Vert^2}{m^2c^2}}} \;\in\; \mathbb{B}_3 \equiv \Big\{ \textbf{x}\in \R^3\;;\; \Vert \textbf{x}\Vert < 1 \Big\}\,,
\end{equation}
where the subscript `$\circ$' marks the entities under the null-curvature limit ${\varkappa}\rightarrow 0$. Finally, the orbit \eqref{omell} becomes the mass shell in $\R^4$:
\begin{equation}
\label{minorb}
\mathcal O(2\ell X_0) \;\underset{{\varkappa} \to 0}{\longrightarrow}\; \mathcal{V}^+_m= \left\{\left( \frac{E}{c}, \pmb{p}\right)\;;\; \frac{E^2}{c^2}-\Vert \pmb{p} \Vert^2 = m^2c^2\right\}\,,
\end{equation}
whose stereographic projection is precisely the open-unit ball $\mathbb{B}_3$ through the map described in \eqref{invmom0}:\footnote{In this context, we encourage readers to see Appendix \ref{appendix:WigPoin}.}
\begin{equation} \label{Epw}
\mathcal{V}^+_m \ni \left( \frac{E}{c}, \pmb{p}\right) \;\mapsto\; \textbf{x}_\circ \in \mathbb{B}_3\,.
\end{equation}
It is proved that the action of the Lorentz subgroup \eqref{solu lorentz} (actually its two-fold covering) on $\pz$ as is defined in \eqref{gpz}:
\begin{eqnarray} \label{lorpz}
\pz^{\prime} = \begin{pmatrix} \xi & 0 \\ 0 & \xi \end{pmatrix}
\begin{pmatrix} \cosh \frac{\nu}{2} & \ii\, \hat{\pmb{v}} \, \sinh \frac{\nu}{2} \\ \ii\,\hat{\pmb{v}} \sinh \frac{\nu}{2} \, & \cosh \frac{\nu}{2} \end{pmatrix} \diamond \pz = \left(\xi\pz \cosh \frac{\nu}{2} + \ii \, \xi\hat{\pmb{v}} \, \sinh \frac{\nu}{2}\right)\left(\ii\, \xi\hat{\pmb{v}} \pz \, \sinh \frac{\nu}{2} +\xi \cosh \frac{\nu}{2}\right)^{-1}\,,
\end{eqnarray}
agrees with the action of the proper orthochronous Lorentz group SO$_0(1,3)$ on the mass hyperboloid $ \mathcal{V}^+_m$ \cite{11Barut1980,22Tung1985,33Yndurain1996}:
\begin{equation} \label{LambdaEp}
\begin{pmatrix} {E^{\prime}}/{c} \\ \pmb{p}^{\prime} \end{pmatrix} = \Lambda(\vartheta,\hat{\pmb{n}})\Lambda(\nu,\hat{\pmb{v}})
\begin{pmatrix} {E}/{c} \\ \pmb{p} \end{pmatrix}\,.
\end{equation}
where $\Lambda(\vartheta,\hat{\pmb{n}})$ is determined by the space rotation ${R}(\vartheta,\hat{\pmb{n}})$ with angle $\vartheta$ around the orientation $\hat{\pmb{n}}$, or equivalently, by $\xi=\left(\cos\frac{\vartheta}{2}\, ,\, \hat{\pmb{n}}\, \sin\frac{\vartheta}{2}\right)\in \mathrm{SU}(2)$:
\begin{equation} \label{lorot}
\Lambda(\vartheta,\hat{\pmb{n}}) = \begin{pmatrix}
1 & {}^{\mathrm{t}}\vec{0} \\ \vec{0} & {R}(\vartheta,\hat{\pmb{n}}) \end{pmatrix}\,, \qquad 
\vec{0} = \begin{pmatrix} 0 \\ 0 \\ 0 \end{pmatrix}\,, \quad {}^{\mathrm{t}}\vec{0} = \begin{pmatrix} 0 & 0 & 0 \end{pmatrix} \,,
\end{equation}
and $\Lambda(\nu,\hat{\pmb{v}})$ is the Lorentz boost determined by the rapidity parameter $\nu$ around the orientation 
$\hat{\pmb{v}}$:
\begin{equation}
\label{lortrgenrvec}
\left\lbrace\begin{array}{rcl}
\ds \frac{E^{\prime}}{c}&=&\ds \frac{E}{c}\,\cosh\nu +   \hat{\pmb{v}} \cdot \pmb{p} \,\sinh\nu \\[0.3cm]
\ds  \pmb{p}^{\prime}
&=&\ds \pmb{p} + \left(\frac{E}{c} \,\sinh\nu- (1- \cosh\nu)\,\hat{\pmb{v}} \cdot \pmb{p}\right)\,\hat{\pmb{v}}
\end{array}\right. \,.
\end{equation}
Explicitly, with the notations of \eqref{solu lorentz}, we have {taken} into account \eqref{invmom0} that:
\begin{equation} \label{agreem}
\textbf{x}_\circ^{\prime} = \,\frac{-\pmb{p}^{\prime}c}{E^{\prime} + mc^2}= r(\xi) \, \lambda(\nu,\hat{\pmb{v}})\diamond\textbf{x}_\circ^{} = r(\xi) \, \lambda(\nu,\hat{\pmb{v}})\diamond \, \frac{-\pmb{p}c}{E + mc^2}\,.
\end{equation}

Besides the space-time contraction AdS$_4\,\xrightarrow[]{}$ Poincar\'e symmetries, there is another one which is physically relevant within the context of our study, namely AdS$_4\,\xrightarrow[contraction]{speed-space}$ Newton-Hooke $\mathrm{NH}_{-4}$ (see Figure 1 of  Ref.~\cite{bacryjmll68}).\footnote{For Newton-Hooke groups and its relevant discussions, see Appendix \ref{appendix:Newton-Hooke}.} The latter contraction limit is implemented through the constrained limits:
\begin{equation} \label{AdSN+}
c \to \infty\,, \qquad {\varkappa} \to 0\,, \qquad c{\varkappa} = \mathrm{constant} \equiv \omega\,.
\end{equation}
Precisely, let us study the behavior of \eqref{E4E2} at large $c$. When $\pmb{l}=\textbf{0}$, we find the approximation:
\begin{equation} \label{Elargec}
E = mc^2 + \frac{\Vert \pmb{p}\Vert^2}{2m} + \frac{m\omega^2 \Vert \pmb{q}\Vert^2}{2} - \frac{1}{2mc^2}\left(\frac{\Vert \pmb{p}\Vert^2}{2m} + \frac{m\omega^2 \Vert \pmb{q}\Vert^2}{2}\right)^2 + o\left(\frac{1}{c^2}\right),
\end{equation}
which shows that the energy of the system is the sum of the mass at rest energy and the energy of a harmonic oscillator with frequency due to the AdS$_4$ curvature. This duality matter-vibration, combined with the equipartition theorem, will take its whole importance in the quantum regime in view of its possible r\^ole in the explanation of the emergence of dark matter \cite{GazeauTannoudji2021, Gazeau2020}.

So far, we have examined the physical meaning of the quantities given in \eqref{dimk0}, \eqref{dimal}, and \eqref{dimbet}. With their analytic descriptions \eqref{tauz}, \eqref{zetz}, and \eqref{betaz}, they have to be understood as basic smooth observables on the six-dimensional K\"{a}hlerian phase-space $\mathcal{D}^{(3)}$ for a free ``massive'' elementary system living in AdS$_4$ spacetime. With the parametrizations \eqref{dimk0}, \eqref{dimal}, and \eqref{dimbet}, they are given a precise physical meaning in terms of the motion of {such a system on AdS$_4$}. The function $k^0$ \eqref{tauz} should be understood as defining the Hamiltonian flow. The vector-valued function $\pmb{\varsigma}$  \eqref{zetz} is associated with the classical orbital angular momentum and spin, while the complex-vector valued function $\pmb{k}$ \eqref{betaz} defines the flows for AdS$_4$ hyperbolic rotations (real part), understood as ``space translations'', and ``Lorentz boost'' (imaginary part). The two constraints defining the orbit class 
$\mathcal O(2\ell X_0)$ in \eqref{omell} are understood as the conservation of the energy and total angular momentum. They extend the familiar conservation laws for the free motion of a massive particle in flat Minkowskian spacetime to the motion in AdS$_4$ spacetime. As basic observables, they form an algebra closed under a Poisson bracket. The latter is determined by the form \eqref{omD3}:
\begin{equation} \label{poissonb}
\{f,g\} = -\frac{\ii}{2} h^{i j}\, \frac{\partial f}{\partial z^i}\frac{\partial g}{\partial \overline{z^j}}\equiv \pmb{\omega}^{i j}\, \frac{\partial f}{\partial z^i}\frac{\partial g}{\partial \overline{z^j}} \,.
\end{equation}
This algebra is in one-to-one correspondence with $\mathfrak{sp}(4,\R)$ (for the latter, see \eqref{sp4RX}) in such a way that, in agreement with the Lie commutators \eqref{AdS4 algebra}, we have:
\begin{eqnarray} \label{crk}
&\left\{k^0,\alpha^{i}\right\} = \beta^{i} \,, \quad \left\{k^0,\beta^{i}\right\} = - \alpha^{i} \,, \quad \left\{\alpha^{j},\varsigma^{k}\right\} = - \alpha^{i} \,,& \nonumber\\
&\left\{\alpha^{i},\beta^{i}\right\} = -k^{0} \,, \quad \left\{\varsigma^{k},\beta^{i}\right\} = \beta^{j} \,, \quad \left\{\varsigma^{k},\varsigma^{j}\right\} = -\varsigma^{i}\,.&
\end{eqnarray}

We finally describe how the quantities $k^{0}$ \eqref{tauz}, $\pmb{\varsigma}$ \eqref{zetz}, and $\pmb{k}$  \eqref{betaz}, i.e., the basic ``classical observables'', transform under the (co-)adjoint representation \eqref{Ad_g AdS4} of Sp$(4,\R)$. Hence, from:
\begin{equation}\label{adjcoadj1}
\mathfrak{sp}(4,\R) \ni \mathsf{X}=
\begin{pmatrix} \ii k^0 + \pmb{\varsigma} \pmb{k} \\ -\overline{\pmb{k}} - \ii k^0 + \pmb{\varsigma} \end{pmatrix} \;\mapsto\;
\mathsf{X}^{\prime} = g\mathsf{X}g^{-1} =
\begin{pmatrix}
\ii k^{\prime 0} + \pmb{\varsigma}^{\prime} & \pmb{k}^{\prime} \\ -\overline{\pmb{k}^{\prime}}   - \ii k^{\prime 0} + \pmb{\varsigma}^{\prime}\end{pmatrix}\,, 
\end{equation}
we get:
\begin{align} \label{adjcoadj2} 
k^{\prime 0} &= k^0(\Vert a\Vert^2 + \Vert b\Vert^2) - \ii \left(a\pmb{\varsigma} a^{\ast}\right)_\mathrm{s} + \ii \left(b\pmb{\varsigma} b^{\ast}\right)_\mathrm{s} + \ii \left(a\pmb{k} b^{\ast}\right)_\mathrm{s} + \ii \left(b\overline{\pmb{k}} a^{\ast}\right)_\mathrm{s}\,, \\[0.05cm]
\label{adjcoadj3} \pmb{\varsigma}^{\prime}&= \ii k^0(a a^{\ast} + b b^{\ast})_{\mathrm{v}} + (a \pmb{\varsigma}a^{\ast})_{\mathrm{v}} - (b \pmb{\varsigma}b^{\ast})_{\mathrm{v}} - (a \pmb{k}b^{\ast})_{\mathrm{v}} - (b \overline{\pmb{k}}a^{\ast})_{\mathrm{v}}\,, \\[0.05cm]
\label{adjcoadj4} \pmb{k}^{\prime} &= 2\ii k^0 (a\widetilde{b})_{\mathrm{v}} +2 (a \pmb{\varsigma}\widetilde{b})_{\mathrm{v}} + a  \pmb{k} \widetilde{a}- b\overline{\pmb{k}}\widetilde{b}\,,
\end{align}
where we have used the notations $(z)_{\mathrm{s}}$ and $(z)_{\mathrm{v}}$ respectively for the scalar and vector parts of the complex quaternion $z$, i.e., $z=\big(z_0\equiv(z)_{\mathrm{s}}, \pmb{z}\equiv (z)_{\mathrm{v}}\big)$. 

Accordingly, the particular actions with respect to the subgroups involved in the time-space-Lorentz decomposition \eqref{Spstdec} of the group read:
\begin{itemize}
\item{Temporal rotations; $h(\theta) \,:\, \mathsf{X} \;\mapsto\; \mathsf{X}^{\prime}$ with:
\begin{equation} \label{temprot}
\begin{array}{rcl}
     k^{\prime 0} &=& k^0 \\[0.051cm]
     \pmb{\varsigma}^{\prime} &=& \pmb{\varsigma} \\[0.05cm]
     \pmb{k}^{\prime} &=& e^{\ii \theta} \pmb{k}
\end{array}\,.
\end{equation}}
\item{Spatial rotations; $r(\xi) \,:\, \mathsf{X} \;\mapsto\; \mathsf{X}^{\prime}$ with:
\begin{equation}\label{sparot}
\begin{array}{rcl}
     k^{\prime 0} &=& k^0 \\[0.05cm]
     \pmb{\varsigma}^{\prime} &=& \xi\pmb{\varsigma}\widetilde{\xi} \\[0.05cm]
     \pmb{k}^{\prime} &=& \xi \pmb{k} \widetilde{\xi}
\end{array}\,.
\end{equation}}
\item{Spatial hyperbolic rotations; $s(\vrh,\hat{\pmb{u}}) \;:\; \mathsf{X} \;\mapsto\; \mathsf{X}^{\prime}$ with:
\begin{equation}\label{spahyprot}
\begin{array}{rcl}
     k^{\prime 0} &=& k^0 \cosh\vrh - \pmb{\beta}\cdot\hat{\pmb{u}}\, \sinh\vrh \\[0.05cm]
     \pmb{\varsigma}^{\prime} &=& \pmb{\varsigma}\, \cosh^2\frac{\vrh}{2} + (-\hat{\pmb{u}}\cdot\pmb{\varsigma}\,\hat{\pmb{u}} + \hat{\pmb{u}}\times\pmb{\varsigma}\times\hat{\pmb{u}})\, \sinh^2\frac{\varrho}{2} + \pmb{\alpha}\times\hat{\pmb{u}}\, \sinh\vrh \\[0.05cm]
     \pmb{k}^{\prime} &=& -\ii \hat{\pmb{u}}\, k^0 \sinh\vrh - \pmb{\varsigma}\times\hat{\pmb{u}}\, \sinh\vrh + \pmb{k}\, \cosh^2\frac{\vrh}{2} + (-\hat{\pmb{u}}\cdot\overline{\pmb{k}}\,\hat{\pmb{u}} + \hat{\pmb{u}}\times\overline{\pmb{k}}\times\hat{\pmb{u}})\, \sinh^2\frac{\vrh}{2}
\end{array}\,.
\end{equation}}
\item{Lorentz boosts; $\lambda(\nu,\hat{\pmb{v}}) \,:\, \mathsf{X} \;\mapsto\; \mathsf{X}^{\prime}$ with:
\begin{equation}\label{lorboost}
\begin{array}{rcl}
     k^{\prime 0} &=& k^0\cosh\nu - \pmb{\alpha}\cdot\hat{\pmb{v}} \, \sinh\nu \\[0.05cm]
     \pmb{\varsigma}^{\prime} &=& \pmb{\varsigma}\, \cosh^2\frac{\nu}{2} + (-\hat{\pmb{v}}\cdot\pmb{\varsigma}\,\hat{\pmb{v}} + \hat{\pmb{v}}\times\pmb{\varsigma}\times\hat{\pmb{v}})\, \sinh^2\frac{\nu}{2} - \pmb{\beta}\times\hat{\pmb{v}}\, \sinh\nu 
     \\[0.05cm]
     \pmb{k}^{\prime} &=& \hat{\pmb{v}}\, k^0 \sinh\nu - \ii\pmb{\varsigma}\times\hat{\pmb{v}}\, \sinh\nu + \pmb{k}\, \cosh^2\frac{\nu}{2} + (\hat{\pmb{v}}\cdot\overline{\pmb{k}}\,\hat{\pmb{v}} - \hat{\pmb{v}}\times\overline{\pmb{k}}\times\hat{\pmb{v}})\, \sinh^2\frac{\nu}{2}
\end{array}\,.
\end{equation}}
\end{itemize}
The above actions should be compared with the representation of Sp$(4,\R)$ as operators acting on the classical observables $f(\pmb{z})$ as:
\begin{equation} \label{fginvz}
\mathrm{Sp}(4,\R) \ni g \,:\, f(\pmb{z}) \;\mapsto\; f^{\prime}(\pmb{z}) \equiv f\left(g^{-1}\diamond\pmb{z}\right)\,,
\end{equation}
with the definition \eqref{gpz}.

\subsection{``Massive" ``spin" (co-)adjoint orbits} \label{Subsec co-A orbit}
Taking steps parallel to the above, in this subsection, we study a class of the Sp$(4,\R)$ (co-)adjoint orbits, each being relevant to the transport of the element  $\big(2\ell X_0 + 2\jmath Y_3\big) = \begin{pmatrix} \ii\,\ell + \jmath\,\pmb{e}_3 & 0 \\ 0 & -\ii\,\ell + \jmath\,\pmb{e}_3 \end{pmatrix}$ of $\mathfrak{sp}(4,\R)$ (see \eqref{X0} and \eqref{Yi}), with $\ell,\jmath \in \R - \{0\}$,\footnote{We will show in the sequel that when $\ell=\jmath$, it corresponds to the situations known as the ``massless" ``spin" cases.} under the (co-)adjoint action \eqref{Ad_g AdS4}. In this case, the stabilizer subgroup is a subgroup of $S_{2\ell X_0 }=K$  \eqref{sgK}:
\begin{equation}\label{stabilizer1}
S_{2\ell X_0 + 2\jmath Y_3} = 
\left\{ h(\theta)\, r\big(\xi^{(\slashed{1}\slashed{2})}\big) 
\,, \, -\pi \leqslant \theta < \pi\,,\, \xi^{(\slashed{1}\slashed{2})}\, \in \mathrm{U}(1)\right\}
\sim \mathrm{S}\big(\mathrm{U}(1)\times \mathrm{U}(1)\big)\,,
\end{equation}
where $\xi^{(\slashed{1}\slashed{2})}:=\xi^4+\xi^3\pmb{e}_3$ (i.e.,  $\xi^1=\xi^2=0$) and since 
$\xi^{(\slashed{1}\slashed{2})}\in \mathrm{U}(1)$ then  $\xi^4,\xi^3\in\R$ and $(\xi^4)^2+(\xi^3)^2=1$.
Therefore, this class of the Sp$(4,\R)$ (co-)adjoint orbits can be realized as a homogeneous space homeomorphic to the group coset $\mathcal O\big(2\ell X_0 + 2\jmath Y_3\big) \sim \mathrm{Sp}(4,\R)/\mathrm{S}\big(\mathrm{U}(1)\times \mathrm{U}(1)\big) \sim \mathcal{D}^{(3)}\times\mathbb{S}^2$. 

Technically, again with respect to the time-space-Lorentz factorization \eqref{Spstdec} of the group, this realization can be achieved by transporting the element ($2\ell X_0 + 2\jmath Y_3$) under the (co-)adjoint action \eqref{Ad_g AdS4}, when $g$ involved in the action belongs to the subgroup 
\begin{equation}\label{subgroup}
L^{(\slashed{3})}_{sbr}=\Big\{s(\vrh,\hat{\pmb{u}})\, \lambda(\nu,\hat{\pmb{v}})\, r\big( \xi^{(\slashed{3})} \big)\;;\;
\vrh,\nu \in \R^+, \; \hat{\pmb{u}},\hat{\pmb{v}} \in \mathbb{S}^2, \; \xi^{(\slashed{3})}
\in \mathrm{SU}(2)/\mathrm{U}(1) \sim \mathbb{S}^2 \Big\} 
\end{equation} 
such that $s(\vrh,\hat{\pmb{u}})$ and $ \lambda(\nu,\hat{\pmb{v}})$ are given in Eq.~\eqref{spaceboots}, and $ r\big( \xi^{(\slashed{3})\big)}$ is given in Eq.~\eqref{sgK1} with $ \xi^{(\slashed{3})}:=\xi^{4}+\xi^1\pmb{e}_1+\xi^2\pmb{e}_2$ such that $\xi^{4},\xi^1,\xi^2\in\R$ and $(\xi^{4})^2+(\xi^1)^2+(\xi^2)^2=1$.

Accordingly, the (co-)adjoint action gives:
\begin{eqnarray}\label{bbbb}
\mbox{Ad}_g\big(2\ell X_0 + 2\jmath Y_3\big) &=& s(\vrh,\hat{\pmb{u}}) \; \lambda(\nu,\hat{\pmb{v}}) \; r\big( \xi^{(\slashed{3})} \big) \; \Big( 2\ell X_0 + 2\jmath Y_3 \Big)\; r^{-1}\big( \xi^{(\slashed{3})} \big) \; \lambda^{-1}(\nu,\hat{\pmb{v}}) \; s^{-1}(\vrh,\hat{\pmb{u}}) \nonumber\\
&=& \begin{pmatrix} \ii k^{0} + \pmb{\varsigma} & \pmb{k} \\ -\overline{\pmb{k}}& - \ii k^{0} + \pmb{\varsigma} \end{pmatrix} = \mathsf{X}( k^{0}, \pmb{\varsigma}, \pmb{k})\,,
\end{eqnarray}
where:
\begin{align}
k^{0} &= \ell\,\cosh\vrh \cosh\nu - \jmath\,\sinh\vrh \sinh\nu\,\big(\hat{\pmb{u}},\hat{\pmb{v}}, \hat{\pmb{\xi}}\big)\,,
\\
\pmb{\varsigma} &= \ell \,\sinh\varrho\sinh\nu\,\big(\hat{\pmb{v}}\times \hat{\pmb{u}}\big)
 +\; \jmath\, \Bigg( \cosh\nu \bigg[ \cosh\rho\, \hat{\pmb{\xi}}_{\perp \hat{\pmb{v}}} - (\cosh\rho - 1)\Big( (\hat{\pmb{u}}\cdot \hat{\pmb{\xi}}) - (\hat{\pmb{v}}\cdot \hat{\pmb{\xi}}) (\hat{\pmb{u}}\cdot \hat{\pmb{v}}) \Big) \hat{\pmb{u}} \bigg] \nonumber\\
& + \,(\hat{\pmb{v}}\cdot \hat{\pmb{\xi}}) \Big( \cosh\rho \, \hat{\pmb{v}}_{\perp \hat{\pmb{u}}} 
+ (\hat{\pmb{u}}\cdot \hat{\pmb{v}})\hat{\pmb{u}} \Big) \Bigg)\,,
\\
\pmb{k} &=\pmb{\alpha}+\ii\pmb{\beta} = \ell\Big( \hat{\pmb{v}}\cosh^2\frac{\varrho}{2}\sinh\nu + \hat{\pmb{u}}\hat{\pmb{v}}\hat{\pmb{u}}\sinh^2\frac{\varrho}{2}\sinh\nu - \ii \hat{\pmb{u}}\sinh{\varrho}\cosh\nu \Big) \nonumber \\
&+  \jmath\,\Bigg( \sinh\vrh \bigg[ \cosh\nu\, (\hat{\pmb{u}}\times \hat{\pmb{\xi}}_{\perp \hat{\pmb{v}}}) + (\hat{\pmb{v}}\cdot \hat{\pmb{\xi}}) (\hat{\pmb{u}}\times \hat{\pmb{v}}) \bigg]
+\; \ii \,\sinh\nu \left((\hat{\pmb{v}}\times \hat{\pmb{\xi}}) + (\cosh\varrho-1)\big(\hat{\pmb{u}},\hat{\pmb{v}},\hat{\pmb{\xi}}\big)\,  \hat{\pmb{u}}\right) \Bigg)\,,
\end{align}
where we have used the notations $\hat{\pmb{\xi}} \equiv \xi^{(\slashed{3})} \,\pmb{e}_3\, \widetilde{\xi^{(\slashed{3})}}\in \mathbb{S}^2$, $\big(\hat{\pmb{u}}, \hat{\pmb{v}}, \hat{\pmb{\xi}}\big) \equiv \hat{\pmb{u}}\cdot\big(\hat{\pmb{v}}\times\hat{\pmb{\xi}}\big)$, $\hat{\pmb{u}}_{\parallel \hat{\pmb{v}}} \equiv \hat{\pmb{v}}\cdot \hat{\pmb{u}}\,\hat{\pmb{v}}$, $\hat{\pmb{u}}_{\perp \hat{\pmb{v}}} \equiv \hat{\pmb{u}} - \hat{\pmb{u}}_{\parallel \hat{\pmb{v}}}$ and the formula $\hat{\pmb{v}}\hat{\pmb{u}}\hat{\pmb{v}}= \hat{\pmb{u}} - 2 \hat{\pmb{u}}_{\parallel \hat{\pmb{v}}}$. 

With respect to the relations that hold between the matrix elements of $\mathsf{X}( k^{0}, \pmb{\varsigma}, \pmb{k})$, there are two constraints that determine the (co-)adjoint orbit class $\mathcal O\big(2\ell X_0 + 2\jmath Y_3\big)$ of Sp$(4,\R)$ in the (dual of) its Lie algebra $\mathfrak{sp}(4,\R)$:\footnote{Again, the first identity is consistent with the constraint issued from the Killing form of the algebra for the generic element of the (co-)adjoint orbit class $\mathcal O\big(2\ell X_0 + 2\jmath Y_3\big)$ and the second one with the fact that the (co-)adjoint action \eqref{Ad_g AdS4} is determinant preserving.}
\begin{eqnarray}
\label{A} \ell^2 + \jmath^2 &=& (k^{0})^2 - \Vert \pmb{k} \Vert^2 + \Vert \pmb{\varsigma} \Vert^2\,, \\
\label{B} \big( \ell^2 - \jmath^2 \big)^2 &=& \det(\ii k^{0} + \pmb{\varsigma}) \det\Big( (-\ii k^{0} + \pmb{\varsigma}) + \overline{\pmb{k}} (\ii k^{0} + \pmb{\varsigma})^{-1}\pmb{k} \Big) = \Big( (k^{0})^2 - \Vert \pmb{k} \Vert^2 - \Vert \pmb{\varsigma} \Vert^2 \Big)^2\,,
\end{eqnarray}
where, to derive the latter equation, we have employed the identity \eqref{deta+b}. Then, $\mathcal O\big(2\ell X_0 + 2\jmath Y_3\big)$ can be viewed as:
\begin{align}\label{hasan}
\mathcal O\big(2\ell X_0 + 2\jmath Y_3\big) = \left\{( k^{0}, \pmb{\varsigma}, \pmb{k} )\,;\, \ell^2 + \jmath^2 = (k^{0})^2 - \Vert \pmb{k} \Vert^2 + \Vert \pmb{\varsigma} \Vert^2 \,, \; \big( \ell^2 - \jmath^2 \big)^2 = \Big( (k^{0})^2 - \Vert \pmb{k} \Vert^2 - \Vert \pmb{\varsigma} \Vert^2 \Big)^2 \right\}\,.
\end{align}

The application of the constraints \eqref{A} and \eqref{B} leads to the existence of three distinct and independent scenarios:
\begin{itemize}
  \item[(i)] Case $l^2=\jmath^2$. Then, in a physically relevant case, we have:
  \begin{equation} \label{leqpmj}
  l^2=\jmath^2 = (k^{0})^2 - \Vert \pmb{k} \Vert^2= \Vert \pmb{\varsigma} \Vert^2\,. 
  \end{equation}
  \item[(ii)] Cases $l^2 > \jmath^2$ and $(k^{0})^2  - \Vert \pmb{k} \Vert^2  - \Vert \pmb{\varsigma} \Vert^2 >0$ or $l^2 < j^2$ and $(k^{0})^2  - \Vert \pmb{k} \Vert^2 - \Vert \pmb{\varsigma} \Vert^2 < 0$. Then: 
  \begin{equation} \label{lneqj1}
  \ell^2 = (k^{0})^2 - \Vert \pmb{k} \Vert^2\,, \qquad \jmath^2 = \Vert \pmb{\varsigma} \Vert^2\,.
  \end{equation}
  \item[(iii)] Cases $l^2 > \jmath^2$ and $(k^{0})^2  - \Vert \pmb{k} \Vert^2  - \Vert \pmb{\varsigma} \Vert^2 <0$ or $l^2 < j^2$ and $(k^{0})^2  - \Vert \pmb{k} \Vert^2  - \Vert \pmb{\varsigma} \Vert^2 > 0$. Then:
  \begin{equation} \label{lneqj2}
  \ell^2 = \Vert \pmb{\varsigma} \Vert^2 \,, \qquad
  \jmath^2 = (k^{0})^2 - \Vert \pmb{k} \Vert^2\,.
  \end{equation}  
\end{itemize}
Each case corresponds to a distinct sub-manifold of the (co-)adjoint orbit $\mathcal O\big(2\ell X_0 + 2\jmath Y_3\big)$. Now, let us proceed with the first two scenarios that are both physically relevant and also consistent with the scalar case \eqref{poori}. As before, we associate appropriate physical dimensions to the involved variables as follows (recall that $ \pmb{k} = \pmb{\alpha} + \ii\pmb{\beta}$):
\begin{eqnarray}
k^{0} &=& \ell \,\frac{E}{mc^2} \,,\\
\pmb{\alpha} &=& \ell \,{\varkappa} \,\pmb{q} \,,\hskip2cm \ell\in\R^+\,,\\
\pmb{\beta} &=& \ell \,\frac{\pmb{p}}{mc} \,, 
\end{eqnarray} 
It is important to note that the parameter $\jmath$ is intentionally left separate as it is equal to $\pm \Vert \pmb{\varsigma} \Vert$, and so is associated with the concept of classical spin. Then, in case (i) (resp. (ii)), the (co-)adjoint orbit class \eqref{leqpmj} (resp. \eqref{lneqj1}) can be interpreted as the phase space for ``massless"  (resp. ``massive'') spin elementary systems living in AdS$_4$ spacetime. These co-adjoint orbits are Cartesian products of a two-dimensional sphere, with radius $\vert \jmath\vert$, with a six-dimensional two-sheeted hyperboloid in $\R^7$, equipped with the metric diag$(+,-,-,-,-,-,-)$ and with radius $\ell$. In this context, the constraint $\ell^2 = (k^{0})^2 - \Vert \pmb{k} \Vert^2$ turns into:
\begin{eqnarray}
E^2 = m^2c^4 + c^2\Vert \pmb{p}\Vert^2 +m^2c^4 {\varkappa}^2\Vert \pmb{q}\Vert^2 \,.
\end{eqnarray}
Under the null-curvature (Poincar\'{e}) limit, as ${\varkappa}\rightarrow 0$, this constraint clearly meets the mass shell hyperboloid \eqref{masshel}, and under non-relativistic (Newtonian) limit, as $c \to \infty$ and ${\varkappa} \to 0$ ($c{\varkappa} = \mathrm{constant} \equiv \omega$), the identity \eqref{Elargec}.

\setcounter{equation}{0} 
\section{Sp$(4,\R)$ representation(s) in the discrete series and contractions} \label{Sec AdS4 reps}
In this section, we study the construction of the AdS$_4$ UIRs relevant to the quantum reading of the described classical systems above. The so-called ``massive" elementary systems living in AdS$_4$ spacetime on the quantum level are associated with the discrete series UIRs $U^{(\varsigma,s)}$ of Sp$(4,\R)$ or of its universal covering $\overline{\mathrm{Sp}(4,\R)}$ \cite{evans67, Fronsdal 2,baelgagi92, Gazeau2022, Dobrev1, Dobrev2}; the parameters $\varsigma \in \R_+$ and $s\in \N/2$ (spin) are dimensionless, and satisfy $\varsigma > s +1$ (the lowest limit $\varsigma = s+1$ is the ``massless" case). For $\varsigma > s +2$, the representation Hilbert spaces are denoted by (Segal-)Bargmann-Fock spaces $\mathcal{F}^{(\varsigma,s)}$ and their elements are holomorphic $(2s+1)$-vector functions:
\begin{equation} \label{FBUIR}
\mathcal{D}^{(3)} \ni \pmb{z} \;\mapsto\; f(\pmb{z}) \in \C^{2s+1}\,,
\end{equation}
that are square integrable with respect to the bilinear form:
\begin{equation} \label{FBinner}
(f_1,f_2)_{(\varsigma,s)} = \mathcal{N}(\varsigma,s) \int_{\mathcal{D}^{(3)}} f_1(\pz)^{\dag}\,D^s\left(\frac{1}{1+\pz\bpz}\right)\,f_2(\pz)\,\big[\det(1+\pz\bpz)\big]^{\varsigma +s-3}\,d{\pz} \,, \quad d{\pz} \equiv \ud^3 \pmb{x}\, \ud^3 \pmb{y}\,.
\end{equation}
Note that: (i) The operator $D^s$ is the holomorphic extension to $\mathrm{M}(2,\C)$ of the irreducible $(2s+1)\times (2s+1)$-matrix representation of the SU$(2)$ group. (ii) The constant $\mathcal{N}(\varsigma,s)$ is chosen in such a way that the particular $(2s+1)$-vector-valued constant function $e_{ss}$ when the general $e_{s\rho}$, with $-s\leqslant \rho \leqslant s$, is defined by:
\begin{equation}\begin{array}{lll} \label{essig}
&\mbox{for integer $s$}:\quad\qquad & e_{s\rho} \equiv \left(\delta_{s,\rho},\, \delta_{s-1,\rho} ,\,   \dots\, ,\, \delta_{1,\rho} ,\, \delta_{0,\rho} ,\, \delta_{-1,\rho},\, \dots\, ,\, \delta_{-s+1,\rho} ,\, \delta_{-s,\rho} \right)^{\mathrm{t}}\,,\\[0.3cm]
&\mbox{for half-integer $s$}:\quad\qquad & e_{s\rho} \equiv \left( \delta_{s,\rho},\, \delta_{s-1,\rho} ,\,\dots \,,\, \delta_{\frac{1}{2},\rho} ,\,\delta_{-\frac{1}{2},\rho},\, \dots\, ,\,\delta_{-s+1,\rho} ,\, \delta_{-s,\rho} \right)^{\mathrm{t}},
\end{array}\end{equation}
has norm one (${\mathrm{t}}$ stands for the transpose). This constant is given by (see, for instance, Ref. \cite{onofri76}):
\begin{equation} \label{Nvzs}
\mathcal{N}(\varsigma,s) = \frac{8}{\pi^3}\,\left(\varsigma + s -\frac{3}{2}\right)\,(\varsigma-s-1)\,(\varsigma-s-2)\,.
\end{equation}
(iii) Modification of the inner product is required for the range $s+1\leqslant\varsigma \leqslant s +2$, where the Shilov boundary $[0,\pi]\times \mathbb{S}^2$ (i.e., the Lie sphere) of $\mathcal{D}^{(3)}$ plays a crucial r\^ole \cite{onofri76}.

The operators of the UIRs $U^{(\vsi,s)}$ of Sp$(4,\R)$ are defined by the following action on the functions $f$ of $\mathcal{F}^{(\vsi,s)}$:
\begin{equation} \label{UIRAdS}
\begin{split}
f^{\prime} (\pz) = \left(U^{(\vsi,s)}(g)\,f\right) (\pz) = \big[\det(-\overline{b}\pz + \overline{a})\big]^{-\vsi -s}\,D^s(\pz b^{\ast} + a^{\ast})\,f\left(g^{-1}\diamond \pz\right)\,,
\end{split}
\end{equation}
with $g^{-1} = \begin{pmatrix} a & b \\ - \overline{b} & \overline{a} \end{pmatrix}$ and $g^{-1}\diamond\pz = (a\pz +b)(-\overline b \pz + \overline a)^{-1}$, where here for convenience we consider $g^{-1}$ in the opposite way to the usual notations following Eq. \eqref{sp4inv}.

When multiplied by $\ii$, the $10$ infinitesimal generators of the action \eqref{UIRAdS} span a representation of the Lie algebra $\mathfrak{sp}(4,\R)$ by self-adjoint operators $L_{\alpha\beta}$ in $\mathcal{F}^{(\vsi,s)}$, with the decomposition in orbital and spin parts:
\begin{equation}\label{Lab}
\mathfrak{sp}(4,\R)\ni N_{\alpha\beta} \;\mapsto\; L_{\alpha\beta} = M_{\alpha\beta} + S_{\alpha\beta}\,.
\end{equation}
According to the Stone theorem \cite{Stone}, their respective expressions read:
\begin{align}
\label{L50} L_{50} &= \left(\pz\cdot \pmb{\nabla}_{\pz} + \vsi\right)\bu_{2s+1}\,, \qquad \pmb{\nabla}_{\pz} = \left(\frac{\partial}{\partial z^1}, \frac{\partial}{\partial z^2},\frac{\partial}{\partial z^3}\right)\,,\\
\label{L12} L_{12} &= \ii \left(z^1\frac{\partial}{\partial z^2} - z^2\frac{\partial}{\partial z^1}\right)\bu_{2s+1} - S_3\,,\\
\label{L23} L_{23} &= \ii \left(z^2\frac{\partial}{\partial z^3} - z^3\frac{\partial}{\partial z^2}\right)\bu_{2s+1} - S_1\,,\\
\label{L31} L_{31} &= \ii \left(z^3\frac{\partial}{\partial z^1} - z^1\frac{\partial}{\partial z^3}\right)\bu_{2s+1} - S_2\,,\\
\label{L51} L_{51} &=\ii \left(\frac{1+\pz\cdot\pz}{2}\frac{\partial}{\partial z^1} - z^1\left(\vsi+s +\pz\cdot \pmb{\nabla}_{\pz}\right)\right)\bu_{2s+1} +z^2S_3-z^3S_2\,,\\
\label{L52} L_{52} &=\ii \left(\frac{1+\pz\cdot\pz}{2}\frac{\partial}{\partial z^2} - z^2\left(\vsi+s +\pz\cdot \pmb{\nabla}_{\pz}\right)\right)\bu_{2s+1} -z^1S_3+z^3S_1\,,\\
\label{L53} L_{53} &=\ii \left(\frac{1+\pz\cdot\pz}{2}\frac{\partial}{\partial z^3}- z^3\left(\vsi+s +\pz\cdot \pmb{\nabla}_{\pz}\right)\right)\bu_{2s+1} +z^1S_2-z^2S_1\,,\\
\label{L01} L_{01} &=-\left(\frac{1-\pz\cdot\pz}{2}\frac{\partial}{\partial z^1} + z^1\left(\vsi+s +\pz\cdot \pmb{\nabla}_{\pz}\right)\right)\bu_{2s+1} -\ii z^2S_3 +\ii z^3S_2\,,\\
\label{L02} L_{02} &=-\left(\frac{1-\pz\cdot\pz}{2}\frac{\partial}{\partial z^2} + z^2\left(\vsi+s +\pz\cdot \pmb{\nabla}_{\pz}\right)\right)\bu_{2s+1} +\ii z^1S_3 -\ii z^3S_1\,,\\
\label{L03} L_{03} &=-\left(\frac{1-\pz\cdot\pz}{2}\frac{\partial}{\partial z^3} + z^3\left(\vsi+s +\pz\cdot \pmb{\nabla}_{\pz}\right)\right)\bu_{2s+1} -\ii z^1S_2 +\ii z^2S_1\,.
\end{align}
The $(2s+1)\times(2s+1)$-matrices $S_1$, $S_2$, and $S_3$ in the above are given respectively by:
\begin{align}
\label{S1} (S_1)_{mm^{\prime}}&= \frac{1}{2} \sqrt{(s+m)(s-m+1)} \,\delta_{m,m^{\prime}+1}+ \frac{1}{2} \sqrt{(s-m)(s+m+1)} \,\delta_{m,m^{\prime}-1} \,,\\
\label{S2} (S_2)_{mm^{\prime}}&= \frac{1}{2\ii} \sqrt{(s-m)(s+m+1)} \,\delta_{m,m^{\prime}-1} - \frac{1}{2\ii} \sqrt{(s+m)(s-m+1)} \,\delta_{m,m^{\prime}+1} \,,\\
\label{S3} (S_3)_{mm^{\prime}}&= m \,\delta_{mm^{\prime}} \,,
\end{align}
for $m$ and $m^{\prime}$ such that $-s\leqslant m,m^{\prime}\leqslant s$. They realize the spin $s$ representation of the Lie algebra $\mathfrak{su}(2)$:
\begin{equation} \label{comsu2}
\left[S_i,S_j\right] = \ii \epsilon_{ij}^{\,\,k}S_k\,, \quad i,j,k \in \{1,2,3\}\,.
\end{equation}
Considering the above, one checks that the generators $L_{\alpha\beta}$  satisfy the quantum version of the commutation rules \eqref{AdS4 algebra} (up to the factor $\ii$):
\begin{equation} \label{qcomrulsp4}
\left[L_{\alpha \beta},L_{\gamma \rho}\right] = \ii \left(\eta^{}_{\alpha \gamma} {L^{}_{\beta \rho}} + \eta^{}_{\beta \rho} {L^{}_{\alpha \gamma}} - \eta^{}_{\alpha \rho} {L^{}_{\beta \gamma}} - \eta^{}_{\beta \gamma} {L^{}_{\alpha \rho}} \right)\,.
\end{equation}

Like for the Poincar\'e group, kinematical group the  Minkowski spacetime, there exist two invariants that characterize each UIR of the AdS$_4$ group. They are the second-order and quartic-order Casimir operators respectively defined by:
\begin{equation} \label{cas12}
Q^{(1)} = - \frac{1}{2} L_{\alpha \beta}L^{\alpha\beta}\,, \qquad Q^{(2)} = - W_{\alpha} W^{\alpha},
\end{equation}
where $W_{\alpha} \equiv - \frac{1}{8}\epsilon_{\alpha \beta \gamma \delta \eta} L^{\beta \gamma}L^{\delta \eta}$. Taking into account the above infinitesimal generators \eqref{L50}-\eqref{L03}, the eigenvalues of the two Casimir operators, completely determining the UIRs $U^{(\vsi,s)}$, read as:
\begin{equation} \label{eigenads1}
\langle Q^{(1)} \rangle = \varsigma(\varsigma - 3) + s(s+1)\,, \qquad \langle Q^{(2)} \rangle = -(\varsigma-1)(\varsigma - 2)s(s+1)\,.
\end{equation}

The very point to be noticed here is that $L_{50}=M_{50}$ is the ladder operator for the UIR parameter $\vsi$. Its eigenvalues are $\vsi,\vsi+1,\vsi+2, \dotsc, \vsi+n, \dotsc$, which means that $U^{(\vsi,s)}$ is a lowest-weight representation in the discrete series. Since $\vsi$ is the lowest value of the discrete spectrum of the generator of ``time'' rotations in AdS$_4$ spacetime, it is naturally given a non-ambiguous meaning of rest energy when it is expressed in the energy AdS$_4$ units $\hbar {\varkappa} c$ ($\hbar$ and $c$ respectively being the Planck constant and the speed of light):
\begin{equation}\label{Erestads}
E^{\mathrm{rest}}_{\text{\tiny{AdS$_4$}}} \equiv \hbar c {\varkappa}\,\vsi\,.
\end{equation}
Therefore, the physical concept of ``energy at rest" survives with the deformation Poincar\'e $\longrightarrow$ AdS$_4$. In a shortcut (see the details in subsection \ref{Subsec Newton contraction}), we would like to point out that, at the difference with the flat spacetime limit, $E^{\mathrm{rest}}_{\text{\tiny{AdS$_4$}}}$ is not restricted to the pure mass energy (proper mass of the system), actually it includes as well a kind of pure quantum vibration energy due to the curvature \cite{GazeauTannoudji2021, Gazeau2020}, as we have already seen in the classical case.

\subsection{Poincar\'e contraction limit}
In this subsection, we describe the way in which the AdS$_4$ discrete series UIRs $ U^{(\vsi,s)}(g)$ become the massive Wigner UIRs {$\mathcal{P}^{(m>0,s)}$ with mass $m>0$} and spin $s$ of the Poincar\'e group in the null-curvature limit, $\varkappa \to 0$.\footnote{The representations $\mathcal{P}^{(m,s)}$ are described in Appendix \ref{appendix:WigPoin}.} To do this, we take parallel steps to those given in Ref. \cite{GazeauHussinJPA} (see also Refs. \cite{del Olmo 1997, De Bievre 1994, De Bievre 1993, De Bievre 1992, Karim 1996}) for such a limit of the AdS$_2$ discrete series UIRs, for which we need to consider the time-space-Lorentz decomposition \eqref{Spstdec} of the group, namely:
\begin{equation} \label{Spstdec2}
\mathrm{Sp}(4,\R) \ni g = j(\theta,\vrh,\hat{\pmb{u}})\,l(\xi,\nu,\hat{\pmb{v}}) = h(\theta)\,s(\vrh,\hat{\pmb{u}})\,r(\xi)\,\lambda(\nu,\hat{\pmb{v}})\,,
\end{equation}
(the explicit expression of the factors $h(\theta),\,r(\xi)),\,s(\vrh,\hat{\pmb{u}})$ and $\lambda(\nu,\hat{\pmb{v}})$ are given in \eqref{sgK1} and \eqref{spaceboots}) and in more detail by introducing physical dimensions in the group parameters. Hence, we put:
\begin{equation} \label{physdimcont}
\theta = {\varkappa} \,c\, t\,, \qquad \varrho = {\varkappa}\, x\,, \qquad \hat{\pmb{v}}\, \tanh\frac{\nu}{2} = \frac{\mathfrak{\textbf{v}}}{c}\,,
\end{equation}
where $c$ is the speed of light, $x\in\R^+$, and $\mathfrak{\textbf{v}}\equiv \big(0, \mathfrak{\textbf{v}}\big) = \big(0, (\text{v}^i)\big)\in\H$. Moreover, we consider the relation between the AdS$_4$ representation parameter $\varsigma$ and the Minkowski mass $m$ as \cite{Gazeau2020,GazeauTannoudji2021,Gazeau2022}:
\begin{eqnarray}\label{varsigmakappa}
\varsigma \equiv \varsigma({\varkappa}) = \frac{mc}{\hbar{\varkappa}} + \frac{3}{2} + \left( s-\frac{1}{2} \right)^2 \left( \frac{\hbar{\varkappa}}{2mc} + {\cal{O}}\left( \frac{{\varkappa}^2}{c^2} \right) \right)\,.
\end{eqnarray}

From the latter identity, when the curvature ${\varkappa}$ goes to zero and subsequently $\varsigma$ goes to infinity, it is apparent that the bilinear form \eqref{FBinner} or equivalently the space $\mathcal{F}^{(\varsigma,s)}$ is not well adapted to such a limiting process. Thus, we introduce a `weighted' (Segal-)Bargmann-Fock space:
\begin{eqnarray}\label{Weighted}
\mathcal{F}^{(\varsigma,s)}_W = \left\{ F(\pz,\overline{\pz}) = \big[\det(1+\pz\bpz)\big]^{\frac{\varsigma +s}{2}} f(\pz) \,,\, f(\pz)\in\mathcal{F}^{(\varsigma,s)} \right\},
\end{eqnarray}
where we have taken into account the {non-negativeness of} $\det(1+\pz\bpz) = \det(1+\bpz\pz)$ {(see \eqref{det1zz})}. This new function space is the Hilbert space of \emph{non-analytic} functions inside $\mathcal{D}^{(3)}$, having the specific form \eqref{Weighted}, and square-integrable with respect to:
\begin{equation} \label{FBinnerWWW}
(F_1,F_2)_{(\varsigma,s)} = \mathcal{N}(\varsigma,s) \, \int_{\mathcal{D}^{(3)}} F_1(\pz,\overline{\pz})^{\dag}\,D^s\left(\frac{1}{1+\pz\bpz}\right)\,F_2(\pz,\overline{\pz})\,\big[\det(1+\pz\bpz)\big]^{-3}\,d{\pz} \,,
\end{equation}
(remember that $d{\pz} \equiv \ud^3 \pmb{x}\, \ud^3 \pmb{y}$). The representation  $U^{(\vsi,s)}_W(g)$ on $\mathcal{F}^{(\varsigma,s)}_W$ is deduced from $U^{(\vsi,s)}(g)$ given by \eqref{UIRAdS}:
\begin{equation} \label{UIRAdSWWW}
\begin{split}
F^{\prime} (\pz,\overline{\pz}) = \left(U^{(\vsi,s)}_W(g)\,F\right) (\pz,\overline{\pz}) = \left[\frac{\det(-b\overline{\pz}+a)}{\det(-\overline{b}\pz + \overline{a})}\right]^{\frac{\vsi +s}{2}}\,D^s(\pz b^{\ast} + a^{\ast})\,F\left(g^{-1}\diamond \pz,\overline{g^{-1}\diamond {\pz}}\right)\,.
\end{split}
\end{equation}
Note that, above, we have used the identity \eqref{footnote 12} and the fact that the above procedure must not change the homogeneity property of the former construction.

Now, in order to eliminate the singularity from $F(\pz,\overline{\pz})$ \eqref{Weighted}, we must impose some constraints on the form of the original analytic function $f(\pz)$. To do this, we factorize $f(\pz)$ as:
\begin{equation}
f(\pz) \equiv f(\pz, {\varkappa}) = N({\varkappa})\, \big[ \det(1+\pz\widetilde{\pz}) \big]^{-\frac{\varsigma({\varkappa}) +s}{2}}\, \check{f}(\pz, {\varkappa})\,,
\end{equation}
where the function $\check{f}$ is analytic in both $\pz\in{\mathcal{D}}^{(3)}$ and ${\varkappa}\in\R^+$, $N({\varkappa})$ is a normalization factor possibly non-analytic in ${\varkappa}$. Note that, in the sequel, normalization will not be imposed to eliminate the term $N({\varkappa})$. The square-integrability condition then takes the form:
\begin{eqnarray}
(\check{f}_1,\check{f}_2)^{\mbox{\small{reg}}}_{(\varsigma,s)} = \mathcal{N}(\varsigma,s) \int_{\mathcal{D}^{(3)}} \check{f}_1(\pz,{\varkappa})^{\dag}\, D^s\left(\frac{1}{1+\pz\bpz}\right)\, \check{f}_2(\pz,{\varkappa})\, \left[\frac{\det(1+\pz\bpz)}{|\det(1+\pz\widetilde{\pz})|}\right]^{\varsigma +s}\, \big[\det(1+\pz\bpz)\big]^{-3}\, d{\pz} \,.
\end{eqnarray}
We accordingly restrict our considerations by working on the subspace of $\mathcal{F}^{(\varsigma,s)}_W$ which consists of functions of the form:
\begin{eqnarray}\label{iii}
F(\pz,\overline{\pz}) = \left[\frac{\det(1+\pz\bpz)}{\det(1+\pz\widetilde{\pz})}\right]^{\frac{\varsigma({\varkappa})+s}{2}} \check{f}(\pz)\,,\qquad \check{f}(\pz)\equiv \check{f}(\pz,\varkappa)\,.
\end{eqnarray}

As the final stage before getting involved with the contraction limit, we set $\pz\mapsto \pz_\ii\equiv -\ii\pz$. Then, the weight regular factor is such that:
\begin{eqnarray}\label{suri}
\lim_{{\varkappa}\rightarrow 0} \left[\frac{\det(1+\pz_\ii \overline{\pz_\ii})}{\det(1+\pz_\ii\widetilde{\pz_\ii})}\right]^{\frac{\varsigma({\varkappa})+s}{2}} = 1 \,.
\end{eqnarray}
The above formula can be easily checked through the parametrization of ${\mathcal{D}}^{(3)}$ in terms of $\pz \equiv \pz(p,q) \equiv \mbox{\eqref{invmom}}$, when we have in mind the dimensionless quantities introduced in Eqs.~\eqref{dimk0}, \eqref{dimal}  and \eqref{dimbet}; $\lim_{{\varkappa}\rightarrow 0}\pz(p,q) \equiv \textbf{x}_\circ(p)$ is a purely real vector quaternion (see Eq.~\eqref{invmom0}), and hence, $\lim_{{\varkappa}\rightarrow 0}\pz_\ii(p,q) = -\ii\textbf{x}_\circ(p) \equiv \textbf{x}_{\circ\ii}(p)$ is a purely imaginary vector quaternion.

Having all the above in mind, we now consider the null-curvature limit of \eqref{UIRAdSWWW} (when $\pz\mapsto\pz_\ii$ and taking  into account  Eqs.~\eqref{iii} and \eqref{suri}) in two steps:
\begin{itemize}
\item{When $g$ is restricted to the ``time-space translations" subgroup:
\begin{itemize}
\item{The ``time-translations" subgroup (see the first expression of \eqref{sgK1} taking $\theta={\varkappa} c t$):
\begin{eqnarray}\label{time}
h^{-1}({\varkappa} c t)= \begin{pmatrix} e^{-\ii {\varkappa} c t/2} & 0 \\ 0 & e^{\ii {\varkappa} c t/2} \end{pmatrix} \equiv \begin{pmatrix} a & b \\ - \overline{b} & \overline{a} \end{pmatrix} = g^{-1}\,,
\end{eqnarray}
we get:
\begin{eqnarray}
\lim_{{\varkappa}\rightarrow 0}\; \left(U_{W}^{\big(\varsigma({\varkappa}),s\big)}\big(h({\varkappa} c t)\big)\,F\right) (\pz_\ii,\overline{\pz_\ii}) &=& \lim_{{\varkappa}\rightarrow 0}\; \left[\frac{\det(e^{-\ii{\varkappa} ct/2})}{\det(e^{\ii{\varkappa} ct/2})}\right]^{\frac{1}{2}\left[ \frac{mc}{\hbar{\varkappa}} + \frac{3}{2} + \left( s-\frac{1}{2} \right)^2 \left( \frac{\hbar{\varkappa}}{2mc} + {\cal{O}}\left( \frac{{\varkappa}^2}{c^2} \right) \right) \right]+\frac{s}{2}} \nonumber\\
&& \quad\quad \times \; D^s\left( e^{\ii{\varkappa} ct/2} \right)\, \check{f}\left(e^{-\ii{\varkappa} ct} \pz_\ii\right) \nonumber\\
&=& e^{-\ii mc^2 t/\hbar}\, \check{f}\big(\textbf{x}_{\circ\ii}(p)\big)\,.
\end{eqnarray}
Note that, above, we have employed the abusive identification $\det(e^{\mp\ii{\varkappa} ct/2}) \equiv \det(e^{\mp\ii{\varkappa} ct/2}(1,\textbf{0}))$.}
\item{The ``space-translations" subgroup  (see the first expression of \eqref{spaceboots} taking $\varrho=\varkappa\,x$):
\begin{eqnarray}\label{space}
s^{-1}( {\varkappa} x, \hat{\pmb{u}}) = \begin{pmatrix} \cosh \frac{{\varkappa} x}{2} & -\hat{\pmb{u}}\, \sinh \frac{{\varkappa} x}{2} \\ \hat{\pmb{u}}\, \sinh \frac{{\varkappa} x}{2} & \cosh \frac{{\varkappa} x}{2} \end{pmatrix} \equiv \begin{pmatrix} a & b \\ - \overline{b} & \overline{a} \end{pmatrix} = g^{-1}\,,
\end{eqnarray}
for which we obtain:
\begin{align}
&\lim_{{\varkappa}\rightarrow 0}\; \left(U_{W}^{\big(\varsigma({\varkappa}),s\big)}\big(s({\varkappa} x, \hat{\pmb{u}})\big)\,F\right) (\pz_\ii,\overline{\pz_\ii}) = \nonumber\\
&\quad\quad \lim_{{\varkappa}\rightarrow 0}\; \left[\frac{\det(\hat{\pmb{u}}\,\overline{\pz_\ii} \, \sinh \frac{{\varkappa} x}{2} + \cosh \frac{{\varkappa} x}{2})}{\det(\hat{\pmb{u}}\,\pz_\ii \, \sinh \frac{{\varkappa} x}{2} + \cosh \frac{{\varkappa} x}{2})}\right]^{\frac{1}{2}\left[ \frac{mc}{\hbar{\varkappa}} + \frac{3}{2} + \left( s-\frac{1}{2} \right)^2 \left( \frac{\hbar{\varkappa}}{2mc} + {\cal{O}}\left( \frac{{\varkappa}^2}{c^2} \right) \right) \right] +\frac{s}{2}} \nonumber\\
&\quad\quad\quad \times \, D^s\left(\pz_\ii\, \hat{\pmb{u}} \,\sinh \frac{{\varkappa} x}{2} + \cosh\frac{{\varkappa} x}{2}\right) \, \check{f}\left(\left(\pz_\ii\, \cosh \frac{{\varkappa} x}{2} - \hat{\pmb{u}}\, \sinh\frac{{\varkappa} x}{2} \right) \left(\hat{\pmb{u}}\, \pz_\ii\, \sinh\frac{{\varkappa} x}{2} + \cosh\frac{{\varkappa} x}{2}\right)^{-1} \right) \nonumber\\
&\qquad = e^{-\big(\hat{\pmb{u}}\textbf{x}_{\circ\ii}(p)+\textbf{x}_{\circ\ii}(p)\hat{\pmb{u}}\big)mcx/2\hbar}\, \check{f}\big(\textbf{x}_{\circ\ii}(p)\big)\,.
\end{align}}
\end{itemize}}

\item{When $g$ is restricted to the Lorentz subgroup (see \eqref{solu lorentz}): 
     \begin{itemize}
     \item{The ``space-rotation" subgroup (see the second expression of \eqref{sgK1}):
           \begin{eqnarray}\label{Lor-xi}
           {r}^{-1}(\xi) = \begin{pmatrix} \widetilde{\xi} & 0 \\ 0 & \widetilde{\xi} \end{pmatrix} \equiv \begin{pmatrix} a & b \\ - \overline{b} & \overline{a} \end{pmatrix} = g^{-1}\,,
           \end{eqnarray}
           based upon which the null-curvature limit reads:
           \begin{eqnarray}
           \lim_{{\varkappa}\rightarrow 0}\; \left(U_{W}^{\big(\varsigma({\varkappa}),s\big)}\big({r}(\xi)\big)\,F\right) (\pz_\ii,\overline{\pz_\ii}) &=& D^s\big(\xi\big)\, \check{f}\left(\widetilde{\xi}\, \textbf{x}_{\circ\ii}(p)\, \widetilde{\xi}^{-1}\right) \,.
           \end{eqnarray}}
      \item{The ``boost-transformations" subgroup (see the second expression of \eqref{spaceboots} and rewriting 
      $(\nu,\hat{\pmb{v}}) =\mathfrak{\textbf{v}}$): 
           \begin{eqnarray}\label{boost}
       \lambda^{-1}(\mathfrak{\textbf{v}}) = \begin{pmatrix} \vartheta(\Vert \mathfrak{\textbf{v}}\Vert) & -\ii\, {\frac{\mathfrak{\textbf{v}}}{c}} \, \vartheta(\Vert \mathfrak{\textbf{v}}\Vert) \\ & \\ -\ii\, {\frac{\mathfrak{\textbf{v}}}{c}} \, \vartheta(\Vert \mathfrak{\textbf{v}}\Vert) \, & \vartheta(\Vert \mathfrak{\textbf{v}}\Vert) \end{pmatrix} \equiv \begin{pmatrix} a & b \\ - \overline{b} & \overline{a} \end{pmatrix} = g^{-1}\,, \qquad \vartheta(\Vert \mathfrak{\textbf{v}}\Vert) = \frac{1}{\sqrt{1-\frac{\mathfrak{\textbf{v}}\cdot \mathfrak{\textbf{v}}}{c^2}}}\,,
           \end{eqnarray}
           based upon which we find:
          \begin{equation} \begin{array}{lll}
          &\ds \lim_{{\varkappa}\rightarrow 0}\; \left(U_{W}^{\big(\varsigma({\varkappa}),s\big)}\big(\lambda(\mathfrak{\textbf{v}})\big)\,F\right) (\pz_\ii,\overline{\pz_\ii}) \\[0.2cm]
           &\hspace{.25cm} \ds= \lim_{{\varkappa}\rightarrow 0}\; \left[\frac{\det\Big(\ii\, {\frac{\mathfrak{\textbf{v}}}{c}\, \overline{\pz}_\ii} \,\vartheta(\Vert \mathfrak{\textbf{v}}\Vert) + \vartheta(\Vert \mathfrak{\textbf{v}}\Vert)\Big)}{\det\Big(-\ii\, {\frac{\mathfrak{\textbf{v}}}{c}\, \pz_\ii}\, \vartheta(\Vert \mathfrak{\textbf{v}}\Vert) + \vartheta(\Vert \mathfrak{\textbf{v}}\Vert)\Big)}\right]^{\frac{1}{2}\left[ \frac{mc}{\hbar{\varkappa}} + \frac{3}{2} + \left( s-\frac{1}{2} \right)^2 \left( \frac{\hbar{\varkappa}}{2mc} + {\cal{O}}\left( \frac{{\varkappa}^2}{c^2} \right) \right) \right] +\frac{s}{2}} \\[0.2cm]
           &\hspace{.5cm} \times\, D^s \left(-\ii\, {\pz_\ii\, \frac{\mathfrak{\textbf{v}}}{c}}\, \vartheta(\Vert \mathfrak{\textbf{v}}\Vert) + \vartheta(\Vert \mathfrak{\textbf{v}}\Vert) \right) \, \check{f}\left(\left({\pz_\ii}\, \vartheta(\Vert \mathfrak{\textbf{v}}\Vert) - \ii\, {\frac{\mathfrak{\textbf{v}}}{c}}\, \vartheta(\Vert \mathfrak{\textbf{v}}\Vert)\right) \left(-\ii\, {\frac{\mathfrak{\textbf{v}}}{c}\, \pz_\ii}\, \vartheta(\Vert \mathfrak{\textbf{v}}\Vert) + \vartheta(\Vert \mathfrak{\textbf{v}}\Vert) \right)^{-1} \right) \\[0.35cm]
           & \hspace{.25cm}\ds =D^s\left(-\ii\, {\textbf{x}_{\circ\ii}(p)\, \frac{\mathfrak{\textbf{v}}}{c}}\, \vartheta(\Vert \mathfrak{\textbf{v}}\Vert) + \vartheta(\Vert \mathfrak{\textbf{v}}\Vert) \right) \, \\[0.2cm]
           &\hspace{1.35cm}\ds\times\, \check{f}\left(\left({\textbf{x}_{\circ\ii}(p)}\, \vartheta(\Vert \mathfrak{\textbf{v}}\Vert) - \ii\, {\frac{\mathfrak{\textbf{v}}}{c}}\, \vartheta(\Vert \mathfrak{\textbf{v}}\Vert)\right) \left(-\ii\, {\frac{\mathfrak{\textbf{v}}}{c}\, \textbf{x}_{\circ\ii}(p)}\, \vartheta(\Vert \mathfrak{\textbf{v}}\Vert) + \vartheta(\Vert \mathfrak{\textbf{v}}\Vert) \right)^{-1} \right)\,.
            \end{array} \end{equation}}
      \end{itemize}}
\end{itemize}
Combining these four elementary representations enables the recovery of the massive Wigner UIRs \eqref{WopUIR} when expressing the mass-shell variables used in the latter in terms of the coordinates in the ball $\mathbb{B}_3$ along \eqref{invmom0}.

\subsection{Dilation of the classical domain and Newton-Hooke contraction limit} \label{Subsec Newton contraction}
Here, to get the Newton-Hooke contraction limit of the AdS$_4$ discrete series UIRs $ U^{(\vsi,s)}(g)$, we take parallel steps to those given in Ref. \cite{GazeauRenaudPLA} for such a limit of the AdS$_2$ discrete series UIRs (see also Ref.~\cite{del Olmo IJTP}).\footnote{The  UIRs of the Newton-Hooke group are described in Appendix \ref{appendix:Newton-Hooke}.} We first dilate the classical domain $\mathcal{D}^{(3)} = \big\{ \pz\,,\, \bu_2 - \pmb{Z}\pmb{Z}^{\dag} >0 \big\}$ \eqref{domC3}, where the one-to-one map $\pz \mapsto \pmb{Z}(\pz)$ is given by Eq.~\eqref{cqumat}. The dilation operation consists of the map:
\begin{equation} \label{mapgam}
\gamma\,:\, \mathcal{D}^{(3)} \,\mapsto\, \mathcal{D}^{(3)}_{\mathfrak{R}} = \Big\{\pmb{\mathfrak{z}}\,;\, {\mathfrak{R}}^2\bu-\pmb{\mathfrak{Z}}\pmb{\mathfrak{Z}}^{\dag}>0 \Big\}\,,
\end{equation}
corresponding to the homographic action:
\begin{equation} \label{matgam}
\mathcal{D}^{(3)} \ni \pz \,\mapsto\, \gamma\diamond \pz = \pmb{\mathfrak{z}} =
\begin{pmatrix} 1 & 0 \\ 0 & 1/{\mathfrak{R}} \end{pmatrix} \diamond\pz = {\mathfrak{R}} \pz\,,
\end{equation}
where ${\mathfrak{R}}$ has $[\mathrm{action}]^{\frac{1}{2}}$ as a physical dimension (e.g.  $\sqrt{2mc/{\varkappa}}$ if we deal with a mass $m$), and $\pmb{\mathfrak{Z}}$ is the matrix associated  
to $\pmb{\mathfrak{z}}$. It then results in the action of $g \in \mathrm{Sp}(4,\R)$ on $\mathcal{D}^{(3)}$ is transformed to the following action on $\mathcal{D}^{(3)}_{\mathfrak{R}}$:
\begin{equation}\label{gamg}
\begin{split}
\pz \,\mapsto\, g\diamond \pz \quad  \Leftrightarrow \quad \pmb{\mathfrak{z}} \,\mapsto\, \gamma g \gamma^{-1}\diamond  \pmb{\mathfrak{z}} =
\begin{pmatrix} a & {\mathfrak{R}} b \\ - \frac{\overline{b}}{{\mathfrak{R}}} & \overline{a} \end{pmatrix}\diamond \pmb{\mathfrak{z}} = (a\pmb{\mathfrak{z}} +{\mathfrak{R}}b)\left(-\frac{\overline{b}}{{\mathfrak{R}}}\pmb{\mathfrak{z}} + \overline{a}\right)^{-1} \equiv g^{}_{\mathfrak{R}} \diamond  \pmb{\mathfrak{z}}\,.
\end{split}
\end{equation}
Then, we consider the new (Segal-)Bargmann-Fock space $\mathcal{F}_{\mathfrak{R}}^{(\varsigma,s)}$ whose elements are holomorphic $(2s+1)$-vector functions:
\begin{equation} \label{FBUIRR}
\mathcal{D}^{(3)}_{\mathfrak{R}} \ni \pmb{\mathfrak{z}} \,\mapsto\, f(\pmb{\mathfrak{z}}) \in \C^{2s+1}\,,
\end{equation}
that are square integrable with respect to the bilinear form:
\begin{equation}\label{innerFockR}
(f_1,f_2)^{\mathfrak{R}}_{(\varsigma,s)} = {\mathfrak{R}}^{-6} \mathcal{N}(\varsigma,s) \int_{\mathcal{D}_{\mathfrak{R}}^{(3)}} f_1(\pmb{\mathfrak{z}})^{\dag}\, D^s\left(\frac{1}{1+{\mathfrak{R}}^{-2}\pmb{\mathfrak{z}}\overline{\pmb{\mathfrak{z}}}}\right)\, f_2(\pmb{\mathfrak{z}})\, \big[\det(1+{\mathfrak{R}}^{-2}\pmb{\mathfrak{z}}\overline{\pmb{\mathfrak{z}}})\big]^{\varsigma +s-3}\, {d{\pmb{\mathfrak{z}}}} \,,
\end{equation}
where $d{\pmb{\mathfrak{z}}}= \mathfrak{R}^{6} \, d\pz$. Using the dilation $\gamma$ introduced in \eqref{matgam} and its representation $\Gamma\,:\, \mathcal{F}^{(\varsigma,s)} \mapsto \mathcal{F}_{\mathfrak{R}}^{(\varsigma,s)}$ defined as:
\begin{eqnarray}
f\mapsto \Gamma f\;:\; \, \big(\Gamma f\big)(\pmb{\mathfrak{z}})= f(\gamma^{-1}\diamond \pmb{\mathfrak{z}}) = f({\mathfrak{R}}^{-1}\pmb{\mathfrak{z}})\,,
\end{eqnarray}
the representation $U^{(\vsi,s)}$ given in \eqref{UIRAdS} now acts on $\mathcal{F}_{\mathfrak{R}}^{(\varsigma,s)}$ as:
\begin{equation} \label{DilUIRAdS}
f^{\prime} (\pmb{\mathfrak{z}}) = \left(U_{\mathfrak{R}}^{(\vsi,s)}(g)\,f\right) (\pmb{\mathfrak{z}}) = \big[\det(-\overline{b}\,\pmb{\mathfrak{z}}/{\mathfrak{R}} + \overline{a})\big]^{-\vsi -s}\, D^s\big(\pmb{\mathfrak{z}} \, b^{\ast}/{\mathfrak{R}} + a^{\ast}\big)\,f\left(\gamma g^{-1}\gamma^{-1}\diamond \pmb{\mathfrak{z}}\right)\,,
\end{equation}
with $g^{-1} = \begin{pmatrix} a & b \\ - \overline{b} & \overline{a} \end{pmatrix}$.

We now describe the way that these representations become the UIRs with spin $s$ of the Newton-Hooke group, in the non-relativistic limit, $c\to \infty$ and ${\varkappa}\to 0$, while $c{\varkappa} \equiv \omega$ remains unchanged. Again, by introducing physical dimensions in the group parameters as \eqref{physdimcont}, we need to consider the time-space-Lorentz decomposition \eqref{Spstdec2}. Moreover, we again set:
\begin{equation}\begin{array}{lll}
\varsigma \equiv \varsigma({\varkappa},c) &=& \ds\frac{mc}{\hbar{\varkappa}} + \frac{3}{2} + \left( s-\frac{1}{2} \right)^2 \left( \frac{\hbar{\varkappa}}{2mc} + {\cal{O}}\left( \frac{{\varkappa}^2}{c^2} \right) \right)\,,\\[0.3cm]
&=&\ds \frac{mc^2}{\hbar\omega} + \frac{3}{2} + \left( s-\frac{1}{2} \right)^2 \left( \frac{\hbar\omega}{2mc^2} + {\cal{O}}\left( \frac{\omega^2}{c^4} \right) \right)\,,
\end{array}\end{equation}
and as already pointed out above we write ${\mathfrak{R}} = \sqrt{\frac{2mc}{{\varkappa}}}$,  where $m$ is a Minkowskian mass. 

Accordingly, the Newton-Hooke contraction limit of the representations $U_{{\mathfrak{R}}=\sqrt{2mc/{\varkappa}}}^{\big(\varsigma({\varkappa},c),s\big)}(g)$ \eqref{DilUIRAdS} reads:
\begin{itemize}
\item{When $g$ is restricted to the ``time-space translations" subgroup:
      \begin{itemize}
      \item{The ``time-translations" subgroup (see Eq.~\eqref{time}) gives:
            \begin{eqnarray}\label{q E}
             \lim_{\substack{{\varkappa}\rightarrow 0,c\rightarrow\infty\\c{\varkappa} = \omega}}\; \left(U_{\sqrt{2mc/{\varkappa}}}^{\big(\varsigma({\varkappa},c),s\big)}\big(h({\varkappa} c t)\big)\,f\right) (\pmb{\mathfrak{z}}) &=& \lim_{\substack{{\varkappa}\rightarrow 0,c\rightarrow\infty\\c{\varkappa} = \omega}}\; \left[\det(e^{\ii{\varkappa} ct/2})\right]^{-\left[ \frac{mc^2}{\hbar\omega} + \frac{3}{2} + \left( s-\frac{1}{2} \right)^2 \left( \frac{\hbar\omega}{2mc^2} + {\cal{O}}\left( \frac{\omega^2}{c^4} \right) \right) \right]-s} 
             \nonumber\\
             && \quad\quad\quad\quad \times \; D^s\left( e^{\ii{\varkappa} ct/2} \right)\, f\left(e^{-\ii{\varkappa} ct} \pmb{\mathfrak{z}}\right) \nonumber\\[0.3cm]
             &=& {{\left(\lim_{c\rightarrow\infty}e^{-\ii mc^2 t/\hbar}\right)}}\, e^{-\ii\frac{3}{2}\omega t}\, f\left(e^{-\ii\omega t}\pmb{\mathfrak{z}}\right)\,.
             \end{eqnarray}
             Note that: (i) Above, we have used the fact that $D^s$ is a homogeneous function of degree $2s$ (see Appendix \ref{appendix:SU(2)}), and hence, $D^s\left( e^{\ii\omega t/2} \right) = e^{\ii\omega st}$. (ii) Clearly, the first factor has \emph{no} specific limit at $c\rightarrow\infty$. To eliminate this term, we put forward a phase factor $e^{\ii mc^2 t/\hbar}$ to be combined with the action \eqref{DilUIRAdS} from the very beginning. Then, we precisely obtain the corresponding Newton-Hooke contraction limit as:
             \begin{eqnarray}
             \lim_{\substack{{\varkappa}\rightarrow 0,c\rightarrow\infty\\c{\varkappa} = \omega}}\; \left(e^{\ii mc^2 t/\hbar}\, U_{\sqrt{2mc/{\varkappa}}}^{\big(\varsigma({\varkappa},c),s\big)}\big(h({\varkappa} c t)\big)\,f\right) (\pmb{\mathfrak{z}}) = e^{-\ii\frac{3}{2}\omega t}\, f\left(e^{-\ii\omega t}\pmb{\mathfrak{z}}\right)\,.
             \end{eqnarray}}
      \item{The ``space-translations" subgroup (see Eq. \eqref{space}):
           \begin{align}
           &\lim_{\substack{{\varkappa}\rightarrow 0,c\rightarrow\infty\\c{\varkappa} = \omega}}\; \left(U_{\sqrt{2mc/{\varkappa}}}^{\big(\varsigma({\varkappa},c),s\big)}\big(s({\varkappa} x, \hat{\pmb{u}})\big)\,f\right) (\pmb{\mathfrak{z}}) \nonumber\\
           & \quad = \lim_{\substack{{\varkappa}\rightarrow 0,c\rightarrow\infty\\c{\varkappa} = \omega}}\; \left[\det(\frac{\hat{\pmb{u}}\,\pmb{\mathfrak{z}} \, \sinh \frac{{\varkappa} x}{2}}{\sqrt{\frac{2mc}{{\varkappa}}}} + \cosh \frac{{\varkappa} x}{2})\right]^{-\left[ \frac{mc^2}{\hbar\omega} + \frac{3}{2} + \left( s-\frac{1}{2} \right)^2 \left( \frac{\hbar\omega}{2mc^2} + {\cal{O}}\left( \frac{\omega^2}{c^4} \right) \right) \right] -s} \nonumber\\
           &\quad\times \, D^s\left(\frac{\pmb{\mathfrak{z}}\, \hat{\pmb{u}} \,\sinh \frac{{\varkappa} x}{2}}{\sqrt{\frac{2mc}{{\varkappa}}}} + \cosh\frac{{\varkappa} x}{2}\right) \, f\left(\left(\pmb{\mathfrak{z}}\, \cosh\frac{{\varkappa} x}{2} - \hat{\pmb{u}}\, \sqrt{\frac{2mc}{{\varkappa}}}\, \sinh \frac{{\varkappa} x}{2} \right) \left(\frac{\hat{\pmb{u}}\,\pmb{\mathfrak{z}}\, \sinh\frac{{\varkappa} x}{2}}{\sqrt{\frac{2mc}{{\varkappa}}}} + \cosh \frac{{\varkappa} x}{2}\right)^{-1} \right) \nonumber\\
           &\quad = e^{- \frac{1}{2\hbar}\big((\hat{\pmb{u}}\pmb{\mathfrak{z}} + \pmb{\mathfrak{z}}\hat{\pmb{u}})\sqrt{\frac{m\omega}{2}}x + \frac{m\omega x^2}{2}\big)} \,f\left(\pmb{\mathfrak{z}}-\hat{\pmb{u}}\, \sqrt{\frac{m\omega}{2}}\, x\right)\,.
           \end{align}}
       \end{itemize}}
 \item{When $g$ is restricted to the Lorentz subgroup (see \eqref{solu lorentz}): 
      \begin{itemize}
      \item{The ``space-rotation" subgroup (see Eq.\eqref{Lor-xi}), for which we have:
           \begin{eqnarray}
           \lim_{\substack{{\varkappa}\rightarrow 0,c\rightarrow\infty\\c{\varkappa} = \omega}}\; \left(U_{\sqrt{2mc/{\varkappa}}}^{\big(\varsigma({\varkappa},c),s\big)}\big(r(\xi)\big)\,f\right) (\pmb{\mathfrak{z}}) = D^s\big(\xi\big)\, f\left(\widetilde{\xi}\, \pmb{\mathfrak{z}}\, \widetilde{\xi}^{-1} \right) \,.
           \end{eqnarray}}
      \item{The ``boost-transformations" subgroup (see Eq.\eqref{boost}), based upon which we find:
           \begin{align}
           &\lim_{\substack{{\varkappa}\rightarrow 0,c\rightarrow\infty\\c{\varkappa} = \omega}}\; \left(U_{\sqrt{2mc/{\varkappa}}}^{\big(\varsigma({\varkappa},c),s\big)}\big(\lambda(\mathfrak{\textbf{v}})\big)\,f\right) (\pmb{\mathfrak{z}}) \nonumber\\
           & \hspace{1cm} = \lim_{\substack{{\varkappa}\rightarrow 0,c\rightarrow\infty\\c{\varkappa} = \omega}}\; \left[\det( -\ii\, {\sqrt{\frac{\varkappa}{2mc}}\, \frac{\mathfrak{\textbf{v}}}{c}\, \pmb{\mathfrak{z}}\, \vartheta(\Vert \mathfrak{\textbf{v}}\Vert)} + \vartheta(\Vert \mathfrak{\textbf{v}}\Vert))\right]^{-\left[ \frac{mc^2}{\hbar\omega} + \frac{3}{2} + \left( s-\frac{1}{2} \right)^2 \left( \frac{\hbar\omega}{2mc^2} + {\cal{O}}\left( \frac{\omega^2}{c^4} \right) \right) \right] -s} \nonumber\\
           & \hspace{3cm} \times\, D^s\left(-\ii\, {\sqrt{\frac{\varkappa}{2mc}}\, \pmb{\mathfrak{z}}\, \frac{\mathfrak{\textbf{v}}}{c}\, \vartheta(\Vert \mathfrak{\textbf{v}}\Vert)} + \vartheta(\Vert \mathfrak{\textbf{v}}\Vert)\right) \nonumber\\
           & \hspace{3cm} \times\, f\left(\left({\pmb{\mathfrak{z}}}\, \vartheta(\Vert \mathfrak{\textbf{v}}\Vert) - \ii\, {\sqrt{\frac{2mc}{{\varkappa}}}\, \frac{\mathfrak{\textbf{v}}}{c}}\, \vartheta(\Vert \mathfrak{\textbf{v}}\Vert)\right) \left(-\ii\, {\sqrt{\frac{\varkappa}{2mc}}\, \frac{\mathfrak{\textbf{v}}}{c}\, \pmb{\mathfrak{z}}\, \vartheta(\Vert \mathfrak{\textbf{v}}\Vert)} + \vartheta(\Vert \mathfrak{\textbf{v}}\Vert) \right)^{-1} \right) \nonumber\\
           & \hspace{1cm} = e^{\frac{1}{\hbar}\big(-\frac{m}{\omega} \mathfrak{\textbf{v}}\cdot \mathfrak{\textbf{v}} + \ii(\mathfrak{\textbf{v}}\pmb{\mathfrak{z}} + \pmb{\mathfrak{z}}\mathfrak{\textbf{v}})\sqrt{\frac{m}{2\omega}}\big)} f\left(\pmb{\mathfrak{z}} - \ii\, \sqrt{\frac{2m}{\omega}}\, \mathfrak{\textbf{v}}\right)\,.
           \end{align}}
      \end{itemize}}
\end{itemize}

As the final remark in this subsection, we would like to draw attention to the Hamiltonian operator, strictly speaking, to the infinitesimal operator issued from the formula \eqref{q E}:
\begin{eqnarray}
H=\ii\hbar \left( \frac{\ud}{\ud t}\, U_{\sqrt{2mc/{\varkappa}}}^{\big(\varsigma({\varkappa},c),s\big)}\big(h({\varkappa} c t)\big) \right)\bigg|_{t=0} = \hbar\omega\, \pmb{\mathfrak{z}}\frac{\ud}{\ud \pmb{\mathfrak{z}}} + mc^2 + \frac{3}{2} \hbar\omega\,,
\end{eqnarray}
which is the quantum counterpart of Eq. \eqref{Elargec}.

In addition, it is crucial to emphasize that this description of the Newton-Hooke contraction limit of the AdS$_4$ discrete series UIRs $U^{(\vsi,s)}(g)$ acting in the Fock-Bargman spaces $\mathcal{F}_\infty^{(\varsigma,s)}$ is consistent with the representations \eqref{uirm}. Since the latter are formulated in terms of wave functions in momentum space, it is necessary to proceed with an integral transform based on the reproducing kernel derived from three-dimensional spin coherent states (for the case of one-dimensional coherent states, refer to Ref.~\cite{GazeauWiley}) to express \eqref{uirm} in terms of the Fock-Bargman spaces $\mathcal{F}_\infty^{(\varsigma,s)}$ of spin holomorphic functions on $\C^3$ (remember that the spaces $\mathcal{F}_\infty^{(\varsigma,s)}$ are the limit of the spaces $\mathcal{F}_{\mathfrak{R}}^{(\varsigma,s)}$  \eqref{FBUIRR} since $\mathfrak{R}$ goes to $\infty$ when $\varkappa\rightarrow 0$ and $c\rightarrow\infty$), equipped with the scalar product limit of \eqref{innerFockR} as $c$ tends to infinity:
\begin{equation} \label{FockC3inf}
(f_1,f_2) = \frac{1}{(\pi\hbar)^3} \int_{\C^3} f_1(\pmb{\mathfrak{z}})^{\dag}\, f^{}_2(\pmb{\mathfrak{z}})\, e^{-\dfrac{\Vert \pmb{\mathfrak{z}}\Vert^2}{\hbar}} \,d{\pmb{\mathfrak{z}}} \,.
\end{equation}

\setcounter{equation}{0} 
\section{Reproducing kernel Hilbert space}
We here briefly study the (Segal-)Bargmann-Fock spaces $\mathcal{F}^{(\vsi,s)}$, which  
are indeed reproducing-kernel spaces. The $(2s+1)\times (2s+1)$-matrix-valued reproducing kernel is given by:
\begin{equation} \label{repkerFB}
K^{(\vsi,s)}(\pz,\overline{\pz^{\prime}}) = \big[\det(1+\pz\overline{\pz^{\prime}})\big]^{-\vsi -s}\,D^s\left(1+\pz\overline{\pz^{\prime}}\right)\,.
\end{equation}
The reproducing property reads as:
\begin{equation} \label{repprop}
f(\pz) = \left({K^{(\vsi,s)}(\pz,\cdot)}^{\dag},f\right)_{(\vsi,s)}\,, \quad \forall\,f\in \mathcal{F}^{(\vsi,s)}\,.
\end{equation}
The following separating expansion of the kernel allows us to determine an orthonormal basis for $\mathcal{F}^{(\vsi,s)}$:
\begin{equation} \label{orthobasisFB}
K^{(\vsi,s)}(\pz,\pz^{\prime}) = \sum_\nu \mathbf{F}^{(\vsi,s)}_{\nu}(\pz)\, {\mathbf{F}^{(\vsi,s)}_{\nu}(\pz^{\prime})}^{\dag}\,,
\end{equation}
where the analytic $\mathbf{F}^{(\vsi,s)}_{\nu}$s are $(2s+1)$-vector-valued, and $\nu$ stands for a set of discrete labels. More details are given in Appendix \ref{appendix:kernelH}.

For instance, in the scalar case, we have the expansion (see Appendix \ref{appendix:SpecialFunctions} for justifications and details):
\begin{equation} \label{expkers0}
\begin{split}
K^{(\vsi,0)}(\pz,\pz^{\prime}) &\equiv \big[\det\left( 1 + \pz\overline{\pz^{\prime}}\right)\big]^{-\vsi} = \sum_{l=0}^{+\infty}\sum_{k = 0}^{\lfloor \frac{l}{2}\rfloor} \sum_{m=2k-l}^{l-2k}\,a_{\vsi,l,k}\,(\pz\cdot\pz)^{k}\, \mathcal{Y}_{l-2k,m}(\pz)\,(\overline{\pz^{\prime}\cdot\pz^{\prime}})^{k}\,\overline{\mathcal{Y}_{l-2k,m}(\pz^{\prime})}\\
&\equiv \sum_{\nu} \mathrm{F}^{\vsi}_{\nu}(\pz)\, \overline{ \mathrm{F}^{\vsi}_{\nu}(\pz^{\prime})}\,.
\end{split}
\end{equation}
with:
\begin{align}
\label{nkm1m2} \nu & \equiv (l,k,m)\,, \qquad l\in \N\,, \;\; 0\leqslant k\leqslant \left\lfloor \frac{l}{2}\right\rfloor\,, \;\; 2k-l\leqslant m \leqslant l-2k\,, \\[0.2cm]
\label{dddddd} a_{\vsi,l,k} &= \frac{2^{2\vsi-1}\pi \,\Gamma(k+\vsi - 1/2)\, \Gamma(\vsi + l - k)}{\Gamma(2\vsi-1)\, k!\, 
\Gamma(l-k + 3/2)}\,.
\end{align}
The functions $\mathcal{Y}_{l-2k,m}(\pz)$ which appear in \eqref{expkers0} play an essential role in establishing all this (Segal-)Bargmann-Fock material. They are holomorphic polynomial extensions to $\C^3$ of the solid spherical harmonics multiplied by even powers of the Euclidean distance in $\R^3$. All detailed expressions and properties are given in Appendix \ref{appendix:SpecialFunctions}. We show there that the polynomials:
\begin{equation} \label{Enus0}
{\mathbf{F}}^{\vsi}_{\nu}(\pz) = \sqrt{a_{\vsi,l,k}}\, (\pz\cdot\pz)^{k}\, \mathcal{Y}_{l-2k,m}(\pz)\,,
\end{equation}
with a suitable normalization encoded by the new coefficients $\sqrt{a_{\vsi,l,k}}$, form an orthonormal basis of the (Segal-)Bargmann-Fock Hilbert space $\mathcal{F}^{(\vsi,0)} \equiv \mathcal{F}^{\vsi}$.

The extension to arbitrary spin $s$ involves the holomorphic extensions of the so-called spinor or vector spherical harmonics. The latter are defined in terms of the spherical harmonics as \cite{edmonds96}:
\begin{equation} \label{spsphar}
Y_{slJM}(\theta,\phi) = \sum_{m\,\rho} (s\rho l m| s l J M)\,e_{s\rho}\,Y_{lm}(\theta,\phi)\,,
\end{equation}
where the vector-coupling (i.e., Clebsch-Gordan or Wigner) coefficients are defined in \eqref{3japc}. By construction, these vector-valued functions are eigenvectors of $\pmb{L}^2 \equiv \sum_{i<j}L_{ij}^2$, $\pmb{S}^2 \equiv \sum_{i<j} S_{ij}^2$, $\pmb{J}^2 \equiv \sum_{i<j}J_{ij}^2$, $J_{12}$, with $J_{ij} = L_{ij}+ S_{ij}$, corresponding to the eigenvalues $l(l+1)$, $s(s+1)$, $j(j+1)$ and $M$, respectively. The vector-valued functions $e_{s\rho}$ are eigenvectors of $\pmb{S}^2$ and $S_{12}$ corresponding to the eigenvalues $s(s+1)$ and $\rho$, respectively. The functions \eqref{spsphar} form an orthonormal basis in $L^2_{\C^{2s+1}}\left(\mathbb{S}^2, d{\hat{\pmb{r}}}\right)$:
\begin{equation} \label{orthoS2s}
\int_{\mathbb{S}^2} \big(Y_{slJM}(\theta,\phi)\big)^{\dag} \,Y_{sl^{\prime}J^{\prime}M^{\prime}}(\theta,\phi)\, d{\hat{\pmb{r}}}= \delta_{ll^{\prime}}\delta_{JJ^{\prime}}\delta_{MM^{\prime}}\,.
\end{equation}
This results from the unitarity property of Clebsch-Gordan coefficients \cite{edmonds96}:
\begin{equation} \label{CGunitary}
\sum_{m_1\, m_2} \overline{(j_1m_1 j_2 m_2 |j_1 j_2 j m)}\,(j_1m_1 j_2 m_2 |j_1 j_2 j^{\prime} m^{\prime})= \delta_{jj^{\prime}}\, \delta_{mm^{\prime}}\,.
\end{equation}

By extension, one then defines the holomorphic spinor or vector solid spherical harmonics, for a given spin $s$, as:
\begin{equation} \label{hsolspsphar}
\mathcal{Y}_{slJM}(\pz) = \sum_{m\, \rho} (s\rho l m| s l J M)\,e_{s\rho}\,\mathcal{Y}_{lm}(\pz)\,.
\end{equation}
For future use, considering the orthogonality relation of Clebsch-Gordan coefficients \cite{edmonds96}:
\begin{eqnarray}\label{9999}
\sum_{j m} \overline{(j^{}_1 m^{}_1 j^{}_2 m^{}_2 |j^{}_1 j^{}_2 j m)}\,(j^{}_1 m^\prime_1 j^{}_2 m^\prime_2 |j^{}_1 j^{}_2 j m) = \delta_{m^{}_1m^{\prime}_1}\, \delta_{m^{}_2m^{\prime}_2}\,.
\end{eqnarray}
we obtain from Eq. \eqref{hsolspsphar}, for a given spin $s$, that:
\begin{equation} \label{converse}
\mathcal{Y}_{lm}(\pz) = \sum_{JM} \overline{(s\rho l m| s l J M)}\, e_{s\rho}^\dagger\, \mathcal{Y}_{slJM}(\pz)\,.
\end{equation}
Note that in the expressions \eqref{CGunitary} and \eqref{9999}, the complex conjugate is not required if, with an appropriate choice of the phase factor, the involved Clebsch-Gordan coefficients are real.

Finally, considering the above, an orthonormal basis of the (Segal-)Bargmann-Fock Hilbert space $\mathcal{F}^{(\vsi,s)}$ reads:
\begin{eqnarray}
\mathbf{F}^{(\vsi,s)}_{\nu}(\pz) \equiv \mathrm{F}^{(\vsi,s)}_{l,k,J,M}(\pz)&=& \sqrt{a_{\varsigma+s,l,k}}\, (\pz\cdot\pz)^{k}\, \mathcal{Y}_{s,l-2k,JM}(\pz) \nonumber\\
&=& \sum_{m\,\rho} (s\rho ,l-2k, m| s ,l-2k, J M)\,e_{s\rho}\, \sqrt{a_{\varsigma+s,l,k}}\, (\pz\cdot\pz)^{k}\,\mathcal{Y}_{l-2k,m}(\pz)\,,
\end{eqnarray}
where,
$\label{nkm1m2}  l\in \N\,, \ 0\leqslant k\leqslant \left\lfloor \frac{l}{2}\right\rfloor\,, \ |l-2k-s| \leqslant J \leqslant l-2k+s\,, \ M=m+\rho\,, \ \mbox{with} \ 2k-l\leqslant m \leqslant l-2k\,, \ -s\leqslant \rho \leqslant s\,,$
and the definition of $a_{\varsigma+s,l,k}$ can be understood from that of $a_{\varsigma,l,k}$ 
\eqref{dddddd}, when $\varsigma$ is replaced by $\varsigma+s$.

In conclusion of this subsection, it is worth noting that the matrix elements of the Sp$(2,\R)$ representations $U^{(\vsi,s)}$, with respect to the aforementioned orthonormal basis, are available in Ref. \cite{M elements}.

\setcounter{equation}{0} 
\section{Concluding remarks} \label{Sec. Conclusion}
This paper explores elementary systems living in AdS$_4$ spacetime whose symmetry group Sp$(4,\R)$ (the two-fold covering of SO$_0(2,3)$) is linked by contraction to the Poincar\'{e} group (see Figure 1 of Ref. \cite{bacryjmll68}), that is, the symmetry group of flat Minkowski spacetime. We have studied the classical level as well as the quantum one. By a classical AdS$_4$ elementary system and its quantum counterpart, we have respectively considered a (co)-adjoint orbit of Sp$(4,\R)$, homeomorphic to the group coset space Sp$(4,\R)/\mathrm{S}\big(\mathrm{U}(1)\times \mathrm{SU}(2)\big)$, and the respective (projective) UIR of Sp$(4,\R)$, which is found in the discrete series of the Sp$(4,\R)$ UIRs.

On the other hand, this work can be viewed in the context of a program of construction of UIRs of the relativity/kinematical Lie group of AdS$_4$ spacetime and their application to quantize physical systems using covariant integral quantization methods (see, for instance, Refs. \cite{bergaz14,aagbook13,gazeauAP16,gazmur16}).

We also would like to stress an interesting feature of AdS$_4$ relativity that has appeared in this paper. Any AdS$_4$ ``massive" elementary system is a deformation of both a Minkowskian-like relativistic free particle (with the rest energy $mc^2$) and an isotropic harmonic oscillator, as a Newton-Hooke elementary system, arising from the AdS$_4$ curvature. Along the lines sketched in Refs. \cite{Gazeau2020, GazeauTannoudji2021}, the appearance of this universal pure vibration, besides the ordinary matter content, may explain the current existence of dark matter in the Universe.

\subsection*{Acknowledgments} 
Mariano A. del Olmo is supported by  MCIN with funding from the European Union-Next Generation (PRTRC17.11), and also by PID2020-113406GB-I00  and PID2023-149560NB-C21 financed by MICIU/AEI/10.13039/501100011033 of Spain.
Hamed Pejhan is supported by the Bulgarian Ministry of Education and Science, Scientific Programme ``Enhancing the Research Capacity in Mathematical Sciences (PIKOM)", No. DO1-67/05.05.2022. J.P. Gazeau would like to thank the University of Valladolid for its hospitality.  This article/publication is based upon work from COST Action CaLISTA CA21109 supported by COST (European Cooperation in Science and Technology).

\appendix

\setcounter{equation}{0} 

\section{Complex representations of Sp$(4,\R)$: complements} \label{appendix:sp4R}
The group Sp$(2n,\R)$, known as the symplectic group, consists of all $2n\times2n$ real matrices that preserve a non-degenerate, skew-symmetric bilinear form. In mathematics, Sp$(2n,\R)$ is the standard notation. In physics, especially in particle physics and string theory, Sp$(n)$ may sometimes be used as a shorthand for the symplectic group Sp$(2n,\R)$, though this can lead to confusion. Formally, the group Sp$(2n,\R)$ is defined as  (we follow the notations of \cite{simonetal95}):
\begin{equation} \label{sp2nRdef}
\mathrm{Sp}(2n,\R) = \left\{ S \in \mathrm{SL}(2n, \R)\,, \, ^{\mathrm{t}}S\,\mathcal{J}\,S= \mathcal{J}\right\}\,, 
\end{equation}
where $ \mathcal{J} = \begin{pmatrix}
   \mathbb{0}_n   &  \mathbbm{1}_n  \\
  - \mathbbm{1}_n   &  \mathbb{0}_n
\end{pmatrix}$  
is the standard \textit{symplectic metric matrix}, which affords a complex structure to $\R^{2n}$ through $ \mathcal{J}^2= - \mathbb{1}_{2n}$, and $\mathbbm{1}_n$ is the $n\times n$ identity matrix. It is a Lie group of real-dimension $n(2n+1)$. More precisely:
\begin{equation} \label{sp2nR}
\mathrm{Sp}(2n,\R) = \left\{\mathrm{SL}(2n, \R)\ni  S= \begin{pmatrix}
  A    &  B  \\
   C   &  D
\end{pmatrix}\, ,\, ^{\mathrm{t}}A\, C =\, ^{\mathrm{t}}C\,A\, , \ ^{\mathrm{t}}B\,D =\, ^{\mathrm{t}}D\,B\, ,\ ^{\mathrm{t}}A\,D - \,^{\mathrm{t}}C\,B=\mathbbm{1}_n\right\}\,. 
\end{equation}
Note that the relations $ A\,^{\mathrm{t}}B = B \,^{\mathrm{t}}A\, , \ C\,^{\mathrm{t}}D = D \,^{\mathrm{t}}C\, ,\ A\,^{\mathrm{t}}D - B \,^{\mathrm{t}}C=\mathbbm{1}_n $ are equivalent to the previous ones of \eqref{sp2nR} and are obtained from the fact that  $S^{-1}=\mathcal{J}\,^{\mathrm{t}}S\,\mathcal{J}^{-1}$ belongs to $\mathrm{Sp}(2n,\R)$ \eqref{sp2nRdef}.

The complex form of $S\in \mathrm{Sp}(2n,\R)$ is given by:
\begin{equation} \label{Sc}
S\mapsto S^\mathrm{c}= \Omega S\Omega^{-1}=
\frac{1}{2}\begin{pmatrix}
A+D + \ii (C-B)      & A-D + \ii (B+C)    \\
  A-D - \ii (B+C)     &  A+D - \ii (C-B) 
\end{pmatrix}\,,  \quad \Omega = \frac{1}{\sqrt{2}}\begin{pmatrix}
   \mathbb{1}_n    & \ii   \mathbb{1}_n   \\
  \mathbb{1}_n      &  -\ii  \mathbb{1}_n 
\end{pmatrix}\,. 
\end{equation}
It is obtained through a Cayley map encoded by $\Omega$.  
Now, let us specify \eqref{Sc} to the case $n=2$. In its complex form, the group Sp$(4,\R)$ is the subgroup of SL$(4,\C)$ made of $2\times2$ block complex matrices as:
\begin{equation}
\label{Sc2}
S^\mathrm{c}= \begin{pmatrix}
   \alpha   &   \beta \\
   \beta^{\mathrm{cc}}   &   \alpha^{\mathrm{cc}}
\end{pmatrix}\, , \quad \alpha, \beta \in \mathsf{M}(2,\C)\, ,\end{equation}
obeying:
\begin{equation}
\label{Sc2inv}
S^{\mathrm{c}}\begin{pmatrix}
   \mathbb{0}_2   &  \mathbbm{1}_2  \\
   -  \mathbbm{1}_2 &  \mathbb{0}_2
\end{pmatrix} \, {}^{\mathrm{t}}S^{\mathrm{c}}= \begin{pmatrix}
   \mathbb{0}_2   &  \mathbbm{1}_2  \\
   -  \mathbbm{1}_2 &  \mathbb{0}_2
\end{pmatrix}\ \Leftrightarrow \    {S^{\mathrm{c}}}^{-1}= \begin{pmatrix}
    \alpha^{\ast}   &    - {}^{\mathrm{t}}\beta\\
 -   \beta^{\ast}   &   {}^{\mathrm{t}} \alpha
\end{pmatrix}\,,
\end{equation}
and where the notation $ \alpha^{\mathrm{cc}} $ stands for the complex conjugate of $\alpha$ and is used to make the difference with the notation $\bar z$ used for complex quaternions in \eqref{ccqcadr}.

Now comparing the expressions \eqref{sp4inv} and \eqref{Sc2inv} for the inverses in complex quaternionic notation and complex notations respectively, one immediately observes that the compact maximal subgroups compare as $K = \left\{k= \begin{pmatrix}
 e^{\ii\theta/2} \xi    &  0  \\
  0    &  e^{-\ii\theta/2}\xi
\end{pmatrix}\, , \, \xi\in \mathrm{SU}(2)\, , \, 0\leq \theta< 2\pi\right\}$ (for \eqref{sgK}) and $K_{\mathrm{sp}} = \left\{k= \begin{pmatrix}
 u     &  0  \\
  0    &  u^{\ast}
\end{pmatrix}\, , \, u\in \mathrm{U}(2)\right\}$ (for \eqref{Sc2}). On the level of their respective Lie algebras bases, this difference reads as 
$\left\{\frac{1}{2}\begin{pmatrix}
    1  &  0  \\
     0 &  -1
\end{pmatrix}\, , \, \begin{pmatrix} {\pmb{e}}_i & 0 \\ 0 & {\pmb{e}}_i \end{pmatrix}\, , \, i=1,2,3\right\}$, for $\mathfrak{k}$ with the notations of \eqref{X0} and \eqref{Yi}, whereas
for $\mathfrak{k}_{\mathrm{sp}}$ we have $\left\{\frac{1}{2}\begin{pmatrix}
    1  &  0  \\
     0 &  -1
\end{pmatrix}\, , \, \frac{1}{2}\begin{pmatrix} {\pmb{e}}_i & 0 \\ 0 &- {\pmb{e}}_i \end{pmatrix}\, , \, i=1,2,3\right\}$. Considering the  Cartan decompositions $\mathfrak{g}=  \mathfrak{p} + \mathfrak{k}$ and $\mathfrak{g}_{\mathrm{sp}}=  \mathfrak{p}_{\mathrm{sp}} + \mathfrak{k}_{\mathrm{sp}}$, the bases for the respective $\mathfrak{p}$ and $\mathfrak{p}_{\mathrm{sp}}$ read as 
$\left\{ \frac{1}{2}\begin{pmatrix}  0& \pmb{e}_i \\ -\pmb{e}_i & 0 \end{pmatrix}\, , \, \frac{1}{2}\begin{pmatrix}  0& \ii\pmb{e}_i \\ \ii\pmb{e}_i & 0 \end{pmatrix}\,,\,i=1,2,3\right\}$ (\eqref{Xi} and \eqref{Zi}) and $\left\{\frac{1}{2}\begin{pmatrix}  0& \pmb{e}_1 \\ \pmb{e}_1 & 0 \end{pmatrix}\, , \, \frac{1}{2}\begin{pmatrix}  0& \pmb{e}_2 \\ -\pmb{e}_2 & 0 \end{pmatrix}\, , \, \frac{1}{2}\begin{pmatrix}  0& \pmb{e}_3 \\ \pmb{e}_3 & 0 \end{pmatrix}\, , \, \frac{1}{2}\begin{pmatrix}  0& \ii\pmb{e}_1 \\ \ii\pmb{e}_1& 0 \end{pmatrix}\,,\,\frac{1}{2}\begin{pmatrix}  0& \ii\pmb{e}_2 \\ -\ii\pmb{e}_2& 0 \end{pmatrix}\,,\,\frac{1}{2}\begin{pmatrix}  0& \ii\pmb{e}_3 \\ \ii\pmb{e}_3& 0 \end{pmatrix}\right\}$.

{Finally, the isomorphism between the two Lie algebras $\mathfrak{g}$ and $\mathfrak{g}_{\mathrm{sp}}$ consists in the replacements of the following 6 basis elements among the ten ones: 
\begin{align}
\label{isom}
\frac{1}{2}\begin{pmatrix} {\pmb{e}}_i & 0 \\ 0 &{\pmb{e}}_i \end{pmatrix}&\mapsto \begin{pmatrix}
   1   &  0  \\
   0   &  -1
\end{pmatrix}\frac{1}{2}\begin{pmatrix} {\pmb{e}}_i & 0 \\ 0 &{\pmb{e}}_i \end{pmatrix}= \frac{1}{2}\begin{pmatrix} {\pmb{e}}_i & 0 \\ 0 &- {\pmb{e}}_i \end{pmatrix}\, , \quad i=1,2,3\, , \\
   \frac{1}{2}\begin{pmatrix}  0& \pmb{e}_i \\ -\pmb{e}_i & 0 \end{pmatrix}  &\mapsto \begin{pmatrix}
   1   &  0  \\
   0   &  -1
\end{pmatrix} \frac{1}{2}\begin{pmatrix}  0& \pmb{e}_i \\ -\pmb{e}_i & 0 \end{pmatrix}=\frac{1}{2}\begin{pmatrix}  0& \pmb{e}_i \\ \pmb{e}_i & 0 \end{pmatrix} \, , \quad i=1,3\, ,   \\
   \frac{1}{2}\begin{pmatrix}  0&\ii \pmb{e}_2 \\ \ii\pmb{e}_2 & 0 \end{pmatrix}  &\mapsto \begin{pmatrix}  1& 0\\ 0 & -1 \end{pmatrix}  \frac{1}{2}\begin{pmatrix}  0&\ii \pmb{e}_2 \\ \ii\pmb{e}_2 & 0 \end{pmatrix} = \frac{1}{2}\begin{pmatrix}  0&\ii \pmb{e}_2 \\ -\ii\pmb{e}_2 & 0 \end{pmatrix}   \,.  
\end{align} }

\section{Massive UIRs of the Poincar\'e group} \label{appendix:WigPoin}
The material of this Appendix A is borrowed from Ref. \cite{Karim 1996}. By full Poincar\'e group we mean the two-fold covering group $\mathcal{P}^{\uparrow}_0(1,3)= \mathrm{T}^4\rtimes \mathrm{SL}(2,\C)$, where $\mathrm{T}^4 \simeq \R^{1,3}$ is the group of space-time translations. Elements of $\mathcal{P}^{\uparrow}_0(1,3)$ are denoted by $(a,A)$, with $a=(a_0,\mathbf{a})\in \R^{1,3}$ and $A\in \mathrm{SL}(2,\C)$. The multiplication law is $(a,A)(a^{\prime},A^{\prime})= (a+ \Lambda a^{\prime},AA^{\prime})$, where $\Lambda \in \mathcal{L}_0^{\uparrow}(1,3)$ (the proper, orthochronous Lorentz group) is the Lorentz transformation corresponding to $A$:
\begin{equation} \label{ALamb}
\Lambda^{\mu}_{\,\,\,\nu}= \frac{1}{2}\,\mathrm{Tr}\big[A\sigma^\mu A^{\dag} \sigma_\nu\big]\, , \quad \mu,\nu = 0,1,2,3\, ,
\end{equation}
where $\sigma_i$ are the three Pauli matrices, $\sigma_0= \bu_2$, and the metric tensor is $g_{00}=1=-g_{11}=-g_{22}=-g_{33}$. Let:
\begin{equation} \label{massH}
\mathcal{V}^+_{m} = \Big\{p=(p_0,\mathbf{p})\in \R^{1,3}\;;\; p^2 = p\cdot p \equiv g_{\mu\nu} p^\mu p^\nu = p_0^2 - \Vert \mathbf{p}\Vert^2 = m^2c^2\, , \,  {m>0\, , \, p_0>0}\Big\}\,,
\end{equation}
be the forward mass hyperboloid. Then:
\begin{equation} \label{Lorp}
p^{\prime}=\Lambda\diamond p \;\Rightarrow\; \sigma\cdot p^{\prime} = A\sigma\cdot p A^{\dag}\,,
\end{equation}
with $\sigma\cdot p= \sigma^{\mu}p_{\mu}= p_0\bu_2- \pmb{\sigma}\cdot\mathbf{p}$ and $\pmb{\sigma}= (\sigma_1,\sigma_2,\sigma_3)$.

In the Wigner realization, the UIR ${\mathcal{P}^{(m>0,s)}}$ of $\mathcal{P}^{\uparrow}_0(1,3)$ for a particle of mass $m>0$ and spin $s=0, \frac{1}{2}, 1, \frac{3}{2}, 2, \dots$ is carried by the Hilbert space:
\begin{equation} \label{HilbertWig}
\mathcal{H}^{(m,s)} = \C^{2s+1}\otimes L^2\left(\mathcal{V}^+_{m}, \frac{\ud \mathbf{p}}{p_0}\right)\,, \quad \ud \mathbf{p}=\ud p_1 \ud p_2 \ud p_3\,,
\end{equation}
of $\C^{2s+1}$-valued functions $\pmb{\phi}$ on $\mathcal{V}^+_{m}$, which are square integrable:
\begin{equation} \label{sqintV }
(\pmb{\phi},\pmb{\phi}) = \Vert \pmb{\phi}\Vert^2 = \int_{\mathcal{V}^+_{m}} \pmb{\phi}(p)^{\dag} \pmb{\phi}(p)\, \frac{\ud \mathbf{p}}{p_0} < \infty\,.
\end{equation}
The Wigner UIR operator is given by:
\begin{equation} \label{WopUIR}
\left(\mathcal{P}^{(m,s)}\,(a,A)\pmb{\phi}\right) (p) = e^{\frac{\ii}{\hbar} p\cdot a}\, {D}^s \Big(\big(h(p)\big)^{-1}Ah\big(\Lambda^{-1}\diamond p\big)\Big)\, \pmb{\phi}\left(\Lambda^{-1}\diamond p\right)\,,
\end{equation}
where ${D}^s$ is the $(2s+1)$-dimensional irreducible spinor representation of SU$(2)$ (see Appendix \ref{appendix:SU(2)}) and:
\begin{equation} \label{phpW}
p \;\mapsto\; h(p) = \frac{mc\bu_2 + \sigma\cdot \widetilde{p}}{\sqrt{2mc(p_0+mc)}}\,, \quad \widetilde{p} = (p_0,-\mathbf{ p})\,,
\end{equation}
is the image in SL$(2,\C)$ of the Lorentz boost $\Lambda_p$, which brings the $4$-vector $(mc,\mathbf{0})$ to the $4$-vector $p$ in $\mathcal{V}^+_{m}$:
\begin{equation} \label{lorboost}
\Lambda_p(mc,\mathbf{0})=p \;\Leftrightarrow\; h(p)\, mc\, \bu_2\, h(p)
= mc\, \big[h(p)\big]^2\\
= \sigma\cdot \widetilde{p}= \begin{pmatrix}
  p_0 + p_3 & p_1-\ii p_2 \\ p_1+\ii p_2 & p_0 - p_3
\end{pmatrix} \,.
\end{equation}
The matrix form of the Lorentz boost is:
\begin{equation} \label{matlorb}
\Lambda_p = \frac{1}{mc}\begin{pmatrix}
   p_0 & \mathbf{p}^{\dag} \\ \mathbf{p} & mc\, V_{p}
\end{pmatrix} = \Lambda_p^{\dag}= \Lambda_{\widetilde{p}}^{-1},
\end{equation}
where $V_{p}$ is the $3\times 3$-symmetric matrix:
\begin{equation} \label{Vpmat}
V_p = \bu_3 + \frac{\mathbf{p} \otimes \mathbf{p}^{\dag}}{mc(p_0 +mc)}=V_{p}^{\dag}\,.
\end{equation}

\setcounter{equation}{0} 
\section{Representations of the Newton-Hooke (3+1) Group} \label{appendix:Newton-Hooke}

\subsection{The Newton-Hooke (3+1) Group}
{The Newton-Hooke ($3+1$) group, referred to as $NH_4$ here, is one of the kinematical groups detailed in the seminal paper by Bacry and L\'evy-Leblond \cite{bacryjmll68}. Despite its cosmological nature, this group remains non-relativistic. Strictly speaking, there exist two distinct Newton-Hooke groups: one, denoted as $NH_{-4}$, corresponds to an oscillating universe, while the other, labeled $NH_{+4}$, represents an expanding universe. Both corresponding Lie groups possess 10 dimensions. The infinitesimal generators are denoted as follows:}
\begin{equation}
H\,,\quad \Pv=(P_1,P_2,P_3)\,,\quad  \Kv=(K_1,K_2,K_3)\,,\quad  \Jv=(J_1,J_2,J_3)\,,
\end{equation}
while the following Lie commutators hold between them \cite{deromedubois,dubois}:
\begin{equation}\begin{array}{llllll}\label{commutatorsnh}
[J_i,H]=0\,,\qquad &[J_i,P_j]=\epsilon_{ij}^{\,\,\,k} P_k\,,\qquad &[J_i,K_j]=\epsilon_{ij}^{\,\,\,k} K_k\,,
\qquad &[J_i,J_j]=\epsilon_{ij}^{\,\,\,k} J_k\,,\\[0.2cm]
[P_i, H]=\pm \frac{1}{\tau^2} K_i\,,\qquad &[P_i,P_j]=0\,,\qquad &[P_i,K_j]=0\,,\\[0.2cm]
[K_i,H]=P_i\,,\qquad &[K_i,K_j]=0\,,&\qquad & i,j,k=1,2,3\,,
\end{array}\end{equation}
where the parameter $\tau$ stands for a proper constant time of the corresponding universe. Note that the commutator $[P_i, H]=\pm \frac{1}{\tau^2} K_i$ corresponds to the Lie algebra of $NH_{\pm4}$, respectively. In the following, we mainly consider the group $NH_{-4}$ associated with an oscillating universe (with oscillating period $\tau$).

The group $NH_{-4}$ can be considered as the transformation group of the $(3+1)$-spacetime $\R^{3,1}$. An arbitrary element of $NH_{-4}$ can be written as:
\begin{equation}
g\equiv (b,\av,\vv,R)=e^{b H}\,e^{\av\cdot\Pv}\,e^{\vv\cdot\Kv}\,R\,, \qquad b\in\R\,,\;\av,\vv\in \R^3\,,\;R\in \mbox{SO}(3)\,.
\end{equation}
The action of $g$ on a point $(t,\xv)$ of the spacetime ($\R^{3,1}$) is given by:
\be \label{action}
(t^\prime,\xv^\prime)\equiv g\diamond\xv=\left(t+b, R\,\xv+\vv \tau \sin\frac{t}{\tau}+\av  \cos\frac{t}{\tau}\right)\,.
\ee
The group law can be easily obtained from \eqref{action}:
\begin{eqnarray}\label{grouplaw-}
g^\prime\,g = (b^\prime,\av^\prime,\vv^\prime,R^\prime)\,(b,\av,\vv,R) = \left( b^\prime+b\,, \, \av^\prime  \cos\frac{b}{\tau}+\vv^\prime \tau \sin\frac{b}{\tau}
+R^\prime\,\av\,, \vv^\prime \cos\frac{b}{\tau}-\frac{\av'}{ \tau} \sin\frac{b}{\tau}+R^\prime\,\vv \,,\,R^\prime\,R\right)\,.
\end{eqnarray}
The inverse element of $g$ is expressed as:
\begin{eqnarray}\label{inverse-}
g^{-1}= (b,\av,\vv,R)^{-1} = \ds \left( -b\,, \, R^{-1}\left[\vv \tau \sin\frac{t}{\tau}-\av \cos\frac{t}{\tau}\right]\,,\,
R^{-1}\left[-\vv  \cos\frac{t}{\tau}-\av^\prime \tau \sin\frac{t}{\tau}\right]\,,\,R^{-1}\right)\,.
\end{eqnarray}

From inspection of formulae \eqref{grouplaw-} and \eqref{inverse-}, we see that the action of $e^{b H}$ on $(\av ^\prime,\vv ^\prime)$ can be given, like it appears in \eqref{grouplaw-}, by:
\be\label{recipe1}
(\av^\prime, \vv^\prime) \,
\left(\begin{array}{ccc}
\cos\frac{b}{\tau}\,\bu_3&-\frac{1}{\tau}\,\sin\frac{b}{\tau}\,\bu_3\\
{\tau}\,\sin\frac{b}{\tau}\,\bu_3&\cos\frac{b}{\tau}\,\bu_3
\end{array}\right)\,.
\ee
It looks to be a rotation of angle $b/\tau$ of SO$(2)$ with the tensor product by $\bu_3$. Similarly, to find the inverse, we consider:
\be\label{recipe2}
(-R^{-1}\av^\prime, -R^{-1}\vv^\prime) \,
\left(\begin{array}{ccc}
\cos\frac{-b}{\tau}\,\bu_3&-\frac{1}{\tau}\,\sin\frac{-b}{\tau}\,\bu_3\\
{\tau}\,\sin\frac{-b}{\tau}\,\bu_3&\cos\frac{-b}{\tau}\,\bu_3
\end{array}\right)\,.
\ee
Following the expressions \eqref{recipe1} and \eqref{recipe2}, we find the group law and the inverse of any element for $NH_{+4}$; it is enough to replace the trigonometric rotation of angle $b/\tau$ by a hyperbolic ``rotation'' of the same argument, i.e., an element of SO$(1,1)$:
\be\label{hyperbolic}
\left(\begin{array}{ccc}
\cosh\frac{b}{\tau}&\frac{1}{\tau}\,\sinh\frac{b}{\tau}\\
{\tau}\,\sinh\frac{b}{\tau}&\cosh\frac{b}{\tau}\,
\end{array}\right)\,.
\ee

The NH group can be factorized as a semidirect product of subgroups:
\be\label{factorization}
{NH}_{\pm4}\sim \big(T_s\otimes H\big)\rtimes \big(\mbox{SO}(3)\otimes T_t\big)\,,
\ee
where $T_s$ is the subgroup of the space-translations ($\sim \R^3$), $H$ the subgroup of the pure Newton-Hooke boosts transformations, SO$(3)$ the group of the space-rotations, and $T_t$ the subgroup of the time-translations. Note that $T_s\otimes H$ is indeed a maximal abelian subgroup.

The second cohomology group of $NH_{\pm4}$ is $H^2(NH_{\pm4})=\R$. The cohomology classes are labeled by $[m]$ with $m\in \R$ ($[P_i,K_j]=-\delta_{ij}\,m$). A lifting of the class $[m]$ is given by \cite{deromedubois}:
\be\label{2factorm}
\omega_m(g',g)=e^{i m\, W(g',g)}\,,
\ee
with:
\be\label{2factor}
W(g^\prime,g)=\frac{1}{2}\,\left(\tau\,\vv^{\prime\,2}-\frac{\av^{\prime\,2}}{\tau}\right)\,\sin\frac{b}{\tau}\,\cos\frac{b}{\tau}
+\left(\vv' \,\cos\frac{b}{\tau}-\frac{\av^{\prime}}{\tau}\,\sin\frac{b}{\tau}\right)\cdot R^{\prime}\,\av-
\frac{\av^{\prime}\cdot \vv^{\prime}}{\tau}\,\sin^2\frac{b}{\tau}\,.
\ee

In order to obtain the projective UIRs of $NH_{\pm4}$, we need to construct new groups called the splitting groups of $NH_{\pm4}$, respectively denoted by $\widehat{NH}_{\pm4}$ \cite{santander}, and which are the central extensions of $NH_{\pm4}$ by $\R$. For that, we first consider the central extensions $NH_{\pm4}$ given by the new commutators \cite{deromedubois, dubois, del Olmo IJTP}:
\be\label{extension}
[P_i,K_j]=-\delta_{ij}\,M\,,
\ee
where $M$ commutes with all the generators of $NH_{\pm4}$. The Lie algebra of $\widehat{NH}_{\pm4}$, $\widehat{\mathfrak{nh}}_{\pm4}$ is  eleven- dimensional with  generators $\{M,H, \Pv,\Kv,\Jv\}$. The corresponding Lie commutators are given by \eqref{commutatorsnh}, at the  exception of  the commutators $[P_i,K_j]=0$ which become $[P_i,K_j]=\delta_{ij}\,M$ and $[M,Y]=0$ for all $Y\in \widehat{\mathfrak{nh}}_{\pm4}$.

The group elements of $\widehat{NH}_{-4}$ are denoted by $\widehat{g}= (\a, g)=(\a,b,\av,\vv,R)$, with $\a\in \R$ and $g\in {NH}_{-4}$. The group law  is given by:
\begin{eqnarray}\label{grouplaw1}
\widehat{g}^\prime\,\widehat{g}=(\a^\prime\,, g^\prime)\,(\a\,, g) = \big(\a^\prime+\a+W(g',g)\,, g^\prime\,g\big)\,,\quad
\end{eqnarray}
where $g^\prime\,g$ is given by \eqref{grouplaw-} and $ W(g',g)$  by \eqref{2factor}.
Note that   we find the group law for $\widehat{NH}_{+4}$, as well from \eqref{recipe1}-\eqref{hyperbolic}. 

The extended groups may be factorized as:
\be\label{factorization-}
\widehat{NH}_{\pm4}\sim \big[(\Theta\otimes T_s)\rtimes (H\rtimes \mbox{SO}(3))\big]\rtimes T_t\,,
\ee
where $\Theta$ is the group generated by the central extension $M$ (see also \eqref{factorization}).

\subsection{From the dS$_4$/AdS$_4$ groups to $NH_\pm$ groups}
The $NH_{\pm4}$ groups can be respectively obtained from the dS$_4$/AdS$_4$ groups through a speed-space contraction \cite{bacryjmll68}. The splitting groups $\widehat{NH}_{\pm4}$ are also obtained by contraction from ${NH}_{\pm4}\otimes \R$ in the same way that the extended Galileo group can be obtained from the Poincar\'{e} one by contraction \cite{del Olmo 1997}.

For the sake of simplicity, let dS$_{\pm4}$ denote the dS$_4$/AdS$_4$ groups, respectively. The Lie commutators dS$_{\pm4}$ are \cite{bacryjmll68}:
\begin{equation}\begin{array}{lllll}\label{commutatosds}
[J_i,H]=0\,,\qquad &[J_i,P_j]=\var_{ijk} P_k\,,\qquad &[J_i,K_j]=\var_{ijk} K_k\,,\qquad &[J_i,J_j]=\var_{ijk} J_k\,,\\[0.3cm]
[P_i, H]=\pm K_i\,,\qquad &[K_j,H]=P_i\,,\qquad  &[P_i,K_j]=H\,,\qquad & i,j,k=1,2,3\,.
\end{array}\end{equation}
In order to perform the contraction towards the extended groups $\widehat{NH}_{\pm4}$, we include the central commutator  $M$ introduced in \eqref{extension}. Then, the basis $\{ M, H,\Pv,\Kv,\Jv \}$ represents the Lie algebras associated with dS$_{\pm4}\otimes \R$. Now, we re-scale the basis as follows:
\be\label{contractionds}
\Big\{M, H,\Pv,\Kv,\Jv \Big\}\quad \longrightarrow \quad \left\{ c^2\, M, c\, H-M,\varkappa\,\Pv,\frac{1}{c\varkappa}\,\Kv,\Jv \right\}\equiv \Big\{ M^\prime, H^\prime,\Pv^\prime,\Kv^\prime,\Jv^\prime \Big\}\,.
\ee
The Lie commutators of the first row of \eqref{commutatosds} remain formally invariant. The second row becomes:
\be\label{commutatosds1}
\begin{array}{lllllll}
&\ds [P^\prime_i, H^\prime]=\varkappa\,[P_i, c\, H-M]=\pm {c\varkappa}\,K_i=\pm \left(c\varkappa\right)^2\,K^\prime_i\,,\\[0.4cm]
& \ds  [K^\prime_j,H^\prime]=\frac{1}{c\varkappa}[K_j, c \,H-M]=\frac{1}{\varkappa}\,P_i=P^\prime_i\,,\\[0.4cm]
&\ds  [P^\prime_i,K^\prime_j]=\frac{1}{c}\,H\,=\frac{1}{c^2}\,H^\prime+M^\prime\,.
\end{array}\ee
Now taking the limits $\varkappa\to 0$, $c\to \infty$, while the ratio ${c\varkappa}\equiv\omega={1}/{\tau}$ remains unchanged, we recover the Lie commutators of the ${NH}_{\pm4}$ groups.

\subsection{The co-adjoint orbits of $NH_{-4}$}
Let $\mathfrak{nh}_{-4}$ be the Lie algebra of $NH_{-4}$ and $\mathfrak{nh}_{-4}^{\circledast}$ its dual with basis $\{m,\E, \vp,\vk,\vj\}$, dual to the basis $\{M,H, \Pv,\Kv,\Jv\}$ of $\mathfrak{nh}_{-4}$. The co-adjoint action of $NH_{-4}$ on $\mathfrak{nh}_{-4}^{\circledast}$ results in \cite{del Olmo IJTP}:
\be\label{co-adjoint}
\begin{array}{lll}
m^\prime=m\,,\\[0.4cm]
\vp^\prime=\ds\cos\frac{b}{\tau}\,R\,\vp-\frac{1}{\tau}\,\sin\frac{b}{\tau}\,R\,\vk
-\frac{m}{\tau}\left(\tau\,\cos\frac{b}{\tau}\,\vv+\sin\frac{b}{\tau}\,\av\right)\,,\\[0.4cm]
\vk^\prime=\ds\tau\,\sin\frac{b}{\tau}\,R\,\vp+ \cos\frac{b}{\tau}\,R\,\vk
+{m}\left(\cos\frac{b}{\tau}\,\av-\tau\,\sin\frac{b}{\tau}\,\vv\right)\,,\\[0.4cm]
\E^\prime=\ds  \E-\vv\cdot R\,\vp+ \frac{1}{\tau^2}\,\av\cdot R\,\vk
+\frac{m}{2}\,\left(\frac{\av^2}{\tau^2}+\vv^2\right)\,,\\[0.4cm]
\vj^\prime=\ds \vj-R\,\vp\wedge\av- R\,\vk\wedge\vv-{m}\vv \wedge \av\,.
\end{array}
\ee

Accordingly, we can classify the orbits of the co-adjoint action using the invariants $m$, $C_1$, and $\mathbf{C}_2$ as follows:
\begin{itemize}
\item{$m\neq 0$: (orbits of dimension 6)
  \be\label{orbitsm}
  C_1=\vp^2+\frac{\vk^2}{\tau^2}-2 m \E\,,\qquad \mathbf{C}_2=\vp\wedge \vk +m\vj\,,
  \ee
  }
\item{$m= 0$:
  \begin{itemize}
  \item{First: (orbits of dimension 6)
  \be
  C_1=\vp^2+\frac{\vk^2}{\tau^2}\,,\qquad \mathbf{C}_2=\vp\wedge \vk \neq \mathbf 0\,,
  \ee
  }
\item{Second: (orbits of dimension 6)
  \be
  C_1=\vp^2+\frac{\vk^2}{\tau^2}\,,\qquad \mathbf{C}_2=\vp\wedge \vk = \mathbf 0\quad \Rightarrow\quad \vk \propto \vp\,,
  \ee
  }
\item{Third: (orbits are points $(\E,\vj)$, i.e, dimension 0)
  \be
  C_1=\vp^2+\frac{\vk^2}{\tau^2}=0\,,\qquad \mathbf{C}_2=\vp\wedge \vk = \mathbf 0\quad \Rightarrow\quad \vk = \vp=\mathbf 0\,,
  \ee
  }
\end{itemize}}
\end{itemize}

The physical meaning of these invariants when $m\neq 0$ is as follows:
\begin{itemize}
\item{The first invariant $C_1=-m U$ is relevant to the internal energy ($U$) of the physical system, provided that $\E$ is the total energy ($E$) of the system. Also, considering the position $\vq=\vk/m$, the position operator $\mathbf Q=\frac{1}{M}\,\Kv$, and $\w=1/\tau$, we get:
    \be
    E=\frac{1}{2m}\,\vp^2+\frac{1}{2m}\,\frac{\vk^2}{\tau^2}+U=\frac{\vp^2}{2m}+\frac12\,m\,\w^2\,{\vq^2}+U\,,
    \ee
    that corresponds to a harmonic oscillator.}

\item{The second invariant $\mathbf C_2=m\, \mathbf s $ is related to the spin of the particle provided that $\vj$ is the total angular momentum of the particle and $\frac{1}{m}\, \vk\wedge\vp$ the angular momentum:
    \be
    \mathbf s=\vj-\frac{1}{m}\,\vk\wedge\vp=\vj-\vq\wedge\vp\,.
    \ee
    Hence, $\mathbf s$ can be seen as the classical spin. Then, the associated UIRs will be labeled by the eigenvalues of $\mathbf S^2$; instead of SO$(3)$ we will use its universal covering SU$(2)$ then we will have $s=1/2,1,3/2,\dots$ .}
\end{itemize}

In order to compare these results with those obtained in subsection~\ref{Subsec Newton contraction}, let us consider the $6$-dimensional co-adjoint orbits with $m\neq 0$, namely, \eqref{orbitsm}:
\be\label{orbitsmo}
O_m=\big\{(\vq=\vk/m,\vp )\big\} \equiv \R^6\,.
\ee
Remembering the theory of the harmonic oscillator, we can rescale and rewrite the coordinates $(p_j, q_j)$ as:
\be\label{newcoordinates}
z_j= p_j+\ii\, m\,\w\,q_j\in \C, \qquad \text{or} \qquad \vz=\vp+\ii \,m\,\w\,\vq\in \C^3\,.
\ee
The co-adjoint action of $NH_{-4}$  on $\vk$ and $\vp$ \eqref{co-adjoint} becomes:
\be\label{co-adjointz}
\vz\qquad \xrightarrow{(b,\av,\vv,R)}\qquad
\vz^{\prime}=e^{\ii\w b} R\,\vz+(- m\vv+\ii m\w\av) \,e^{\ii\w b}\,.
\ee

\subsection{The UIRs of $NH_{-4}$}
The (projective) UIRs associated with the above orbits can be obtained by following the theory of induced representations by Kirillov \cite{Kirillov2} and Mackey \cite{mackey}, based upon which the following cases come to the fore:
\begin{itemize}
\item{$m\neq 0$. 
In this case, we consider the elements $(0,g)\equiv {g}\in \widehat{NH}_{-4}$ such that $g\in NH_{-4}$ but now $R$ denotes the elements of SU$(2)$. The UIRs are labelled by the invariants $m, U \in \R$ and $s$ such that $s\in \N/2$:
    \be\label{uirm}
    \left(\U(g)_{m}^{s,U}\psi\right)^\a(\vp)= e^{\ii(Eb-\vp\cdot\av)}\,\big[D^s(R)\big]^{\a}_{\;\,\b}\,\psi^\beta \big(R^{-1}\diamond(\vp-m \vv)\big)\,,
    \ee
    where $D^s$ denote the UIRs of SU$(2)$ with $s\in \N/2$, $\a,\b=-s,-s+1,\dots,s-1,s$, and $E$ is the operator:
    \be
    E=\frac{p^2}{2m}-\frac{m}{2\tau^2}\,\nabla^2_{\vp}\,.
    \ee
    Note that $\big(\U(g)_{m}\big)^{s,U}$ is projectively equivalent to $\big(\U(g)_{m}\big)^{s,0}\,,$ since
    $
    \U(g)_{m}^{s,U}= e^{\ii Ub}\,\U(g)_{m}^{s,0}\,.
    $}

\item{$m= 0$. Here, we present the true UIRs of $NH_{-4}$ and we can distinguish the following cases:
  \begin{itemize}
  \item{The vectors $\vp$ and $\vk$ are not collinear, hence we get:
  \be\label{uir01}
  \left(\U(g)^{C_1,{\mathbf C}^2_2}\psi\right)\,(\vp,\vk)= e^{\ii(\vp\cdot\av+\vk\cdot\vv)}\,\psi(g^{-1}\diamond\vp,g^{-1}\diamond\vk)\,.
  \ee
  From \eqref{inverse-} and \eqref{co-adjoint}, we obtain:
  \begin{eqnarray}\label{actionspk}
  \vp^\prime=\cos\frac{b}{\tau}\,R^{-1}\,\vp + \frac{1}{\tau}\,\sin\frac{b}{\tau}\,R^{-1}\,\vk\,, \quad \vk^\prime= -\tau\,\sin\frac{b}{\tau}\,R^{-1}\,\vp+ \cos\frac{b}{\tau}\,R^{-1}\,\vk\,.
  \end{eqnarray}}

\item{The vectors $\vp$ and $\vk$ are collinear, hence ${\mathbf C}=\mathbf 0$. In this case, the UIRs are induced by the UIRs of SO$(2)$, that is, the little group of a fixed point of the orbit ($\vp_0$). The UIRs of SO$(2)$ are labeled by an integer $h$ (helicity), that is, the eigenvalue of the 
generator of the rotations (helicity operator) $\mathbf J\cdot\Pv/\vert \Pv \vert $. Then:
    \be\label{uir01}
    \left(\U(g)^{C_1,h}\psi\right)\,(\vp,\vk)= e^{\ii(\vp\cdot\av+\vk\cdot\vv-h \theta(\vp,R))}\,\psi(g^{-1}\diamond\vp,g^{-1}\diamond\vk)\,,
    \ee
    where the expressions of $(g^{-1}\diamond\vp,g^{-1}\diamond\vk)$ are given in \eqref{actionspk} and $\theta(\vp,R)$ is the angle corresponding to the rotation $R^{-1}_{\vp^\prime}\,R\,R^{-1}_{\vp}$ of SO$(2)$ associated to the following process:
    \be
    \vp_0\;\;\xrightarrow{R_\vp}\;\; \vp\;\;\xrightarrow{R}\;\; \vp^\prime\;\;\xrightarrow{R^{-1}_{\vp^\prime}}\;\; \vp_0 \,.
    \ee}
 \item{The case with orbits points $(0,\mathbf 0,\mathbf 0,E,\vj)$. The little group of $(\vp=\mathbf 0,\vk=\mathbf 0)$ is SU$(2)\otimes T_t$. Hence, the UIRs are induced by the UIRs of SU$(2)\otimes T_t$, namely:
     \be
      \U(g)^{E}_j=e^{\ii b E}\,D^j(R)\,.
      \ee
      These representations are unfaithful or degenerate.}
 \end{itemize}}
\end{itemize}

\setcounter{equation}{0} 
\section{SU$(2)$ UIR matrix elements and some useful expansions} \label{appendix:SU(2)}

\subsection{Matrix elements of the SU$(2)$ UIRs}
For $\xi = \begin{pmatrix} \xi^4 + \ii \xi^3 & - \xi^2 + \ii \xi^1 \\ \xi^2 + \ii \xi^1 & \xi^4 - \ii \xi^3 \end{pmatrix} \in \mathrm{SU}(2)$ viewed as the unit norm real quaternion $\xi = \big(\xi^4, \pmb{\xi}= (\xi^i)\big) \in \H \sim \R^4$, matrix elements of the SU$(2)$ UIRs, in agreement with Talman \cite{talman68}, read as:
\begin{align} \label{matelsu2}
\nonumber D^j_{m_1 m_2} (\xi) &= (-1)^{m_1-m_2} \sqrt{(j+m_1)! \; (j-m_1)! \; (j+m_2)! \; (j-m_2)!} \\
&\qquad\times \sum_{t} \frac{(\xi^4 + \ii \xi^3)^{j-m_2 - t}}{(j - m_2 -t)!} \, \frac{(\xi^4 - \ii\xi^3)^{j + m_1 - t}}{(j + m_1 -t)!}\,\frac{(- \xi^2 + \ii \xi^1)^{t+m_2 - m_1}}{(t + m_2 - m_1)!} \frac{(\xi^2 + \ii \xi^1)^{t}}{t!} \,,
\end{align}
where $2 j\in \N$  
and $-j \leqslant m_1,m_2 \leqslant j$, while the values of $t$ to be summed over are those for which the arguments of the factorial functions remain non-negative, i.e., the values compatible with $m_1-m_2\leqslant t \leqslant j+m_1$ and $0\leqslant t \leqslant j-m_2$. The holomorphic extension of these elements to $\C^4 \sim \H_{\C}$ is denoted by 
$D^j_{m_1 m_2} (z)$,  ($z\in \mathbb{H}_{\C}$), and reads as:
\begin{align} \label{matelHC}
\nonumber D^j_{m_1 m_2} (z) &= (-1)^{m_1-m_2} \sqrt{(j+m_1)! \; (j-m_1)! \; (j+m_2)! \; (j-m_2)!} \\
&\qquad \times \sum_{t} \frac{(z^4 + \ii z^3)^{j-m_2 - t}}{(j - m_2 -t)!} \, \frac{(z^4 - \ii z^3)^{j + m_1 - t}}{(j + m_1 -t)!}\,\frac{(- z^2 + \ii z^1)^{t+m_2 - m_1}}{(t + m_2 - m_1)!} \frac{(z^2+ \ii z^1)^{t}}{t!} \,.
\end{align}
These elements are harmonic polynomials in the sense that:
\begin{equation}\label{harmpol4}
\sum_{a=1}^4\frac{\partial^2}{\partial (z^a)^2} D^j_{m_1 m_2} (z) = 0\,.
\end{equation}
Note that, considering the above, in the simplest case ($j=1/2$)  we have:
\begin{equation}\label{D12nz}
{D}^{1/2}_{m_1m_2}(z) =
\begin{pmatrix} z^4-\ii z^3 & -z^2 - \ii z^1 \\ z^2-\ii z^1 & z^4+\ii z^3 \end{pmatrix}.
\end{equation}
The polynomials \eqref{matelHC} in the case of pure-vector complex or real quaternions (i.e., when $z^4=0$) take the form:
\begin{align} \label{matelVHC}
\nonumber D^j_{m_1 m_2} (\pz) &= (-1)^{-m_2} \sqrt{(j+m_1)! \; (j-m_1)! \; (j+m_2)! \; (j-m_2)!} \\
&\qquad\times \sum_{t} (-1)^{-t}\; \frac{(z^3)^{2j-2t+m_1-m_2}}{(j - m_2 -t)! \; (j + m_1 -t)!}\,\frac{( z^1 + i z^2)^{t+m_2 - m_1}}{(t +m_2 - m_1)!} \frac{( z^1- i z^2)^{t}}{t!} \,.
\end{align}

\subsection{Three key expansion theorems for polynomials ${D}^{j}_{m_1m_2}(z)$}
Let us introduce the following notations:
\begin{equation} \label{sjm1}
\sigma_m^j = \frac{1}{\sqrt{(j-m)!\,(j+m)!}}\, , \qquad \sigma_{m_1 m_2}^j = \sigma_{m_1 }^j \, \sigma_{ m_2}^j \,.
\end{equation}
We then have three key expansion formulae for the $D^j$ functions  ($z,z^{\prime} \in \H_{\C}$): 
\begin{align}
\label{addexp1}
&\sigma_{m_1 m_2}^j\, D^j_{m_1 m_2} (z + z^{\prime}) = \sum_{j^\prime m^\prime_1 m^\prime_2} \sigma_{m_1-m^\prime_1\, m_2 -m^\prime_2}^{j-j^\prime} \,D_{m_1-m^\prime_1\, m_2 -m^\prime_2}^{j-j^\prime}(z) \,\sigma_{m^\prime_1 m^\prime_2}^{j'} \,D_{m^\prime_1 m^\prime_2}^{j^\prime} (z^{\prime})\,,
\\[0.25cm]
\label{addexp2}
& \left(\sigma_{m_1 m_2}^j\right)^{-1}\, \big[\det(z+z^{\prime})\big]^{-1} \,D^j_{m_1 m_2} \left((z + z^{\prime})^{-1}\right) = \nonumber \\
&\quad\sum_{j^\prime m^\prime_1 m^\prime_2} (-1)^{2j^\prime} \sigma_{m^\prime_1 m^\prime_2}^{j^\prime} \,D_{m^\prime_1 m^\prime_2}^{j^\prime} (z)\,\left(\sigma_{m_1 +m^\prime_2\, m_2 +m^\prime_1}^{j +j^\prime}\right)^{-1} \,(\det z^{\prime})^{-1}\,D_{m_1 +m^\prime_2\, m_2 +m^\prime_1}^{j +j^\prime}({z^{\prime}}^{-1}) \,, \quad \Vert z \Vert < \Vert z^{\prime} \Vert\,, \\[0.25cm]
\label{multexp}  
&D^j_{m_1 m_2} (z z^{\prime}) = \sum_{m^\prime}D^j_{m_1 m^\prime} (z)\,D^j_{m^\prime m_2} ( z^{\prime}) \, ,
\end{align}
where the values of indices to be summed over are those that meet the allowed ranges of indices of the SU$(2)$ representations. Note that the first two expansions are of the type of addition theorems whereas the latter one just results from the group representation. Justifications are given in Ref. \cite{gazeau78}.

\subsection{Wigner $3$-$j$ symbols}
In connection with the reduction of the tensor product of two UIRs of SU$(2)$, we have the following equivalent formulae involving the so-called $3$-$j$ symbols (proportional to Clebsch-Gordan coefficients), in the  notation of Ref.~\cite{talman68}:
\begin{align}
D^j_{m_1 m_2}(\xi)\,D^{j^\prime}_{m^\prime_1 m^\prime_2} (\xi) & = \sum_{j^{\prime\prime} m^{\prime\prime}_1 m^{\prime\prime}_2}(2j^{\prime\prime} + 1)
\begin{pmatrix} j & j^\prime & j^{\prime\prime} \\ m_1 & m^\prime_1 & m^{\prime\prime}_1 \end{pmatrix}
\begin{pmatrix} j & j^\prime & j^{\prime\prime} \\ m_2 & m^\prime_2 & m^{\prime\prime}_2 \end{pmatrix}\, \overline{D^{j^{\prime\prime}}_{m^{\prime\prime}_1 m^{\prime\prime}_2} (\xi)}\\
&= \sum_{j^{\prime\prime} m^{\prime\prime}_1 m^{\prime\prime}_2}(2j^{\prime\prime} + 1) (-1)^{m^{\prime\prime}_1 - m^{\prime\prime}_2}\,
\begin{pmatrix} j & j^\prime & j^{\prime\prime} \\ m_1 & m^\prime_1 & - m^{\prime\prime}_1 \end{pmatrix}
\begin{pmatrix} j & j^\prime & j^{\prime\prime} \\ m_2 & m^\prime_2 & - m^{\prime\prime}_2 \end{pmatrix} \, D^{j^{\prime\prime}}_{m^{\prime\prime}_1 m^{\prime\prime}_2} (\xi)\,.
\end{align}

One of the multiple expressions of the $3$-$j$ symbols of Wigner (in the convention that they are all real) is given by:
\begin{align} \label{3japc}
\nonumber \begin{pmatrix} j & j^\prime & j^{\prime\prime} \\ m & m^\prime & m^{\prime\prime} \end{pmatrix} &= (-1)^{j-j^\prime-m^{\prime\prime}} \sqrt{\frac{(j +j^\prime-j^{\prime\prime})!\; (j -j^\prime+j^{\prime\prime})!\; (-j +j^\prime +j^{\prime\prime})!}{(j +j^\prime +j^{\prime\prime} +1)!}}\\
\nonumber&\qquad\times \sqrt{(j+m)!\; (j-m)!\; (j^\prime+m^\prime)!\; (j^\prime-m^\prime)!\; (j^{\prime\prime}+m^{\prime\prime})!\; (j^{\prime\prime}-m^{\prime\prime})!} \\
\nonumber&\quad\qquad\times \sum_s (-1)^s  \frac{1}{s!\; (j^\prime+m^\prime -s)!\; (j-m-s)!\; (j^{\prime\prime}-j^\prime+m +s)!}\\
&\mbox{\hskip 1.35cm}\times \frac{1}{(j^{\prime\prime} -j -m^\prime +s)!\; (j + j^\prime -j^{\prime\prime}-s)!} \nonumber\\[0.25cm]
&\equiv (-1)^{j-j^{\prime}-m^{\prime\prime}}(2j^{\prime\prime}+1)^{-1/2}\,\left(j m j^{\prime} m^{\prime}|j j^{\prime} j^{\prime\prime} -m^{\prime\prime}\right)\,,
\end{align}
with the constraints:
\be
\label{cons 1} m + m^{\prime} + m^{\prime\prime}=0\,, \qquad 
\vert j-j^{\prime}\vert \leqslant j^{\prime\prime}\leqslant j+ j^{\prime}\,,\qquad
j +j^{\prime} - j^{\prime\prime}\in \Z\,,
\ee
while the values of $s$ to be summed over are those for which the arguments of the factorial functions remain non-negative. Note that the notation $\left(j m j^{\prime} m^{\prime}|j j^{\prime} j^{\prime\prime} -m^{\prime\prime}\right)$ stands for the so-called \textit{vector-coupling} or \textit{Clebsch-Gordan} coefficients \cite{edmonds96}.

Let us recall two fundamental summation rules resulting from the unitary equivalence between couplings \cite{edmonds96}:
\begin{equation}\begin{array}{lll} \label{3jmm}
\ds \sum_{j_3,m_3} (2j_3 +1) \begin{pmatrix} j_1 & j_2 & j_3 \\ m_1 & m_2 & m_3 \end{pmatrix} \begin{pmatrix} j_1 & j_2 & j_3 \\m^{\prime}_1 & m^{\prime}_2 & m_3 \end{pmatrix} &=&\ds \delta_{m_1m^{\prime}_1}\delta_{m_2m^{\prime}_2}\,.
\\[0.4cm]
\ds \sum_{m_1,m_2} (2j_3 +1)\begin{pmatrix} j_1 & j_2 & j_3 \\ m_1 & m_2 & m_3 \end{pmatrix} \begin{pmatrix} j_1 & j_2 & j^{\prime}_3 \\ m_1 & m_2 & m^{\prime}_3 \end{pmatrix} &=& \ds\delta_{j_3j^{\prime}_3}\delta_{m_3m^{\prime}_3} \delta(j_1j_2j_3)\,,
\end{array}\end{equation}
where $\delta(j_1j_2j_3)=1$ if $j_1,j_2,j_3$ satisfy the triangular condition, and is zero otherwise.

\setcounter{equation}{0} 
\section{Reproducing kernel Hilbert space} \label{appendix:kernelH}

Here, we follow Refs.~\cite{horzela12,horzela12A}. Suppose that the sequence of holomorphic functions $\mathrm{F}^{\lambda}_{\nu}(\pz)$ satisfies the two conditions:
\begin{itemize}
\item Square summability in the following sense:
  \begin{equation}\label{sqsum}
  \sum_{\nu}\left\vert\mathrm{F}^{\lambda}_{\nu}(\pz)\right\vert^2 < \infty\,, \quad \pz\in \mathcal{D}^{(3)} \,.
  \end{equation}
\item Completeness:
  \begin{equation} \label{cpletn}
  (a_\nu) \in \ell^2 \quad \mbox{and}\quad \sum_{\nu} a_\nu \,\mathrm{F}^{\lambda}_{\nu}(\pz)=0\,\, \mbox{(for all $\pz\in\mathcal{D}^{(3)}$)}\, \quad \Rightarrow \quad a_\nu = 0\;\,\forall \nu \,.
  \end{equation}
\end{itemize}
Condition \eqref{sqsum} allows to view $ \left[\det\left( 1 + \pz\,\overline{\pz^{\prime}}\right)\right]^{-\lambda}$ as a positive definite kernel (i.e. $\mathrm{K}^{\lambda}(\pz,\pz^{\prime}) = \left[\det\left(1 + \pz\,\overline{\pz^{\prime}}\right)\right]^{-\lambda}$)
on $\mathcal{D}^{(3)}$ and to uniquely determine a Hilbert space $\mathcal{K}^{\lambda}$ of holomorphic functions on this domain, with the inner product $(\cdot,\cdot)_{\lambda}\equiv \big( \mathrm{K}^{\lambda}(\pz,\pz) \mathrm{K}^{\lambda}(\pz^\prime,\pz^{\prime})\big)^{-1/2}\; \mathrm{K}^{\lambda}(\pz,\pz^{\prime})$, where, to avoid any further `renormalization', it is assumed that $\mathrm{K}^{\lambda}(\pz,\pz)=1$ ($\pz\in\mathcal{D}^{(3)}$) \cite{horzela12,horzela12A}. It follows that the set of functions
$\big\{\mathrm{K}^{\lambda}_{\pz}\big\}_{\pz\in \mathcal{D}^{(3)}}\,$, 
with $\mathrm{K}^{\lambda}_{\pz} \equiv \mathrm{K}(\cdot,\pz)_\lambda\,$,
is complete in $\mathcal{K}^{\lambda}$ and the reproducing kernel property reads as:
\begin{equation} \label{reprop}
f(\pz) = \left(\mathrm{K}^{\lambda}_{\pz},f\right)_{\lambda}\,, \quad f\in \mathcal{K}^{\lambda}\,.
\end{equation}
Due to the conditions \eqref{sqsum} and \eqref{cpletn} the sequence $\big\{\mathrm{F}^{\lambda}_{\nu}\big\}$ is a basis in $\mathcal{K}^{\lambda}$ \cite{aronszajn50}.

\setcounter{equation}{0} 
\section{{Special functions material}} \label{appendix:SpecialFunctions}

\subsection{Fundamental kernel expansion for complex quaternions}
We start from the generating function for Gegenbauer polynomials \cite{hua63}:
\begin{equation} \label{expgeg}
(1 + u^2 - 2 u t)^{-\lambda} = \sum_{l=0}^{\infty} u^l \, C_l^{\lambda}(t)\,, \quad \vert u \vert <1 \,,
\end{equation}
and we apply it to the expansion of $\left[\det\left(1 - z{z^{\prime}}^{\ast}\right)\right]^{-\lambda}$ when $z$ lies in the classical domain: \begin{equation}\label{domC4}
\mathcal{D}^{(4)}= \Big\{ Z\,;\, \bu_2 - ZZ^{\dag} >0 \Big\}\,,
\end{equation}
where we remind the one-to-one map \eqref{cqumat}. Then:
\begin{align}\label{popo}
\left[\det\left( 1 - z{z^{\prime}}^{\ast}\right)\right]^{-\lambda} &=
\left[1 - 2 z\cdot \overline{z^{\prime}}+ \left(z\cdot z\right)\,\overline{\left(z^{\prime}\cdot z^{\prime}\right)}\right]^{-\lambda} \nonumber\\
&=\left[1 - 2 \left[\left(z\cdot z\right)\overline{\left(z^{\prime}\cdot z^{\prime}\right)}\right]^{1/2} \frac{z\cdot \overline{z^{\prime}}}{\left[\left(z\cdot z\right)\overline{\left(z^{\prime}\cdot z^{\prime}\right)}\right]^{1/2}}+ \left(z\cdot z\right)\overline{\left(z^{\prime}\cdot z^{\prime}\right)}\right]^{-\lambda} \nonumber\\
&= \sum_{l=0}^{\infty} \left[\left(z\cdot z\right)\overline{\left(z^{\prime}\cdot z^{\prime}\right)}\right]^{l/2} \, C_l^{\lambda}\left(\frac{z\cdot \overline{z^{\prime}}}{\left[\left(z\cdot z\right)\overline{\left(z^{\prime}\cdot z^{\prime}\right)}\right]^{1/2}}\right) \,,
\end{align}
where $z\cdot \overline{z^{\prime}}$ stands for the analytic continuations of the Euclidean inner product in $\R^4$. Here, the square root of the holomorphic quadratic term $\left(z\cdot z\right)$ is understood as lying in the first Riemann sheet, i.e., becomes $\Vert x \Vert$ as $y = 0$.

We now make use of the expansion formula concerning Gegenbauer polynomials \cite{hua63} and the holomorphic extensions \eqref{matelHC} of the matrix elements of the SU$(2)$ UIRs:
\begin{align} \label{gegen}
&C_l^{\lambda}(t) = \frac{1}{\Gamma(\lambda)\;\Gamma(\lambda-1)}\,\sum_{k = 0}^{\lfloor \frac{l}{2}\rfloor} c_k\, C_{l-2k}^{1} (t)\,, \qquad
c_k = \frac{(l-2k + 1)\; \Gamma(k+\lambda - 1)\; \Gamma(\lambda + l - k)}{k!\; \Gamma(l-k + 2)} \,,\\
\label{geghyp} &\left[\left(z\cdot z\right)\overline{\left(z^{\prime}\cdot z^{\prime}\right)}\right]^{j} \, C_{2j}^{1}\left(\frac{z\cdot \overline{z^{\prime}}}{\left[\left(z\cdot z\right)\overline{\left(z^{\prime}\cdot z^{\prime}\right)}\right]^{1/2}}\right) = \sum_{m_1 m_2} D_{m_1m_2}^j(z)\, \overline{D_{m_1m_2}^j(z^\prime)}\,, \quad z\,, z^{\prime} \in \mathbb{H}_{\C} \,.
\end{align}
Combining Eqs.~\eqref{popo}, \eqref{gegen} and \eqref{geghyp}, while we have in mind \eqref{matelHC}, we get the expansion formula:
\begin{equation} \label{expdetlambd}
\left[\det\left(1 - z{z^{\prime}}^{\ast}\right)\right]^{-\lambda} = \frac{1}{\Gamma(\lambda)\;\Gamma(\lambda-1)}\, \sum_{l=0}^{\infty}\sum_{k = 0}^{\lfloor \frac{l}{2}\rfloor} \sum_{m_1 m_2}c_k\,(z\cdot z)^{k}\, D_{m_1m_2}^{{l}/{2}-k}(z)\,\overline{(z^{\prime}\cdot z^{\prime})}^{k}\,\overline{D_{m_1m_2}^{{l}/{2}-k}(z^{\prime})} \,.
\end{equation}
Hence, provided the above series converges in the quadratic sense, the set of holomorphic functions:
\begin{equation}\label{holfctslambd}
\mathrm{F}^{\lambda}_{l,k,m_1,m_2}(z)\equiv \mathrm{F}^{\lambda}_{\nu}(z) \equiv \sqrt{\frac{c_k}{\Gamma(\lambda)\;\Gamma(\lambda-1)}}\,(z\cdot z)^{k}\, D_{m_1m_2}^{{l}/{2}-k}(z) \,,
\end{equation}
with $ z\in \mathcal{D}^{(4)}$ and : 
\begin{equation}\label{nkmm}
\nu \equiv (l,k,m_1,m_2)\,, \qquad l\in \N\,, \qquad 0\leqslant k\leqslant \left\lfloor \frac{l}{2}\right\rfloor\,, \qquad -\frac{l}{2}+k \leqslant m_1,m_2 \leqslant \frac{l}{2}-k \,,
\end{equation}
allow to build a reproducing kernel space as will be presented in the sequel.

\subsection{Holomorphic solid spherical harmonics}
The spherical harmonics are defined in Ref.~\cite{edmonds96} as functions of spherical coordinates of $\hat{\pmb{r}}$ in the unit sphere $\mathbb{S}^2$ by:
\begin{equation} \label{spherharm}
Y_{lm}(\theta,\phi) = (-1)^m\left[\frac{(2l+1)(l-m)!}{4\pi(l+m)!}\right]^{1/2}\,P_l^m(\cos\theta)\,e^{\ii m \phi}\,,
\end{equation}
where $(l,m)\in \N \times \Z$, with $-l\leqslant m\leqslant l$, and the associated Legendre polynomials $P_l^m$ are given for $0\leqslant m\leqslant l$ by:
\begin{equation} \label{Plmg0}
P_l^m(x) = \frac{(1-x^2)^{m/2}}{2^l\,l!}\,\frac{\ud^{l+m} }{\ud x^{l+m}}(x^2-1)^l\, ,
\end{equation}
and extended to negative $m$ thanks to the relation:
\begin{equation}\label{Plms0}
P_l^{m}(x) = (-1)^m\frac{(l+m)!}{(l-m)!}\, P_l^{-m}(x)\, .
\end{equation}
There results in:
\begin{equation} \label{ccYlm}
\overline{Y_{lm}(\theta,\phi)} = (-1)^{m}\,Y_{l,-m}(\theta,\phi)\,.
\end{equation}
A closed form of $P_l^{m}(x)$, for $x=\cos\theta$, is given by:
\begin{equation} \label{Plmcf}
P_l^m(\cos \theta) = 2^l (\sin\theta)^m\sum_{k=m}^l \frac{\Gamma\left(\frac{k+l+1}{2}\right)}{\Gamma\left(\frac{k-l+1}{2}\right)}\frac{(\cos\theta)^{k-m}}{(l-k)!\, (k-m)!}\,.
\end{equation}
Three important features of spherical harmonics are:
\begin{itemize}
\item Orhonormality:
  \begin{equation} \label{sphHarOrth}
  \int_{\mathbb{S}^2}Y_{lm}(\hat{\pr})\,\overline{Y_{l^{\prime}m^{\prime}}(\hat{\pr})}\,d{\hat{\pr}}= \int_0^{\pi}\sin\theta\,\ud\theta\int_0^{2\pi}\ud\phi\,Y_{lm}(\theta,\phi)\,\overline{Y_{l^{\prime}m^{\prime}}(\theta,\phi)}= \delta_{ll^{\prime}}\,\delta_{mm^{\prime}}\,.
  \end{equation}
\item Legendre polynomial as a reproducing kernel:
  \begin{equation} \label{sphHarrepker}
  C_l^{1/2}\left(\hat{\pr}\cdot \hat{\pr}^{\prime}\right)=P_l\left(\hat{\pr}\cdot \hat{\pr}^{\prime}\right)= \frac{2l+1}{4\pi}\sum_{m=-l}^{l}Y_{lm}(\hat{\pr})\,\overline{Y_{lm}(\hat{\pr}^\prime)}\,.
  \end{equation}
\item Product as a sum:
  \begin{equation} \label{YY3j}
  \begin{split}
  Y_{l_1m_1}(\theta,\phi)\,Y_{l_2m_2}(\theta,\phi)&= (-1)^{m_3}\sum_{l_3}\left[\frac{(2l_1+1)(2l_2+1)(2l_3+1)}{4\pi}\right]^{1/2} \\
 &  \qquad \times\, \begin{pmatrix} l_1 & l_2 & l_3 \\ m_1 & m_2 & - m_3 \end{pmatrix}\, \begin{pmatrix} l_1 & l_2 & l_3 \\ 0 & 0 & 0 \end{pmatrix}\, Y_{l_3m_3}(\theta,\phi)\,.
  \end{split}
  \end{equation}
\end{itemize}
Also, note that one finds in other definitions of Eqs.~\eqref{Plmg0} and \eqref{Plmcf}, like those given in Ref. \cite{Olver}, an extra factor $(-1)^m$ which is absent in our convention.

In Eq.~\eqref{Plmcf} the ratio $\frac{\Gamma\left(\frac{k+l+1}{2}\right)}{\Gamma\left(\frac{k-l+1}{2}\right)}$ vanishes for all $(k,l)$ when $l-k$ is odd. This crucial point allows us to extend spherical harmonics to $\R^3$ to define the so-called solid spherical harmonics and to get their closed form:
\begin{equation} \label{solspharmp1111111}
\begin{split}
\mathcal{Y}_{lm}(\vec{r}) &\equiv r^l\,Y_{lm}(\theta,\phi)\\
&= (-1)^m\left[\frac{(2l+1)(l-m)!}{4\pi(l+m)!}\right]^{1/2}\,2^l\, (x^1+\ii x^2)^m\,
\sum_{k=m}^l \frac{\Gamma\left(\frac{k+l+1}{2}\right)}{\Gamma\left(\frac{k-l+1}{2}\right)}\frac{(x^3)^{k-m}}{(k-m)!}\,\frac{r^{l-k}}{(l-k)!}\,,
\end{split}
\end{equation}
where $\vec{r} = (x^1,x^2,x^3)\in \R^3$ and $0\leqslant m\leqslant l$. This has to be completed by the relation:
\begin{equation} \label{solspharmn}
\mathcal{Y}_{l,-m}(\vec{r}) = (-1)^m\frac{(l-m)!}{(l+m)!}\, \mathcal{Y}_{lm}(\vec{r})\,.
\end{equation}
These homogenous polynomials are harmonic, i.e., obey the Laplace equation $\triangle \mathcal{Y}_{lm}(\vec{r})=0$. With this extension, Eq. \eqref{YY3j} still holds with the necessary homogeneity adaptation:
\begin{equation}\label{SYSY3j}
\begin{split}
\mathcal{Y}_{l_1m_1}(\vec{r})\,\mathcal{Y}_{l_2m_2}(\vec{r}) &= (-1)^{m_3} \sum_{l_3}\left[\frac{(2l_1+1)(2l_2+1)(2l_3+1)}{4\pi}\right]^{1/2} \\
&\qquad\times r^{l_1 +l_2 -l_3} \begin{pmatrix} l_1 & l_2 & l_3 \\ m_1 & m_2 & -m_3 \end{pmatrix}\,\begin{pmatrix} l_1 & l_2 & l_3 \\ 0 & 0 & 0 \end{pmatrix}\,\mathcal{Y}_{l_3m_3}(\vec{r})\,.
\end{split}
\end{equation}
Note that the particular $3$-$j$ coefficient:
\begin{equation}\label{part3j}
\begin{split}
\begin{pmatrix} l_1 & l_2 & l_3 \\ 0 & 0 & 0 \end{pmatrix} &= (-1)^{J/2} \left[\frac{(J-2l_1)!(J-2l_2)!(J-2l_3)!}{(J+1)!}\right]^{1/2}\\ 
&\qquad \times\frac{\left(\frac{J}{2}\right)!}{\left(\frac{J}{2}-l_1\right)!\left(\frac{J}{2}-l_2\right)!\left(\frac{J}{2}-l_3\right)!}\,, 
\qquad J\equiv l_1+l_2+l_3\,,
\end{split}
\end{equation}
vanishes, if $J$ is odd.

We then extend Eq.~\eqref{solspharmp1111111} to the domain $\mathcal{D}^{(3)}$ as:
\begin{equation}\label{solspharmp}
\begin{split}
\mathcal{Y}_{lm}(\pz)&= (-1)^m\left[\frac{(2l+1)(l-m)!}{4\pi(l+m)!}\right]^{1/2}\\
&\qquad\times 2^l(z^1+\ii z^2)^m\, \sum_{k=m}^l \frac{\Gamma\left(\frac{k+l+1}{2}\right)}{\Gamma\left(\frac{k-l+1}{2}\right)}\frac{(z^3)^{k-m}}{(k-m)!}\,\frac{(\pz\cdot\pz)^{\frac{l-k}{2}}}{(l-k)!}\,,
\end{split}
\end{equation}
with $ \pz = (z^1,z^2,z^3) \in \mathcal{D}^{(3)}$. Consistently, Eq.~\eqref{SYSY3j} generalises as:
\begin{equation}\label{HYHY3j}
\begin{split}
\mathcal{Y}_{l_1m_1}(\pz)\,\mathcal{Y}_{l_2m_2}(\pz) &= (-1)^{m_3}\sum_{l_3}\left[\frac{(2l_1+1)(2l_2+1)(2l_3+1)}{4\pi}\right]^{1/2} \\
&\qquad\times (\pz\cdot\pz)^{\frac{l_1 +l_2 -l_3}{2}} \begin{pmatrix} l_1 & l_2 & l_3 \\ m_1 & m_2 & -m_3 \end{pmatrix}\,\begin{pmatrix} l_1 & l_2 & l_3 \\ 0 & 0 & 0 \end{pmatrix}\,\mathcal{Y}_{l_3m_3}(\pz)\,,
\end{split}
\end{equation}
where we recall that $l_1 +l_2 -l_3$ has to be even due to the property of the particular $3$-$j$-coefficient $\begin{pmatrix} l_1 & l_2 & l_3 \\ 0 & 0 & 0 \end{pmatrix}$.

Note the following formula relating the holomorphic solid spherical to the SU$(2)$ UIR matrix elements \cite{hassan80}:
\begin{equation} \label{YlDl2}
\big(\sigma^l_m\big)^{-1}\,\mathcal{Y}_{lm}(\pmb{z}) = 2^{-l}\left(\frac{2l+1}{4\pi}\right)^{1/2} \sum_{\stackrel{m_1 m_2}{m_2-m_1=m}}(-1)^{m_1}\; \big(\sigma^{l/2}_{m_1m_2}\big)^{-1}\; D^{l/2}_{m_1m_2}[(z^4,\pmb{z})]\,,
\end{equation}
where $z^4$ is arbitrary. Conversely (writing in an abuse of notation $(0,\pmb{z})\equiv \pmb{z}$):
\begin{equation} \label{DlYl2}
\begin{split}
{D}^{l/2}_{m_1m_2}(\pmb{z})
&= (4\pi)^{1/2}\sum_{\stackrel{l^{\prime}}{l-l^{\prime} = \mathrm{even}}}(-1)^{l^{\prime}+m_2}\; 2^{l^{\prime}}\; (2l^{\prime}+1)^{1/2} \left(\frac{(l-l^{\prime})!}{(l+l^{\prime}+1)!}\right)^{1/2} \frac{\left[\frac{1}{2}(l+l^{\prime})\right]!}{\left[\frac{1}{2}(l-l^{\prime})\right]!}\\
&\qquad\qquad\quad \times \begin{pmatrix} \frac{l}{2} & \frac{l}{2}& l^{\prime} \\ m_1 & -m_2 & m^{\prime} \end{pmatrix} (\pz\cdot\pz)^{\frac{l-l^{\prime}}{2}}\,\mathcal{Y}_{l^{\prime}m^{\prime}}(\pmb{z})\,.
\end{split}
\end{equation}

\subsection{Fundamental kernel expansion for pure-vector complex quaternions}
We start again from the generating function for Gegenbauer polynomials \cite{hua63}:\footnote{\label{footnote negative lambda0}Here, for later use, it is worthwhile noting that, for negative values of $\lambda\in\mathbb{Z}$, the left-hand side of Eq. \eqref{expgeg1} turns to a finite sum as:
$\ds (1 + u^2 - 2 u t)^{-\lambda} = \sum_{l=0}^{2|\lambda|} u^l \, C_l^{\lambda}(t)$.}
\begin{equation} \label{expgeg1}
(1 + u^2 - 2 u t)^{-\lambda} = \sum_{l=0}^{\infty} u^l \, C_l^{\lambda}(t)\,, \quad \vert u \vert <1\,, \quad \lambda \neq 0\,,
\end{equation}
and we apply it to the expansion of $\left[\det\left(1 + \pz\,\overline{\pz^{\prime}}\right)\right]^{-\lambda}$:
\begin{align}\label{opop}
\left[\det\left( 1 + \pz\,\overline{\pz^{\prime}}\right)\right]^{-\lambda}&=  \left[1 - 2 \pz\cdot\overline{\pz^{\prime}} + \left(\pz\cdot\pz\right)\overline{\left(\pz^{\prime}\cdot\pz^{\prime}\right)}\right]^{-\lambda} \nonumber\\
&= \left[1 - 2 \left[\left(\pz\cdot\pz\right)\overline{\left(\pz^{\prime}\cdot\pz^{\prime}\right)}\right]^{1/2} \frac{\pz\cdot\overline{\pz^{\prime}}}{\left[\left(\pz\cdot\pz\right)\overline{\left(\pz^{\prime}\cdot\pz^{\prime}\right)}\right]^{1/2}}+ \left(\pz\cdot\pz\right)\overline{\left(\pz^{\prime}\cdot\pz^{\prime}\right)}\right]^{-\lambda} \nonumber\\
&= \sum_{l=0}^{\infty} \left[\left(\pz\cdot\pz\right)\overline{\left(\pz^{\prime}\cdot\pz^{\prime}\right)}\right]^{l/2} \, C_l^{\lambda}\left(\frac{\pz\cdot\overline{\pz^{\prime}}}
{\left[\left(\pz\cdot\pz\right)\overline{\left(\pz^{\prime}\cdot\pz^{\prime}\right)}\right]^{1/2}}\right)\,,
\end{align}
where $\pz\cdot\overline{\pz^{\prime}}$ stands for the analytic continuations of the Euclidean inner product in $\R^3$. Above, the square root of the holomorphic quadratic $\left(\pz\cdot\pz\right)$ is understood as lying in the first Riemann sheet, i.e., becomes $\Vert \px \Vert$ as $\py = 0$.

Then, we make use of the expansion formula concerning Gegenbauer polynomials \cite{hua63} and the holomorphic extensions \eqref{solspharmp} of the solid spherical harmonics:
\begin{align}\label{gegen1}
& C_l^{\lambda}(t) = \frac{\sqrt{\pi}}{\Gamma(\lambda)\; \Gamma\left(\lambda-\frac{1}{2}\right)}\,\sum_{k = 0}^{\lfloor \frac{l}{2}\rfloor} d_k\,  C_{l-2k}^{1/2} (t)\,, \qquad 
d_k = \frac{\left(l-2k + \frac{1}{2}\right)\, \Gamma\left(k+\lambda - \frac{1}{2}\right)\, \Gamma(\lambda + l - k)}{k! \,\Gamma\left(l-k + \frac{3}{2}\right)} \,,\\[0.25cm]
\label{geghyp1} &\left[\left(\pz\cdot\pz\right)\overline{\left(\pz^{\prime}\cdot\pz^{\prime}\right)}\right]^{l/2} \, C_l^{1/2}
\left(\frac{\pz\cdot\overline{\pz^{\prime}}}{\left[\left(\pz\cdot\pz\right)\overline{\left(\pz^{\prime}\cdot\pz^{\prime}\right)}\right]^{1/2}}\right) = \frac{4\pi}{2l + 1} \sum_{m=-l}^l \mathcal{Y}_{lm}(\pz)\,\overline{\mathcal{Y}_{lm}(\pz^{\prime})}\,.
\end{align}
For future use, considering the following expansions (note that $\Gamma(x)=(x-1)\, \Gamma(x-1)$):
\begin{align}
&\mbox{for}\quad k\neq0 \;:\; \quad \Gamma\big(k+\lambda-{1}/{2}\big) = \big(k-1+\lambda-{1}/{2}\big)\, \big(k-2+\lambda-{1}/{2}\big)\, \dotsc \,\underbrace{ \big(k-k+\lambda-{1}/{2}\big)}_{=(\lambda-{1}/{2})}\, \Gamma(\lambda-{1}/{2}\big)\,, \\
&\mbox{for}\quad l-k\neq0 \;:\; \quad \Gamma\big(\lambda + l-k\big) = \big(\lambda + l-k-1\big)\, \big(\lambda + l-k-2\big)\, \dotsc \, \underbrace{\big(\lambda + l-k-(l-k)\big)}_{(\lambda)}\, \Gamma\big(\lambda\big)\,,
\end{align}
we also rewrite Eq.~\eqref{gegen1} as:
\begin{align}
\label{234} C_l^{\lambda}(t) = {\sqrt{\pi}}\,\sum_{k = 0}^{\lfloor \frac{l}{2}\rfloor} d^\prime_k\, C_{l-2k}^{1/2} (t)\,,
\end{align}
where:
\begin{align}
\label{234'} d^\prime_k = \frac{\left(l-2k + \frac{1}{2}\right)}{k! \,\Gamma\left(l-k + \frac{3}{2}\right)}\, \underbrace{\big[\big(k-1+\lambda-{1}/{2}\big)\, \big(k-2+\lambda-{1}/{2}\big)\, \dotsc\, \big(\lambda-{1}/{2}\big)\big]}_{\mbox{should be replaced by $1$ for $k=0$}}\, \underbrace{\big[\big(\lambda + l-k-1\big)\, \big(\lambda + l-k-2\big)\, \dotsc \, \big(\lambda\big)\big]}_{\mbox{should be replaced by $1$ for $l-k=0$}} \,.
\end{align}
Note that, contrary to the former identity \eqref{gegen1}, the absence of the terms $\Gamma(\lambda-1/2)$ and $\Gamma(\lambda)$ in the latter allows one to take into account the negative values of $\lambda$ as well.

Now, combining Eqs.~\eqref{opop}, \eqref{gegen1} and \eqref{geghyp1}, while we have in mind the \emph{Legendre duplication formula}: 
\begin{align}
\Gamma(x) \; \Gamma(x+1/2) = 2^{1-2x} \sqrt{\pi} \; \Gamma(2x)\,,
\end{align}
the following expansion (for $\lambda \neq 0$) comes to the fore:
\begin{equation} \label{expdetlambd'}
\left[\det\left( 1 + \pz\,\overline{\pz^{\prime}}\right)\right]^{-\lambda} = \sum_{l=0}^{\infty}\sum_{k = 0}^{\lfloor \frac{l}{2}\rfloor} \sum_{m=2k-l}^{l-2k}\,a_{\lambda,l,k}\,(\pz\cdot\pz)^{k}\, \mathcal{Y}_{l-2k,m}(\pz)\,\overline{(\pz^{\prime}\cdot\pz^{\prime})}^{k}\,\overline{\mathcal{Y}_{l-2k,m}(\pz^{\prime})}\,,
\end{equation}
where:
\begin{equation} \label{normlk}
a_{\lambda,l,k} = \frac{2^{2\lambda-1} \,\pi \,\Gamma\left(k+\lambda - \frac{1}{2}\right) \Gamma(\lambda + l - k)}{\Gamma(2\lambda-1)\,k! \,\Gamma\left(l-k + \frac{3}{2}\right)}\,.
\end{equation}
Hence, provided the series \eqref{expdetlambd'} converges in the quadratic sense, the set of holomorphic functions:
\begin{eqnarray}\label{D3 basis1111111} 
\mathrm{F}^{\lambda}_{l,k,m}(\pz) \equiv \mathrm{F}^{\lambda}_{\nu}(\pz) &\equiv& \sqrt{a_{\lambda,l,k}}\,(\pz\cdot\pz)^{k}\, \mathcal{Y}_{l-2k,m}(\pz) \,,
\end{eqnarray}
with $\pz \in \mathcal{D}^{(3)}$ and:
\begin{equation}
\label{nkmm''}
\nu \equiv (l,k,m)\,, \qquad l\in \N\,, \quad 0\leqslant k\leqslant \left\lfloor \frac{l}{2}\right\rfloor\,, \qquad 2k-l\leqslant m \leqslant l-2k\,,
\end{equation}
allow to build a reproducing kernel space as is presented in the next subsection.

\subsection{Bergman kernel and holomorphic basis for $\mathcal{D}^{(3)}$}
The Bergman kernel for $\mathcal{D}^{(3)}$ is deduced from \eqref{Kpzbpz} and
\eqref{VD3}:
\begin{equation} \label{KpzbpzApp}
K(\pz,\overline{\pz^{\prime}}) = \frac{1}{V}[\det (1 + \pz\,\overline{\pz^{\prime}})]^{-3} \,, \qquad 
V = \int_{\mathcal{D}^{(3)}}\,d{\pz} = \frac{\pi^3}{24}\,.
\end{equation}
For the general case, considering the above materials/notations of this appendix \ref{appendix:SpecialFunctions}, in particular, 
\eqref{D3 basis1111111}, we obtain:
\begin{equation} \label{expbergker}
K(\pz,\overline{\pz^{\prime}}) = \frac{1}{V} \sum_{\nu} \mathrm{F}^{\lambda=3}_{\nu}(\pz) \, \overline{\mathrm{F}^{\lambda=3}_{\nu}(\pz^\prime)}\,,
\end{equation}
with:
\begin{equation} \label{holfphinu}
\mathrm{F}^{\lambda=3}_{\nu}(\pz) \equiv \mathrm{F}^{\lambda=3}_{l,k,m}(\pz) \equiv \sqrt{a_{\lambda=3,l,k}} \, (\pz\cdot\pz)^{k}\, \mathcal{Y}_{l-2k,m}(\pz)\,.
\end{equation}
Therefore, we write the orthogonality formula for the holomorphic spherical harmonics on $\mathcal{D}^{(3)}$:
\begin{eqnarray} \label{orthiholsh}
\int_{\mathcal{D}^{(3)}} \overline{\left(\pz\cdot\pz\right)}^{k}\,\overline{ \mathcal{Y}_{l-2k,m}(\pz)}\, \left(\pz\cdot\pz\right)^{k^{\prime}}\mathcal{Y}_{l^{\prime}-2k^{\prime},m^{\prime}}(\pz)\,d{\pz} &=& \frac{V}{a^{}_{\lambda=3,l,k}} \; \delta_{ll^{\prime}}\,\delta_{kk^{\prime}}\,\delta_{mm^{\prime}} \nonumber\\
&=& \frac{\pi^2}{32}\,\frac{k!\,\Gamma\left(l-k+\frac{3}{2}\right)}{\Gamma\left(k+\frac{5}{2}\right)\,\Gamma(l-k+3)} \delta_{ll^{\prime}}\,\delta_{kk^{\prime}}\,\delta_{mm^{\prime}} \,.
\end{eqnarray}
From the orthogonality of the spherical harmonics, it is straightforward to check that the functions $\mathrm{F}^{\lambda=3}_{\nu}(u)$, with $u=e^{\ii \theta}\,\hat{\bm{\xi}}$ ($\theta\in[0,\pi]$ and $\hat{\bm{\xi}}\in\mathbb{S}^2$), form an orthonormal system in the Hilbert space $L^2(\mathfrak{S}^{(3)}, d{u})$ of square-integrable functions on the characteristic manifold (Shilov boundary) $\mathfrak{S}^{(3)} = [0,\pi]\times \mathbb{S}^2$ of the domain $\mathcal{D}^{(3)}$.

\subsection{About the Onofri's basis}
In the present discussion, we are going to examine the relation between the chosen basis above, given in terms of the holomorphic functions \eqref{D3 basis1111111}, and its counterpart already introduced by Onofri in his seminal paper \cite{onofri76}. We start again from the expansion \eqref{opop} of the kernel $\left[\det\left( 1 + \pz\,\overline{\pz^{\prime}}\right)\right]^{-\lambda} $ in terms of the Gegenbauer polynomials. 
Here, we appeal to the addition theorem for Gegenbauer polynomials for $\textbf{u}=(\sin\theta\cos\phi,\sin\theta\sin\phi,\cos\theta)$ and $\textbf{u}^\prime = (\sin\theta^\prime\cos\phi^\prime,\sin\theta^\prime\sin\phi^\prime,\cos\theta^\prime)$ belonging to 
${\mathbb{S}}^3$,  with $0\leqslant \theta,\theta^\prime\leqslant\pi$ and $0\leqslant \phi,\phi^\prime<2\pi$, \cite{Magnus}:
\begin{eqnarray}
C_l^{\lambda}(\textbf{u}\cdot\textbf{u}^\prime) &=& C_l^{\lambda}\big(\cos\theta\cos\theta^\prime + \sin\theta\sin\theta^\prime\cos(\phi-\phi^\prime)\big) \nonumber\\
&=& \frac{\Gamma(2\lambda-1)}{\big[ \Gamma(\lambda) \big]^2} \sum_{m=0}^l 2^{2m}\, \frac{\Gamma(l-m+1)\, \big[ \Gamma(\lambda+m)\big]^2}{\Gamma(l+2\lambda+m)}\, (2\lambda+2m-1)\, (\sin\theta)^m\, (\sin\theta^\prime)^m \nonumber\\
&&\hspace{4cm} \times\, C^{\lambda+m}_{l-m}(\cos\theta)\, C^{\lambda+m}_{l-m}(\cos\theta^\prime)\, C^{\lambda-1/2}_{m}\big(\cos(\phi-\phi^\prime)\big)\,,
\end{eqnarray}
while we have in mind that:
\begin{eqnarray}
C^{\lambda-1/2}_{m}\big(\cos(\phi-\phi^\prime)\big) &=& \frac{1}{\big[ \Gamma(\lambda-\frac{1}{2}) \big]^2} \sum_{n=0}^m \frac{\Gamma(\lambda-\frac{1}{2}+n)\, \Gamma(\lambda-\frac{1}{2}+m-n)}{n!\, (m-n)!} \nonumber\\
&& \hspace{3.5cm} \times \, \frac{1}{2} \left( e^{\ii(m-2n)(\phi-\phi^\prime)} + e^{-\ii(m-2n)(\phi-\phi^\prime)} \right)\,, \\
&& \mbox{or equivalently:} \nonumber\\
&=& \frac{1}{\big[ \Gamma(\lambda-\frac{1}{2}) \big]^2} \sum_{n=0}^m \frac{\Gamma(\lambda-\frac{1}{2}+n)\, \Gamma(\lambda-\frac{1}{2}+m-n)}{n!\, (m-n)!} \, e^{\ii(m-2n)(\phi-\phi^\prime)}\,.
\end{eqnarray}
Next, the holomorphic solid extension goes through the replacements:
\begin{align}
\cos\theta \;\mapsto\; \frac{z^3}{(\pz\cdot\pz)^{1/2}}\,, \qquad \sin\theta\, e^{\pm\ii\phi} \;\mapsto\; \frac{z^1 \pm \ii z^2}{(\pz\cdot\pz)^{1/2}}\,, \qquad \sin^2\theta \;\mapsto\; \frac{(z^1+\ii z^2)(z^1-\ii z^2)}{(\pz\cdot\pz)}\,, \qquad e^{2\ii\phi} \;\mapsto\; \frac{z^1+\ii z^2}{z^1-\ii z^2}\,,
\end{align}
based upon which we obtain:
\begin{eqnarray}
C_l^{\lambda}\left(\frac{\pz\cdot\pz^\prime}{\left[(\pz\cdot\pz)(\pz^\prime\cdot\pz^\prime)\right]^{1/2}}\right) &=& \frac{\Gamma(2\lambda-1)}{\big[ \Gamma(\lambda)\, \Gamma(\lambda-\frac{1}{2}) \big]^2} \nonumber\\
&& \quad\times\, \sum_{m=0}^l \sum_{n=0}^m 2^{2m}\, \frac{\Gamma(l-m+1)\, \big[ \Gamma(\lambda+m)\big]^2\, \Gamma(\lambda-\frac{1}{2}+n)\, \Gamma(\lambda-\frac{1}{2}+m-n)}{\Gamma(l+2\lambda+m)\, n!\, (m-n)!} \nonumber\\
&&\quad \times\, (2\lambda+2m-1)\, (z^1+\ii z^2)^{m-n}\, (z^1-\ii z^2)^{n}\, (z^{\prime\,1}-\ii z^{\prime\,2})^{m-n}\, (z^{\prime\,1}+\ii z^{\prime\,2})^{n}\, \nonumber\\
&&\quad \times\, (\pz\cdot\pz)^{-m/2}\, C^{\lambda+m}_{l-m}\left(\frac{z^3}{(\pz\cdot\pz)^{1/2}}\right)\, (\pz^\prime\cdot\pz^\prime)^{-m/2}\, C^{\lambda+m}_{l-m}\left(\frac{z^{\prime\,3}}{(\pz^\prime\cdot\pz^\prime)^{1/2}}\right) \,.
\end{eqnarray}
Considering the above, the expansion \eqref{opop} can be rewritten as:
\begin{align}
\left[\det\left( 1 + \pz\,\overline{\pz^{\prime}}\right)\right]^{-\lambda} = \sum_{l=0}^{\infty} \sum_{m=0}^l \sum_{n=0}^m \, \varphi^{}_{lmn}(\pz) \, \overline{\varphi^{}_{lmn}(\pz^\prime)}\,,
\end{align}
where:
\begin{eqnarray} \label{Onofri'}
\varphi^{}_{lmn}(\pz) &=& 2^{m}\, \frac{\Gamma(\lambda+m)}{\Gamma(\lambda)\, \Gamma(\lambda-\frac{1}{2})}\, \left(\frac{\Gamma(2\lambda-1)\, \Gamma(l-m+1)\, \Gamma(\lambda-\frac{1}{2}+n)\, \Gamma(\lambda-\frac{1}{2}+m-n)\, (2\lambda+2m-1)}{\Gamma(l+2\lambda+m)\, n!\, (m-n)!}\right)^{1/2} \nonumber\\
&&\quad \times\, (z^1+\ii z^2)^{m-n}\, (z^1-\ii z^2)^{n}\, (\pz\cdot\pz)^{(l-m)/2}\, C^{\lambda+m}_{l-m}\left(\frac{z^3}{(\pz\cdot\pz)^{1/2}}\right)\,.
\end{eqnarray}
Here, if one desires to compare the above result with the Onofri's (see Ref. \cite{onofri76}, Eqs. (54) and (55)):
\begin{itemize}
\item{First, one should invoke the Legendre duplication formula, based upon which we have:
\begin{eqnarray}
\Gamma(\lambda) \, \Gamma(\lambda-1/2) = 2^{2-2\lambda}\, \sqrt{\pi}\, \Gamma(2\lambda-1)\,.
\end{eqnarray}}
\item{Second, one should consider the following map:
\begin{eqnarray}
\lambda \,\mapsto\, \acute{l}+\frac{1}{2}\,, \quad l \,\mapsto\, \acute{s}\,, \quad m \,\mapsto\, \acute{k}\,, \quad n \,\mapsto\, \frac{\acute{k}-\acute{m}}{2}\,, \quad \big(\mbox{while,}\,\, \acute{k}-\acute{m} = \mbox{even} \big)\,,
\end{eqnarray}
where, with respect to the allowed ranges of the former set of parameters, the latter set of parameters takes the values such that:
\begin{align}\label{1}
\acute{s}=0,1,2,\dotsc\,, \qquad \acute{k}=0,1,2, \dotsc, \acute{s}\,, \qquad \acute{m}= -\acute{k}, -\acute{k}+1, \dotsc, -1,0,1,\dotsc, \acute{k}-1, \acute{k}\,,
\end{align}
or, equivalently and in consistency with the Onofri's result, such that:
\begin{align}\label{2}
\acute{s}=0,1,2,\dotsc\,, \qquad \acute{m}= -\acute{s}, -\acute{s}+1, \dotsc, -1,0,1,\dotsc, \acute{s}-1, \acute{s}\,, 
\qquad \acute{k}=0,1,2, \dotsc, |\acute{m}|\,, 
\end{align}
in both cases, \eqref{1} and \eqref{2}, $\acute{k}-\acute{m} = \mbox{even} $. Note that the equivalence of \eqref{1} and \eqref{2} can be easily checked numerically for a given value of $\acute{s}$; let set $\acute{s}=1$, then:
\begin{eqnarray}
\mbox{from \eqref{1}}&:& \;\; \acute{k}=0 \,\Rightarrow\, \acute{m}=0\,, \quad \mbox{and} \quad  \acute{k}=1 \,\Rightarrow\, \acute{m}=-1,1\,,\nonumber\\
\mbox{from \eqref{1}}&:& \;\; \acute{m}=0 \,\Rightarrow\, \acute{k}=0\,, \quad \mbox{and} \quad \acute{m}=-1,1 \,\Rightarrow\, \acute{k}=1\,. \nonumber
\end{eqnarray}}
\end{itemize}
Accordingly, one can bring Eq. \eqref{Onofri'} into the form already introduced by Onofri (Eq. (55)):
\begin{eqnarray} \label{Onofri}
\varphi^{}_{\acute{s} \acute{m} \acute{k}}(\pz) &=& \frac{2^{2\acute{l}+\acute{k}-1}}{\sqrt{\pi\, \Gamma(\,\underline{\underline{2}}\,\acute{l}\,)}}\, \Gamma(\acute{l}+1/2+\acute{k})\, \left(\frac{(\acute{s}-\acute{k})!\, 2(\acute{l}+\acute{k})}{\Gamma(\acute{s}+\acute{k}+2\acute{l}+1)}\right)^{1/2} \, \left(\frac{\Gamma\big(\acute{l}+(\acute{k}-\acute{m})/2\big)\, \Gamma\big(\acute{l}+(\acute{k}+\acute{m})/2\big)}{\big[(\acute{k}-\acute{m})/2\big]!\, \big[(\acute{k}+\acute{m})/2\big]!}\right)^{\underline{\underline{1/2}}} \nonumber\\
&&\quad \times\, (z^1+\ii z^2)^{(\acute{k}+\acute{m})/2}\, (z^1-\ii z^2)^{(\acute{k}-\acute{m})/2}\, (\pz\cdot\pz)^{(\acute{s}-\acute{k})/2}\, C^{\acute{l}+\acute{k}+1/2}_{\acute{s}-\acute{k}}\left(\frac{z^3}{(\pz\cdot\pz)^{1/2}}\right)\,.
\end{eqnarray}
Note that this result exactly coincides with the Onofri's, with the exception of two items identified above by drawing a double line below them.




\end{document}